%% file: dusty_young_universe_catalog_revised.tex
\documentclass[iop,revtex4,numberedappendix,letterpaper]{emulateapj}
\usepackage[pass]{geometry}
\usepackage{lscape}
\usepackage{comment}

\shorttitle{Dust emission of QSOs at $z>5$}
\shortauthors{Leipski et al.}

\begin{document}


\title{Spectral Energy Distributions of QSOs at $z>5$: common 
AGN-heated dust and occasionally strong star-formation. }


\author{C. Leipski\altaffilmark{1}}
\author{K. Meisenheimer\altaffilmark{1}}
\author{F. Walter\altaffilmark{1}}
\author{U. Klaas\altaffilmark{1}}
\author{H. Dannerbauer\altaffilmark{2}}
\author{G. De Rosa\altaffilmark{3}}
\author{X. Fan\altaffilmark{4}}
\author{M. Haas\altaffilmark{5}}
\author{O. Krause\altaffilmark{1}}
\author{H.-W. Rix\altaffilmark{1}}

\altaffiltext{1}{Max-Planck Institut f\"ur Astronomie (MPIA),
     K\"onigstuhl 17, D-69117 Heidelberg, Germany; email: {\tt leipski@mpia-hd.mpg.de}}
\altaffiltext{2}{Universit\"at Wien, Institut f\"ur Astrophysik, T\"urkenschanzstra{\ss}e 17, 1180 Wien, Austria}
\altaffiltext{3}{Department of Astronomy, The Ohio State University, 140 West 18th Avenue, Columbus, OH 43210, USA}
\altaffiltext{4}{Steward Observatory, University of Arizona, Tucson, AZ 85721, USA}
\altaffiltext{5}{Astronomisches Institut Ruhr-Universit\"at Bochum, Universit\"atsstra{\ss}e
     150, D-44801 Bochum, Germany}

\begin{abstract}

We present spectral energy distributions (SEDs) of 69 QSOs at
$z>5$, covering a rest frame wavelength range of 0.1\,$\mu$m to
$\sim$80\,$\mu$m, and centered on new {\it Spitzer} and {\it Herschel} observations. 
The detection rate of the QSOs with {\it Spitzer} 
is very high (97\% at $\lambda_{\rm rest}\,\lesssim\,4\,\mu$m), but drops
towards the {\it Herschel} bands with 30\%  detected in 
PACS (rest frame mid-infrared) and 15\% additionally in the
SPIRE (rest frame far-infrared; FIR).  
We perform multi-component SED fits 
for {\it Herschel}-detected objects and confirm that to match the
observed SEDs, a clumpy torus model needs to be complemented by a hot
($\sim$1300K) component and, in cases with prominent FIR emission,
also by a cold ($\sim$50K) component. 
In the FIR detected cases the
luminosity of the cold component is on the order
of $10^{13}$ $L_{\odot}$ which is likely heated by star formation. 
From the SED fits we also determine that the AGN dust-to-accretion disk
luminosity ratio declines with UV/optical luminosity. 
Emission from hot ($\sim$1300K) dust is common in our sample, showing that nuclear dust 
is ubiquitous in luminous QSOs out to redshift 6.  However, about 
15\% of the objects appear under-luminous in the near infrared compared to
their optical emission and seem to be deficient in (but not 
devoid of) hot dust. Within our full sample, the QSOs detected with {\it Herschel} are found at the
high luminosity end in $L_{\rm UV/opt}$ and $L_{\rm NIR}$ and show low
equivalent widths (EWs) in H$\alpha$ and in Ly$\alpha$. In the
distribution of H$\alpha$ EWs, as determined from the {\it Spitzer} 
photometry, the high-redshift QSOs show little difference to low redshift AGN.

\end{abstract}





\section{Introduction}

%\begin{comment}

High-redshift quasars are powerful probes for the early evolution of
black holes and their host galaxies. Even less than a billion
years after the Big Bang they already have inferred black-hole  masses of
the order of 10$^8$ to 10$^9$ M$_{\odot}$
\citep[e.g.,][]{wil03,kur07,jia07}.  The metallicities of their nuclear
emission-line gas is about solar, without significant redshift
evolution \citep[e.g.,][]{mai03,fre03,jia07,jua09,der11}, which
indicates fast metal enrichment of the interstellar gas, at least in
the circumnuclear region of the quasar host galaxy.

The remarkable similarity in the rest frame UV spectra with their
lower-redshift analogs  appears to extend into the near-infrared
(NIR): {\it Spitzer}  observations of a number of high-redshift
quasars revealed the presence  of hot dust, which indicates that the
nuclear structures governing the  shape of the optical/NIR spectral
energy distribution (SED) of luminous  quasars are in place already at
$z\sim6$ \citep[e.g.,][]{hin06,jia06,jia10}.  At the long wavelength
end of the thermal dust emission spectrum,  $\sim$30\% of the the
known quasars at $z\gtrsim5.7$ show prominent  submm/mm emission
\citep[e.g.,][]{ber03,wan08b,wan10}, which has been attributed to 
dust heated by star formation.

However, comprehensive studies of the dust SED in $z>5$ QSOs, including the 
diagnostically important rest frame mid-infrared (MIR), 
have been missing so far. 
 The spectral shape in the NIR and MIR  may hold clues on
the range of dust temperatures and the dust distribution in the
central parsecs of the objects and may provide insight into the
heating source  of the cooler dust (AGN versus star formation). In
order to explore these questions  we here present {\it Spitzer} 
\citep{wer04} and
{\it Herschel}\footnote{{\it Herschel} is an ESA space  observatory
with science instruments provided by European-led Principal
Investigator consortia and with important participation from NASA.}
\citep{pil10} observations of 69 quasars  at $z>5$. In combination 
with literature data, these new observations provide comprehensive 
SEDs of luminous quasars in the early universe covering the rest 
frame wavelengths  from 0.1\,$\mu$m to $\sim$80\,$\mu$m.

In Section 2 we present our sample and outline the available data as 
well as the observations and the data reduction. The detection rates in the 
{\it Spitzer} and {\it Herschel} bands are described in Section 3. In 
Section 4 we focus on the analysis and discussion. A summary and conclusions 
follow in Section 5. Throughout the paper we use a $\Lambda$CDM 
cosmology with $H_0=71$~km~s$^{-1}$~Mpc$^{-1}$, $\Omega_{{\rm m}} = 0.27$, and
$\Omega_{\Lambda} = 0.73$.

\begin{large}
\input{latex_table_obslog_astroph.tex}
\end{large}
%\end{comment}

\section{Data}

\subsection{Sample}
The parent sample for this study consisted of all quasars with 
redshift $z>5$  that were known at the time of submission of the 
original {\it Herschel} proposal (early 2007). Due to various factors (e.g. revised 
sensitivity estimates after the launch of {\it Herschel}, sparse 
supplemental data coverage, revised redshifts, uncertain identifications) a 
small number of sources was subsequently removed from the target list. The 
final sample includes 69 quasars at $z>5$, all of which have been observed 
with {\it Herschel} in five bands. For 68 of them we also 
present {\it Spitzer} photometry in five bands. 

Most of the quasars in our sample come from the Sloan Digital Sky Survey 
(SDSS), either from the main survey or from the deeper Stripe 82. A 
small complement of objects consists of serendipitously discovered 
high-redshift quasars \citep[][]{sha01,rom04,mah05,mcg06}. The 
final sample is presented in Table\,\ref{obslog}, along with an observation 
log for the {\it Herschel} data. In this table we 
give the full name of the source. For the following tables, figures and 
in the text we use abbreviated source names in the format of J$hhmm \pm ddmm$.

\input{latex_table_nir_magnitudes.tex}

\subsection{UV continuum flux}
The UV continuum brightness of high-redshift quasars is typically indicated by 
their monochromatic flux at a rest frame wavelength of 1450\,\AA, often 
expressed in terms of apparent AB magnitudes. We have compiled these values 
for our quasars from the literature and report them in Table\,\ref{obslog}. 
Two objects (J0841+2905 and J2245+0024) had only their 
absolute magnitude at 1450\,\AA~(rest frame) given \citep[][]{sha01,got06}. 
For 
these cases we calculated the apparent magnitudes using the world models 
cited in the respective papers.

For objects that did not have mag(1450\AA) available in the literature, we 
retrieved the spectrum from the SDSS data base, corrected it for galactic 
foreground extinction using the map of \citet{sch98} and 
determined mag(1450\AA) from the corrected spectrum 
following the procedure of \citet{fan04}. This approach has been adopted for 
31 objects with $z\leq5.41$ (see Table\,\ref{obslog}).

Where no values for the 1450\,\AA~flux were provided in the literature and no 
spectra where available in electronic form (6 sources, 
see Table\,\ref{obslog}), we scaled a redshifted version 
of the SDSS quasar template spectrum \citep{van01} to match the (extinction 
corrected) z-band magnitude (taking into account the filter curve). From 
the redshifted and scaled template spectrum we then determined 
mag(1450\AA) as in \citet{fan04}. 

\begin{figure*}[b!]
\centering
\includegraphics[angle=0,scale=.15]{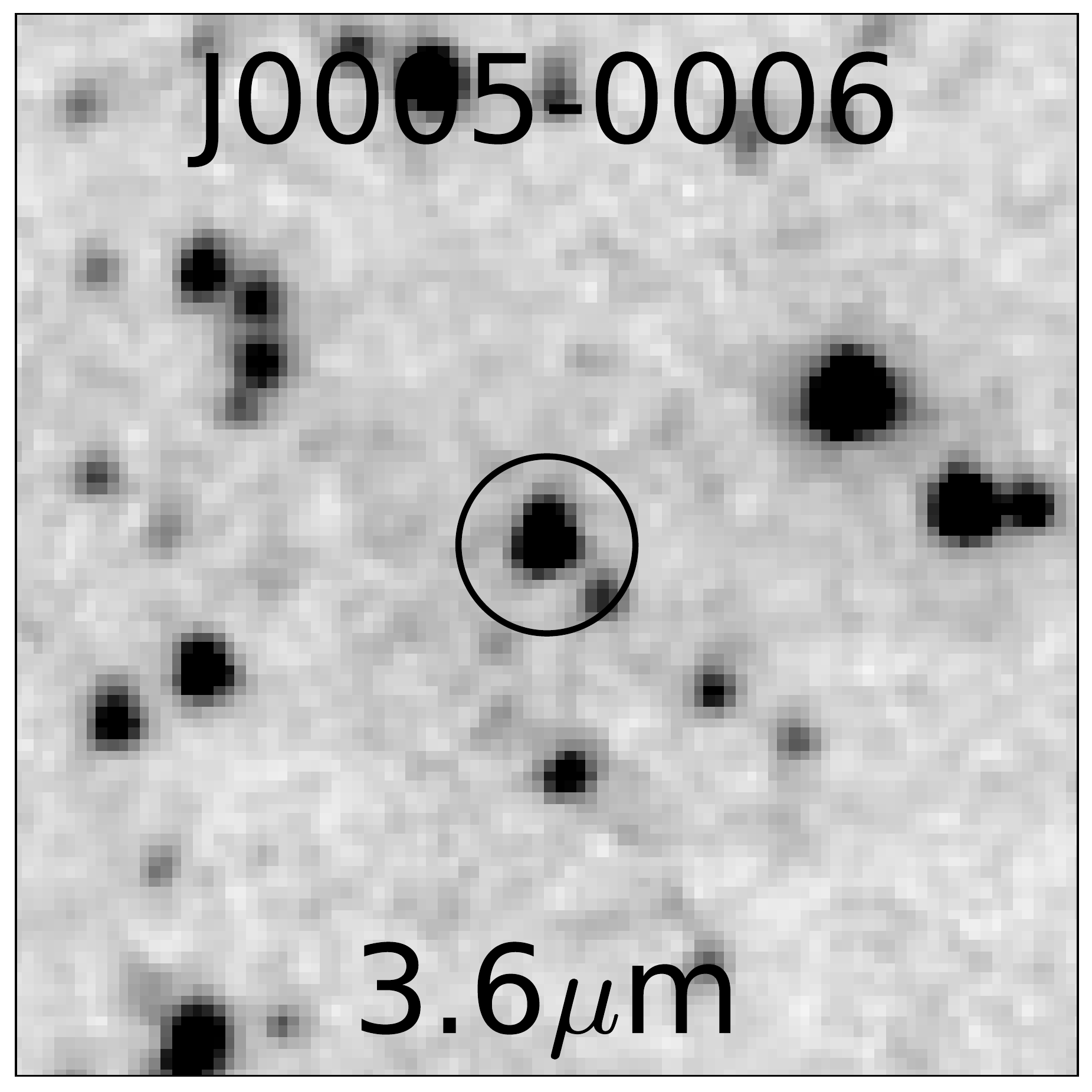}
\includegraphics[angle=0,scale=.15]{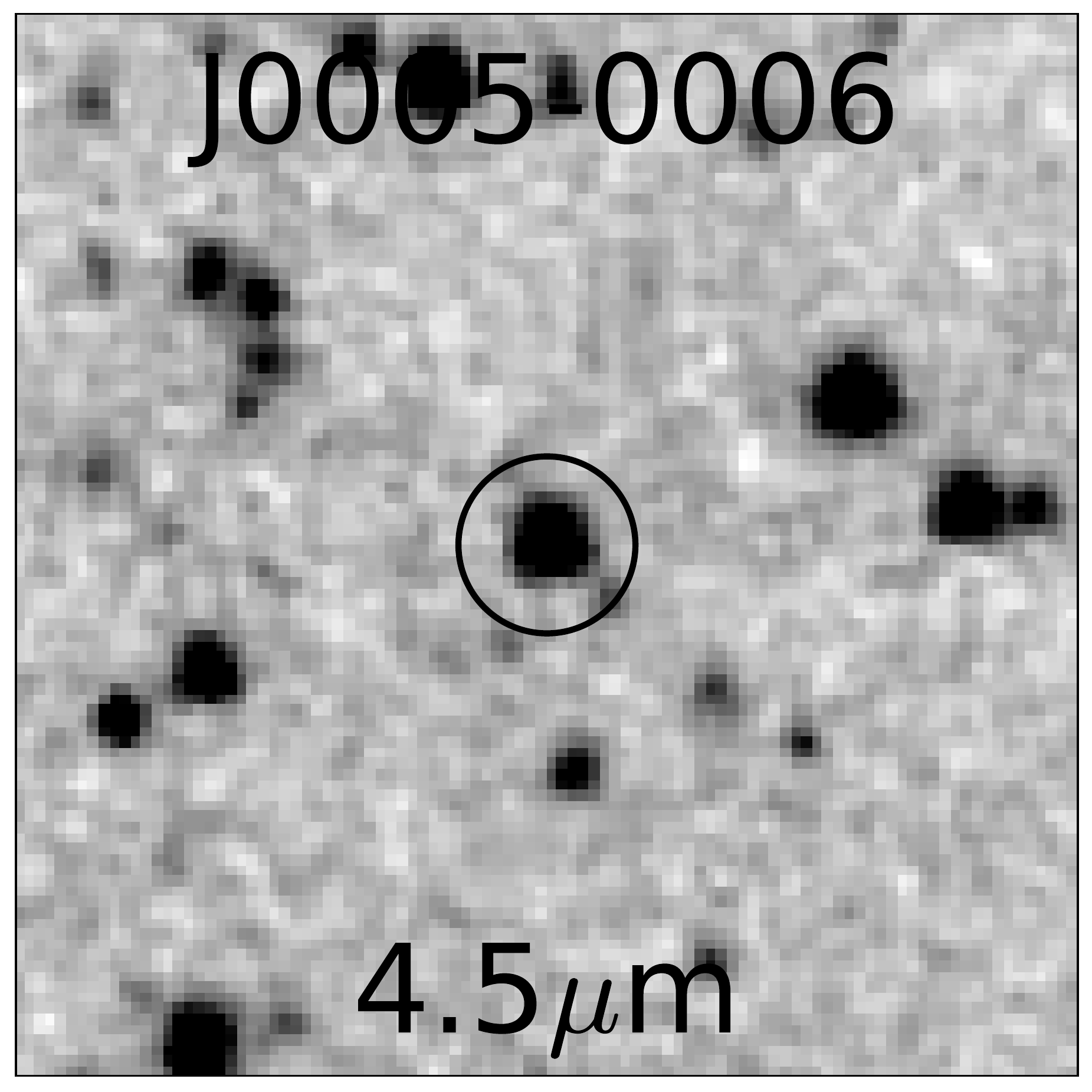}
\includegraphics[angle=0,scale=.15]{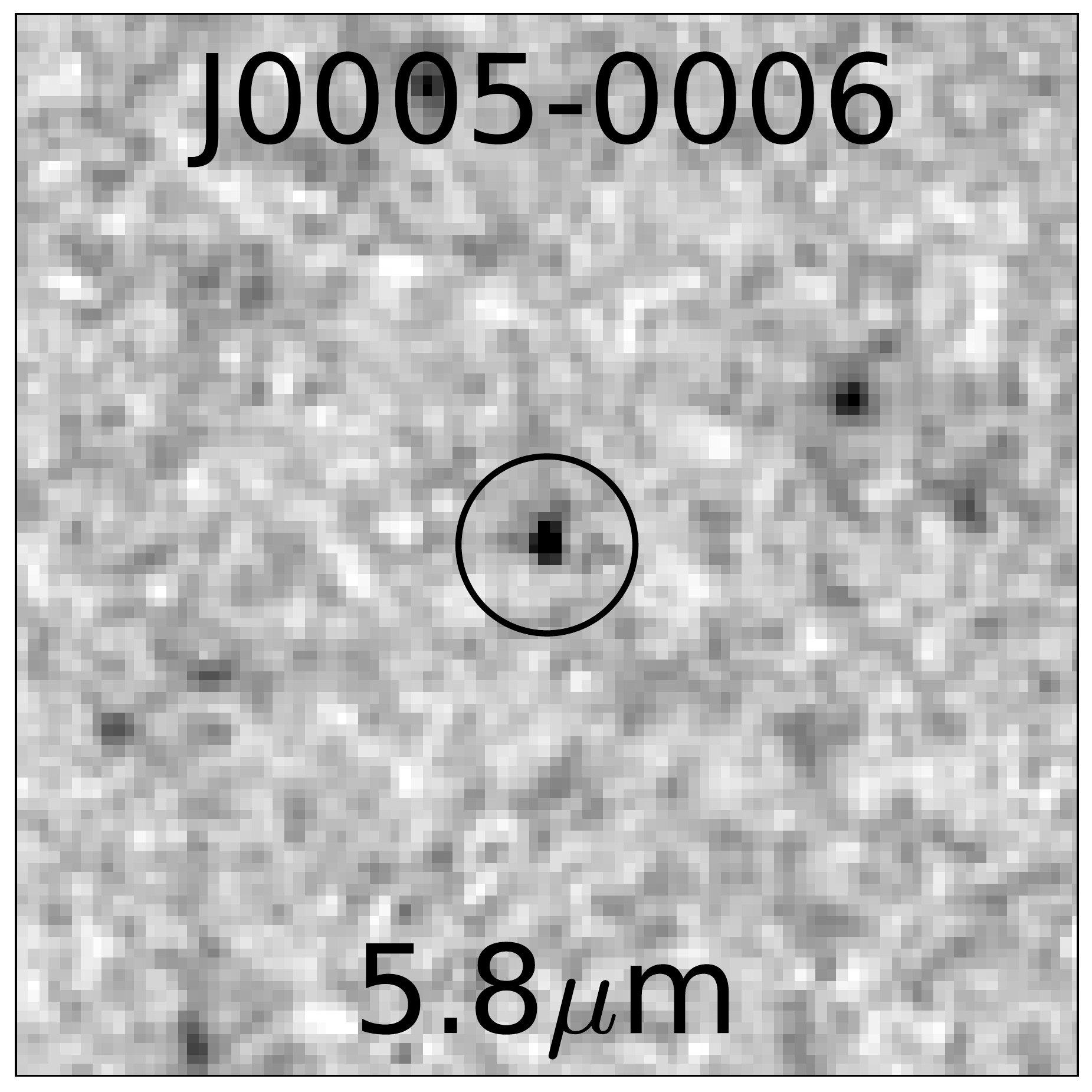}
\includegraphics[angle=0,scale=.15]{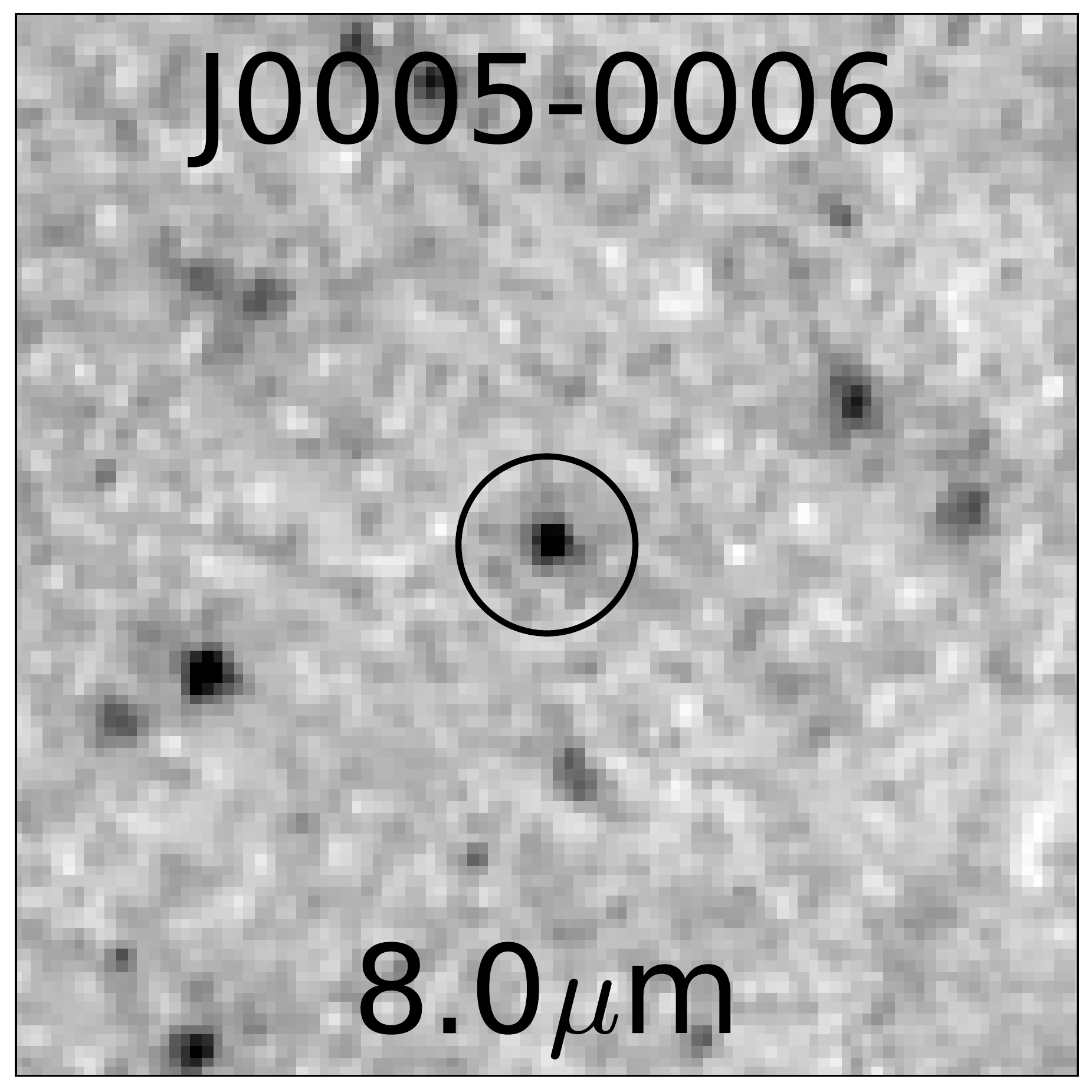}
\includegraphics[angle=0,scale=.15]{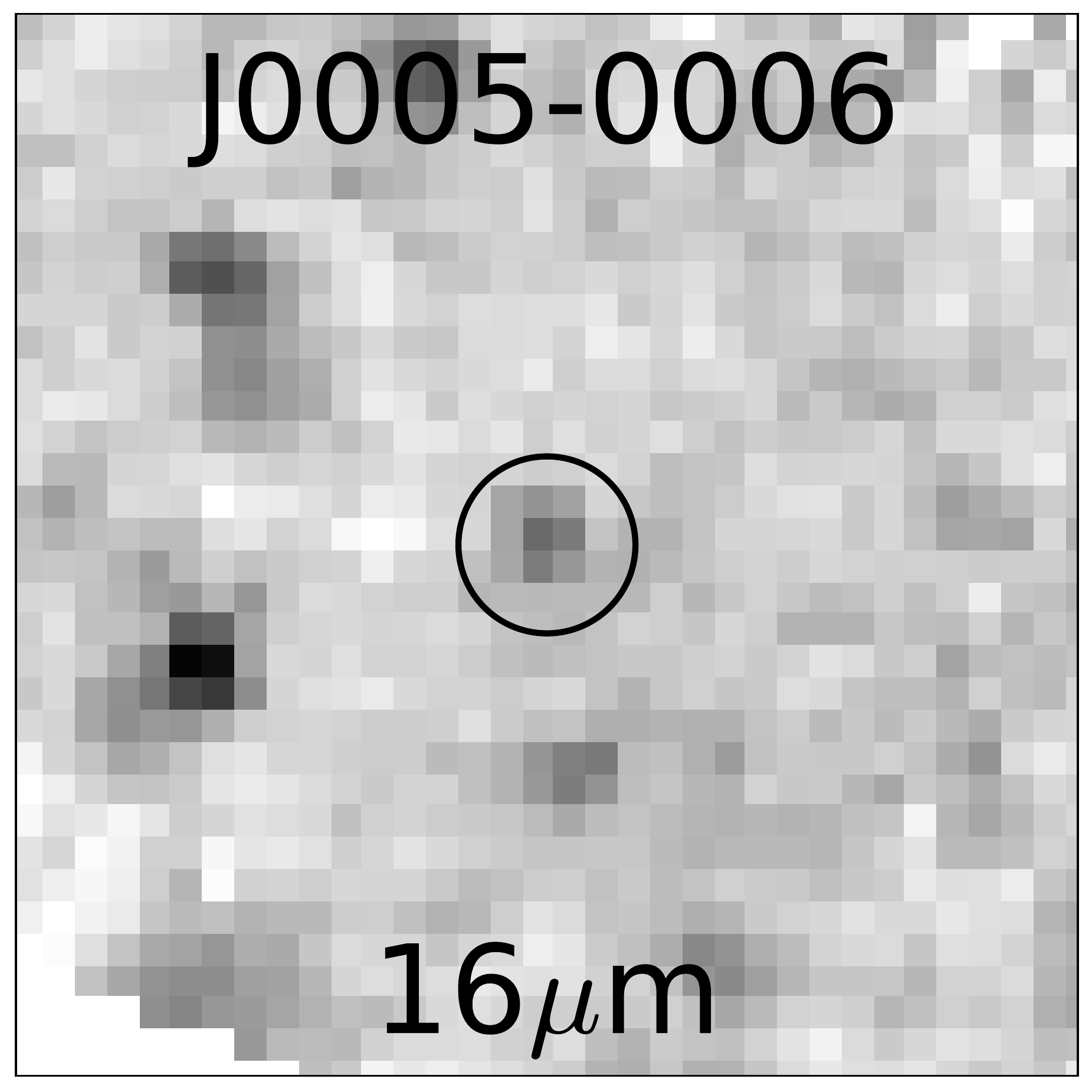}
\includegraphics[angle=0,scale=.15]{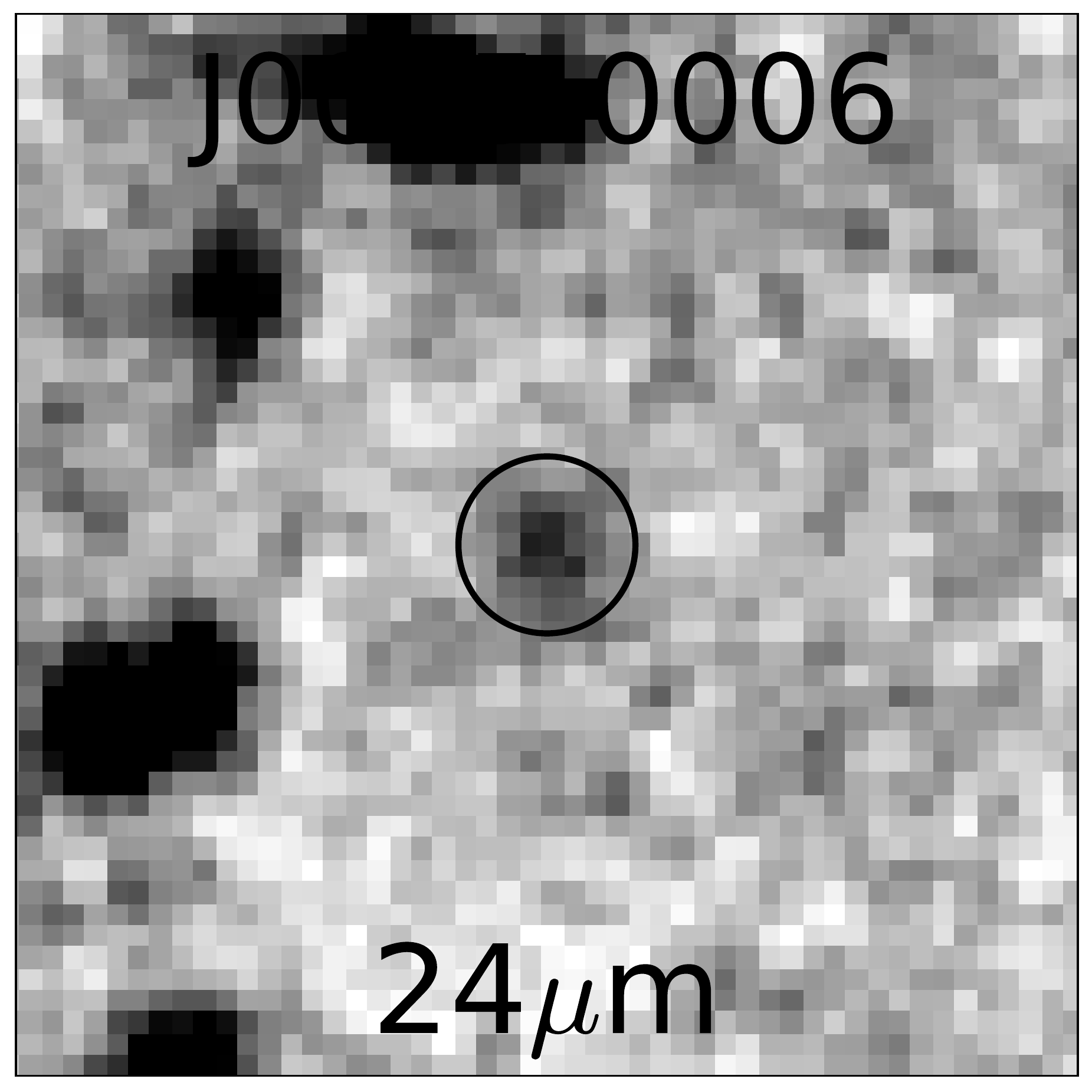}\\
\includegraphics[angle=0,scale=.15]{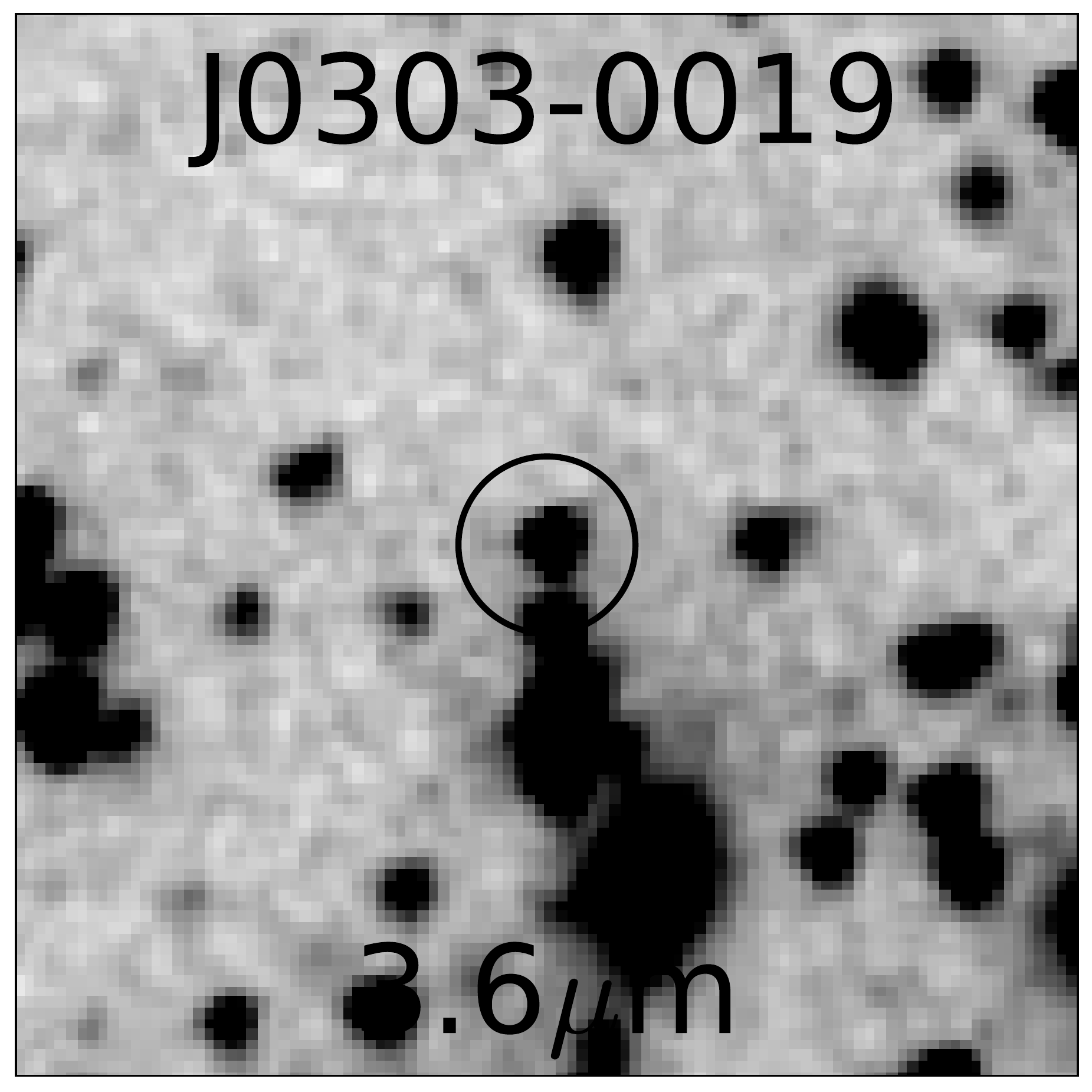}
\includegraphics[angle=0,scale=.15]{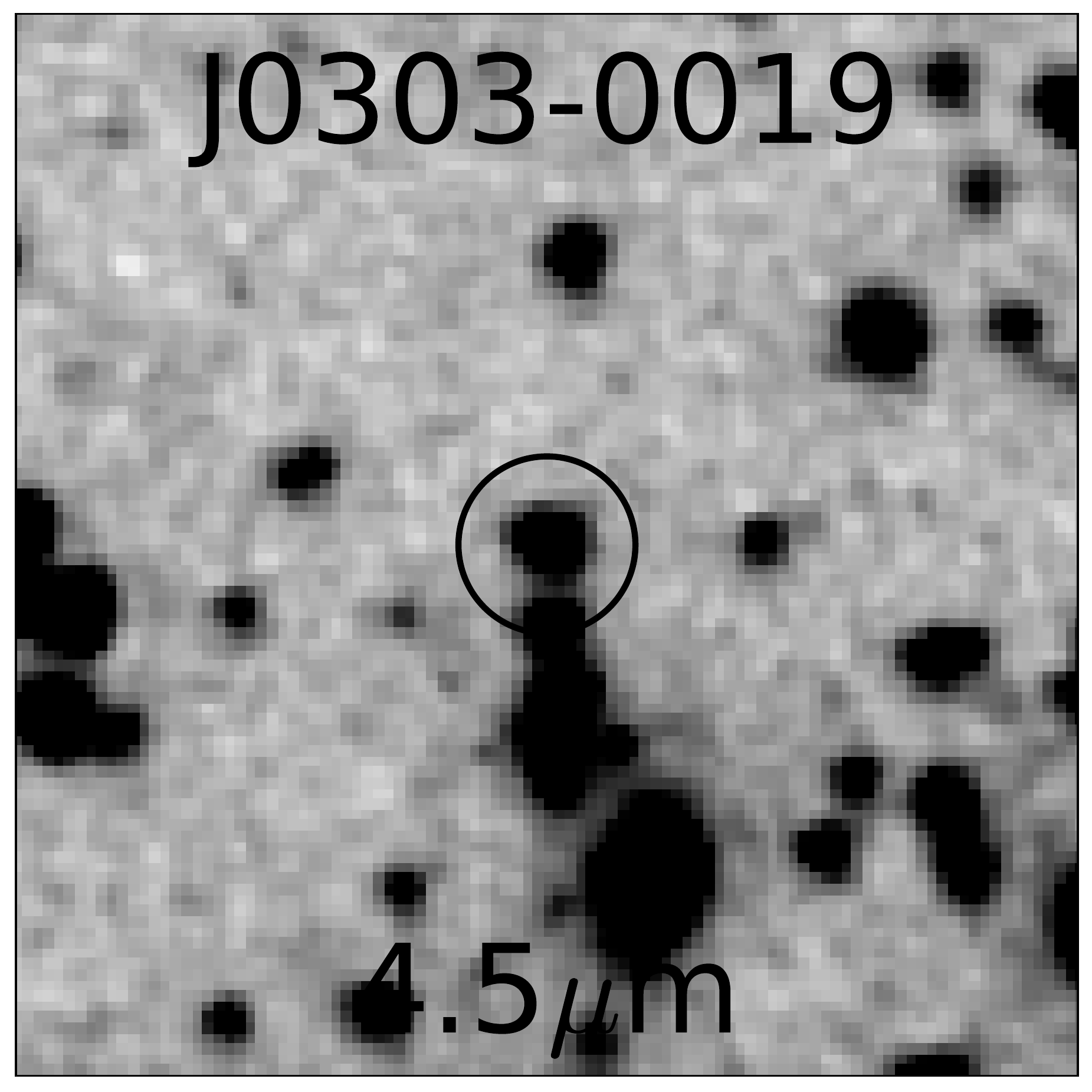}
\includegraphics[angle=0,scale=.15]{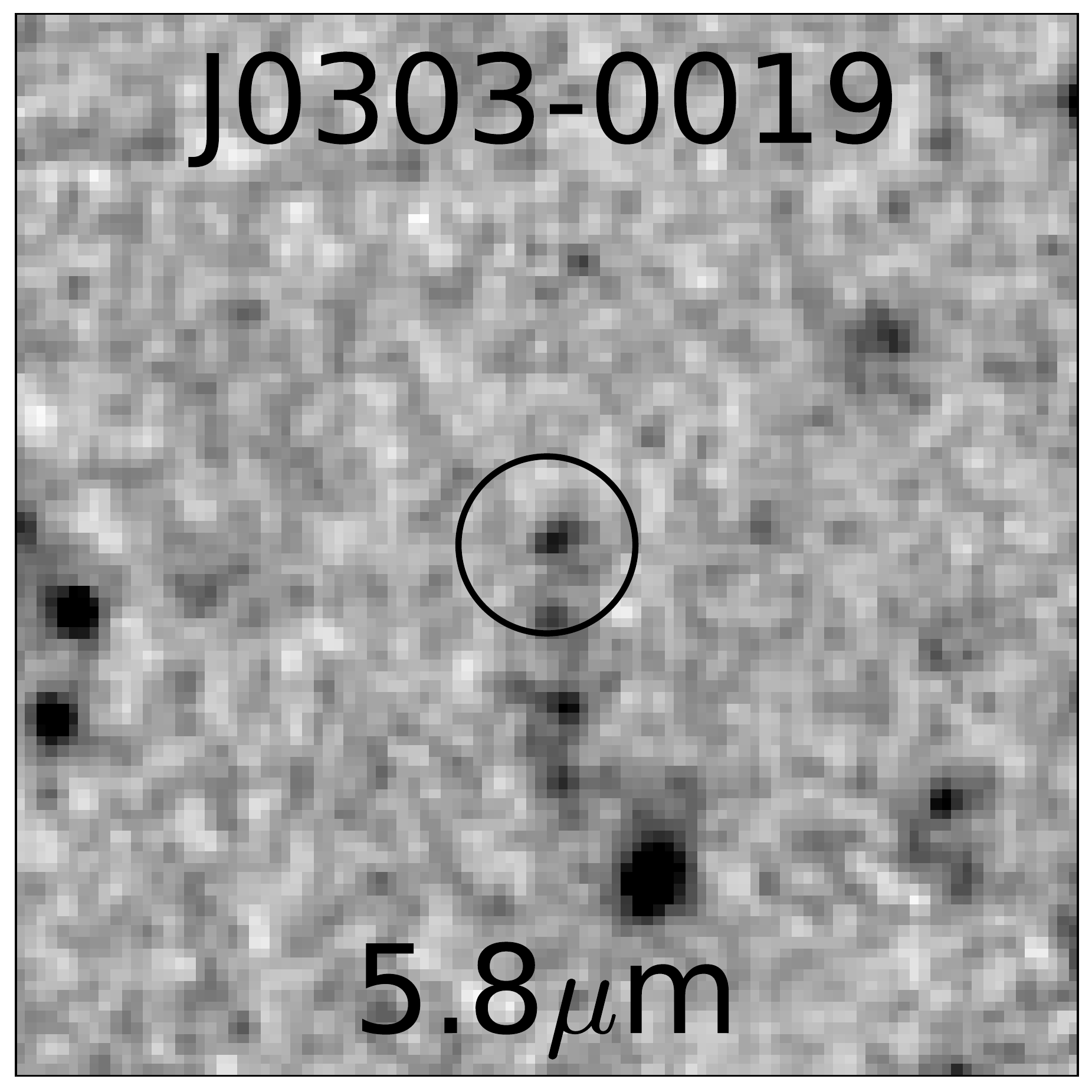}
\includegraphics[angle=0,scale=.15]{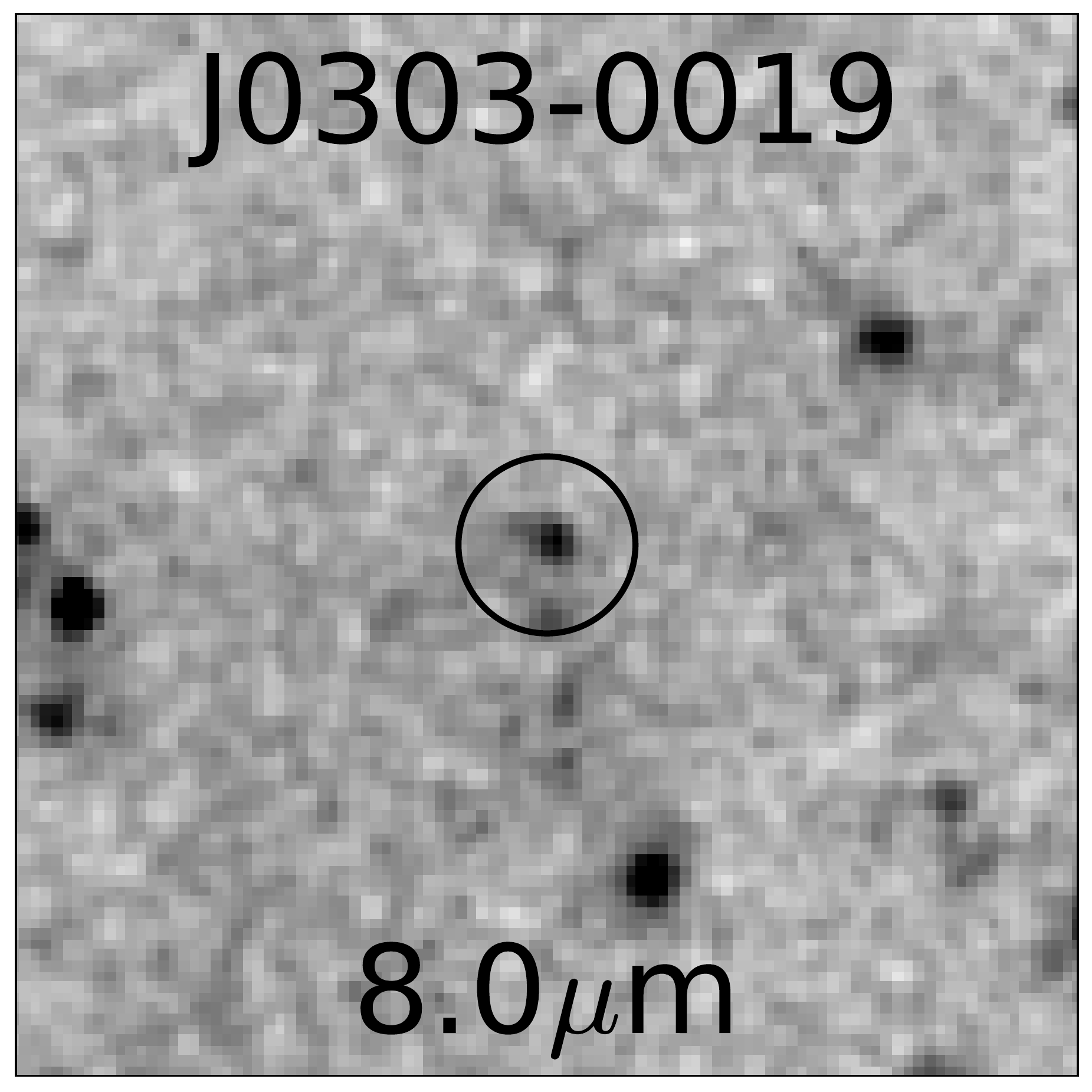}
\includegraphics[angle=0,scale=.111]{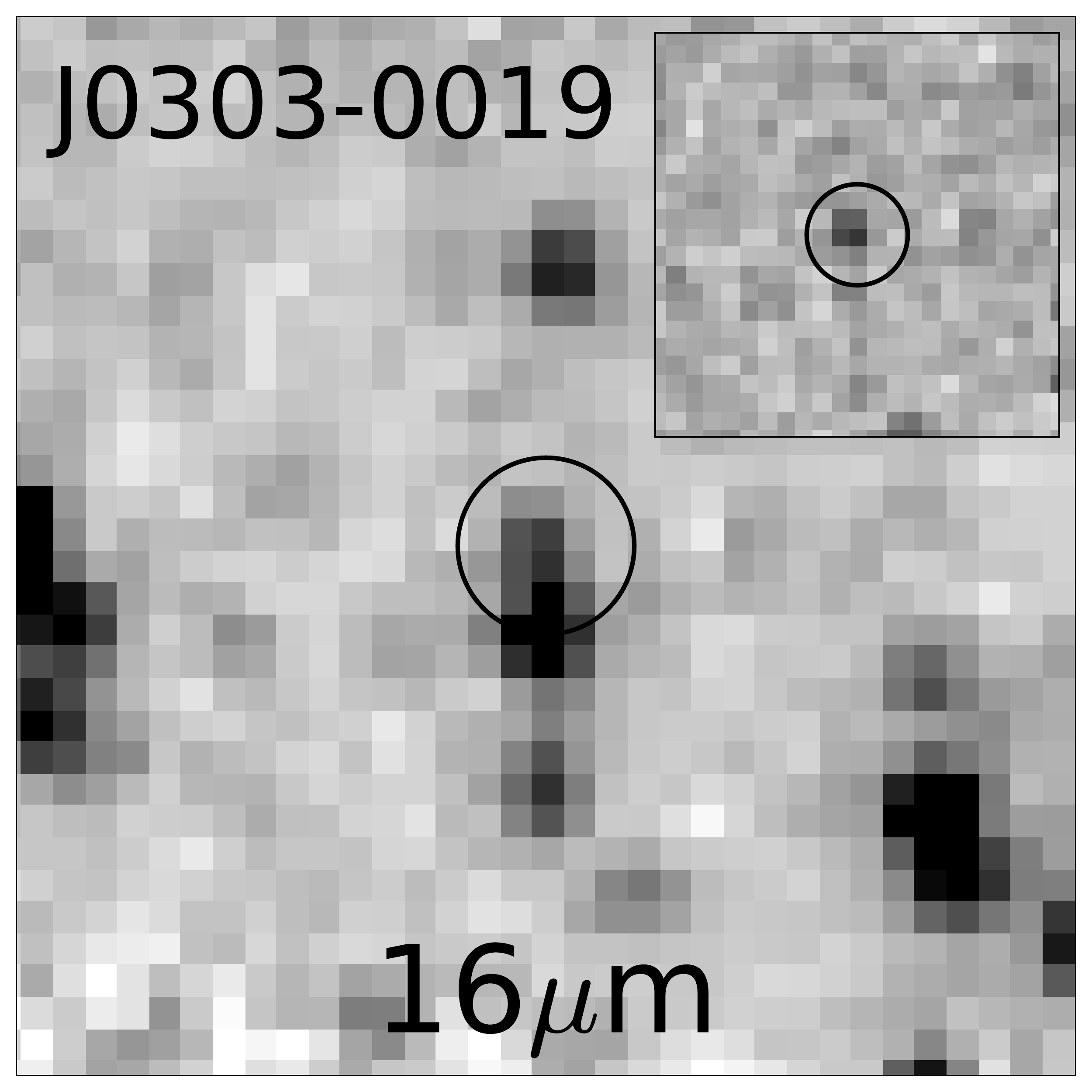}
\includegraphics[angle=0,scale=.111]{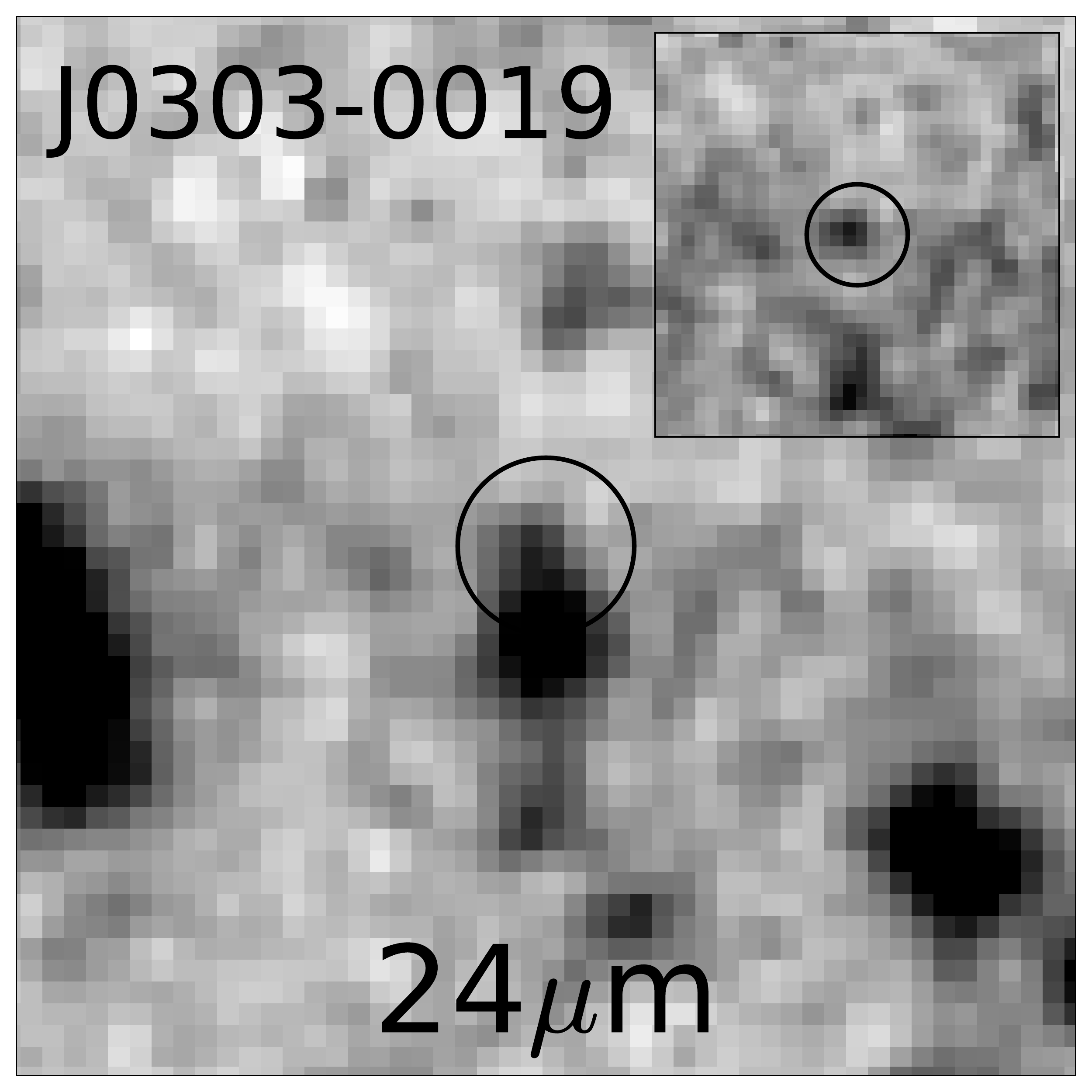}
\caption{{\it Spitzer} images of the two quasars previously undetected in the 
longer {\it Spitzer} bands: J0005$-$0006 (top) and J0303$-$0019 (bottom). 
All panels show an area of 1\arcmin$\times$1\arcmin. The circle 
indicating the quasar position has a diameter of 10\arcsec. For J0303$-$0019 the 
inset at the two longest wavelengths show a 40\arcsec$\times$40\arcsec~subimage 
around the quasar position after subtracting the bright confusing source (see text). 
This confusing source is also visible in the IRAC frames where it is well 
separated from the quasar itself.\label{dust_free} }
\end{figure*}

\subsection{Photometry in z and y bands}

We also compiled z-band and y-band photometry for the majority of the 
sample (see Table\,\ref{nir_fluxes}). For most of the quasars, 
the z-band photometry is taken from SDSS or from the discovery papers, which 
sometimes presented deeper observations. 

The y-band photometry was mainly provided by Pan-STARRS
\citep{kai10}, complemented by data from UKIDSS \citep{law07}. 
For objects where the NIR photometry (see below) was taken from UKIDSS, we 
also used the y-band flux from this survey, for consistency. In most
cases where y-band data exist 
from both surveys they agree within the combined errors.

%Note that at redshifts $z \gtrsim 5.5$ the onset of the Gunn-Peterson trough 
%moves into the passband of the z-band filter. For the redshift range of our 
%objects ($5.0 \lesssim z \lesssim 6.4$), the y-band is not affected by this. 

\subsection{NIR photometry}

An important source of photometry in the $J$, $H$, and/or $K$ bands were the 
discovery papers (or follow-up work on those). In the majority of cases, 
magnitudes were given in a Vega-based system and were obtained with a 
multitude of instruments across our sample. We here consistently use the 
values given in \citet{hew06} to convert all the Vega-based magnitudes 
into the AB system. The NIR photometry from the literature was 
complemented by photometry from UKIDSS for a sizable fraction of our sample. 
For additional nine objects we obtained $J$-band photometry using Omega2000 
at the 3.5m telescope of the Calar Alto observatory. For the data reduction 
and photometry we followed standard procedures. Magnitudes and the 
corresponding references are reported in Table\,\ref{nir_fluxes}.

\begin{table}[t!]
\begin{center}
\caption{{\it Spitzer} IRS photometry at 16\,$\mu$m.\label{irs}}
\begin{tabular}{l r@{$\pm$}l}
\tableline\tableline
Source & \multicolumn{2}{c}{flux} \\
     & \multicolumn{2}{c}{$\mu$Jy}  \\
(1)  & \multicolumn{2}{c}{(2)}  \\
\tableline
J0005$-$0006    & 24&8    \\
J0303$-$0019    & 59&20   \\
J0353+0104      & 225&30   \\
J0818+1722      & 666&63   \\
J0842+1218      & 610&66   \\
J1137+3549      & 306&31   \\
J1250+3130      & 696&31   \\
J1411+1217      & 120&26   \\
J1427+3312      & 128&15   \\
J2315$-$0023    & 107&20   \\
\tableline
\end{tabular}
%\tablecomments{(1) SDSS name ordered by R.A.; (2) flux measured at 16$\mu$m in $\mu$Jy.}
\end{center}
\end{table}

\subsection{{\it Spitzer}}\label{sec:spitzer}

Mid-infrared imaging from {\it Spitzer} exists for all {\it
  Herschel} targets, with the exception of J2054$-$0005. These data 
consist of observations at 3.6, 4.5,
5.8, and 8.0\,$\mu$m with IRAC \citep{faz04} as well as at
 24\,$\mu$m with MIPS \citep{rie04}. A small number 
of objects was also observed at 16\,$\mu$m with the peak-up array 
of IRS \citep{hou04}.

The {\it Spitzer} data were processed in a standard manner using 
procedures within the MOPEX software package provided by the 
{\it Spitzer} Science Center (SSC). The resulting maps from IRAC and MIPS 
are presented in the Appendix in Figure\,\ref{all_images}. Aperture photometry on the final 
images was performed in IDL. In most cases we used apertures with a
radius of 3.6, 5.4, and 7 arcseconds in IRAC, IRS, and MIPS, respectively. For
some objects the aperture size was reduced to avoid contamination from 
nearby objects. Appropriate aperture corrections were taken from the 
respective instrument handbooks (also available from the SSC website).

Errors on the photometry were determined by measuring the
fluxes in 500 apertures (with sizes identical to the science target
aperture) which were randomly placed on 
source-free regions of the background, avoiding area of low coverage. 
The distribution of these 500 fluxes was fit by a Gaussian. The
sigma of this Gaussian was taken as the 1$\sigma$ uncertainty on the 
photometry. The measured fluxes and uncertainties are presented in 
Table\,\ref{tab_all} for IRAC and MIPS. The additional IRS photometry 
for a small subset of objects is presented in Table\,\ref{irs}. 
We note that some of the {\it Spitzer} data have been published previously 
\citep[e.g.,][]{hin06,jia06,jia10} and in these cases our photometry 
is consistent with the earlier results.

For a few objects, direct aperture photometry (even with smaller
apertures) was difficult to obtain
due to severe blending with neighboring sources. In such cases we
used the point source extraction tool APEX in the MOPEX software
package to subtract the 
confusing source from the science image. We then performed aperture
photometry as described above on the residual image (i.e. where the 
confusing source has been removed) for consistency with the rest of 
the sample.

\subsection{{\it Herschel}}

\subsubsection{PACS}\label{sec:pacs}

We observed all objects at 100\,$\mu$m (green channel) and 160\,$\mu$m 
(red channel) with PACS \citep{pog10}. We employed the
mini-scan map observing template with parameters as recommended in 
the corresponding Astronomical Observation Template 
release note, which includes a combination of two scans with different
scan directions. For each scan direction, five repetitions were 
executed. The total on-source integration time was $\sim$\,900\,s for 
each object.\footnote{For J0005$-$0006 and J0303$-$0019, which had
  previously been dubbed 'dust-free quasars' \citep{jia10}, we chose
  to execute nine repetitions for each scan direction, which
  translates into $\sim$\,1620\,s on-source time.}

\begin{table}[th!]                  
\begin{center}
\caption{Millimeter photometry from the literature.\label{mmphot}}
\begin{tabular}{l r@{$\pm$}l c}
\tableline\tableline
Source & \multicolumn{2}{c}{$F_{250\,{\rm GHz}}$} & reference\\
     & \multicolumn{2}{c}{mJy}  & \\
(1)  & \multicolumn{2}{c}{(2)} & (3)  \\
\tableline
J0002+2550   & \multicolumn{2}{c}{$<2.6$}   &  5 \\ %  0.20 +/- 0.88
J0005$-$0006 & \multicolumn{2}{c}{$<1.4$}   &  4 \\ %  0.36 +/- 0.48
J0203+0012   & 1.85&0.46                    &  1 \\  
J0231$-$0728 & \multicolumn{2}{c}{$<3.5$}   &  3 \\   % < 3.5
J0303$-$0019 &\multicolumn{2}{c}{$<1.5$}    &  4 \\ %  0.23 +/- 0.51
J0338+0021   & 3.7&0.3                      &  2 \\
J0353+0104   & \multicolumn{2}{c}{$<1.4$}   &  4 \\ %  1.20 +/- 0.46
J0756+4104   & 5.5&0.5                      &  3 \\
J0818+1722   & 1.19&0.38\tablenotemark{a}   &  4 \\
J0836+0054   & \multicolumn{2}{c}{$<2.9$}   &  3 \\ % -0.39 +/- 0.96
J0840+5624   & 3.20&0.64                    &  5 \\
J0841+2905   & \multicolumn{2}{c}{$<1.3$}   &  4 \\ %  1.00 +/- 0.43
J0842+1218   & \multicolumn{2}{c}{$<1.7$}   &  4 \\ %  0.11 +/- 0.55
J0913+5919   & \multicolumn{2}{c}{$<2.8$}   &  3 \\  % < 2.8
J0927+2001   & 4.98&0.75                    &  4 \\ 
J1030+0524   & \multicolumn{2}{c}{$<3.4$}   &  5 \\ % -1.15 +/- 1.13
J1044$-$0125 & 1.82&0.43                    &  4 \\
J1048+4637   & 3.0&0.4                      &  6 \\
J1137+3549   & \multicolumn{2}{c}{$<3.4$}   &  5 \\ %  0.10 +/- 1.13
J1148+5251   & 5.0&0.6                      &  6 \\
J1204$-$0021 & \multicolumn{2}{c}{$<1.8$}   &  2 \\ %  0.6  +/- 0.4
J1208+0010   & \multicolumn{2}{c}{$<3.1$}   &  3 \\  % <3.1
J1250+3130   & \multicolumn{2}{c}{$<2.7$}   &  5 \\ %  0.07 +/- 0.90
J1306+0356   & \multicolumn{2}{c}{$<3.1$}   &  3 \\ % -1.05 +/- 1.04
J1335+3533   & 2.34&0.50                    &  5 \\
J1411+1217   & \multicolumn{2}{c}{$<1.9$}   &  5 \\ %  1.00 +/- 0.62
J1427+3312   & \multicolumn{2}{c}{$<2.0$}   &  4 \\ %  0.39 +/- 0.66	
J1436+5007   & \multicolumn{2}{c}{$<3.4$}   &  5 \\ % -0.21 +/- 1.14
J1602+4228   & \multicolumn{2}{c}{$<1.6$}   &  4 \\ %  1.41 +/- 0.54
J1623+3112   & \multicolumn{2}{c}{$<2.4$}   &  5 \\ %  0.17 +/- 0.80
J1630+4012   & \multicolumn{2}{c}{$<1.8$}   &  4 \\ %  0.80 +/- 0.60	
J2054$-$0005 & 2.38&0.53                    &  4 \\
J2315$-$0023 &\multicolumn{2}{c}{$<1.8$}    &  4 \\ %  0.28 +/- 0.60
\tableline
\end{tabular}
\tablenotetext{a}{The measured 250\,GHz flux may be contaminated by a galaxy 
located close to the quasar \citep{lei13}.}
\tablecomments{(1) source name; (2) observed 250\,GHz flux in mJy. Errors are 1$\sigma$, upper limits are 3$\sigma$; (3) references for column (2)}
\end{center}
\tablerefs{  
  (1) \citealt{wan11};
  (2) \citealt{car01};
  (3) \citealt{pet03};
  (4) \citealt{wan08b};
  (5) \citealt{wan07};
  (6) \citealt{ber03}}
\end{table}

The data were processed within the {\it Herschel} Interactive
Processing  Environment (HIPE, \citealt{ott10}), version 10. We
followed standard procedures for deep field data reduction, including
source masking and high-pass filtering. The two scan directions were
processed individually and later combined into a final map. The half-width of the
high-pass filter was set to 12 and 16 samples in green and red,
respectively. Considering the scan speed of 20\,\arcsec/s used for 
our observations and the effective sampling of 10\,Hz of the bolometer 
pixels, this corresponds to a total high-pass filter window of 50\arcsec~(green) 
and 66\arcsec~(red) on sky. Source masking was performed via 
circular masks of typically 6 to 8\arcsec~size (or larger if needed 
given the source structure). The mask was created by hand through
visual inspection of the mosaicked maps. For this purpose we first
created a map (with both scan directions combined) without source
masking. On this map we masked all visible sources and source
structures that could lead to artifacts during high-pass
filtering. This proved to be more reliable than a strict sigma cut, as
it also allowed the masking of fairly faint features which could 
potentially influence the measured fluxes of our faint science targets, if
located nearby. The data were then reprocessed including the
source mask. Only in a few cases it was necessary to improve the mask
 using this new map. The frames contributing to the final map 
 were selected based on
the scan speed and we adopted a limit of $\pm$\,5\,\arcsec/s around the 
nominal scan speed of 20\,\arcsec/s. During map projection, the pixel 
fraction parameter \citep[e.g.,][]{fru02} was set to 0.6 to take 
advantage of the moderate redundancy in our data provided by the 
repetition factor of five. We show images around the QSOs position at 
100\,$\mu$m and 160\,$\mu$m in the Appendix (Figure\,\ref{all_images}).

\input{latex_table_fluxes.tex}

Source fluxes or upper limits were determined via aperture
photometry in IDL. We used apertures of 6\arcsec~and 9\arcsec~radius
in green and red, respectively. A residual sky was measured in a sky
annulus between 20\arcsec~and 25\arcsec~(green) or 24\arcsec~and 
28\arcsec~(red). 
Appropriate aperture corrections were determined
from the encircled energy fraction of unresolved sources provided as
part of the calibration data.

The uncertainties of the {\it Herschel} maps was determined in a
similar fashion as for the {\it Spitzer} data: For a given map we 
performed aperture photometry at 500 random positions across the
map. The placement of the apertures was limited to regions in the scan map with at
least 75\,\% of the coverage compared to the science target. The
aperture radius was fixed to the value used for the quasar
photometry. 
The distribution of these 500 flux measurements was then fitted with a
Gaussian, the sigma of which we take as the 1$\sigma$ uncertainty on
the photometry \citep[e.g.,][]{lut11,pop12}. For a few objects, slight
changes in the photometry scripts 
resulted in revised error estimates compared to \citet{lei13}.

We have also re-observed six sources with unusual MIR ($\sim$4--15\,$\mu$m rest frame) 
SEDs during 
{\it Herschel}'s second open time cycle. These targets were undetected 
in our standard {\it Herschel} observations and we selected sources 
with unusually small 100\,$\mu$m/24\,$\mu$m flux limits for deeper observations. 
The observational layout and data reduction procedure was similar to that of the 
standard observations. We executed additional three visits for each source with the 
same parameters as before, essentially quadrupling the
on-source integration time. The new fluxes and deeper flux limits are 
included and marked in Table\,\ref{tab_all} where we report the full 
{\it Herschel} photometry results. Note that the error estimates in 
Table\,\ref{tab_all} do not include the $\sim$5\% uncertainty on the absolute 
flux calibration \citep{bal13}.

\subsubsection{SPIRE}\label{sec:spire}

The SPIRE \citep{gri10} instrument on board {\it Herschel} was used to observe all quasars in our sample 
at 250, 350, and 500\,$\mu$m. The observations were carried out in 
small scan map mode with five repetitions for each objects, totaling
$\sim$\,190\,s on-source integration time per source. This
observational set-up ensured that our maps are dominated 
by confusion noise which is on the order of $6-7$\,mJy\,beam$^{-1}$ in 
the SPIRE photometric bands \citep{ngu10}.

\begin{figure}[t!]
\centering
\includegraphics[angle=0,scale=.45]{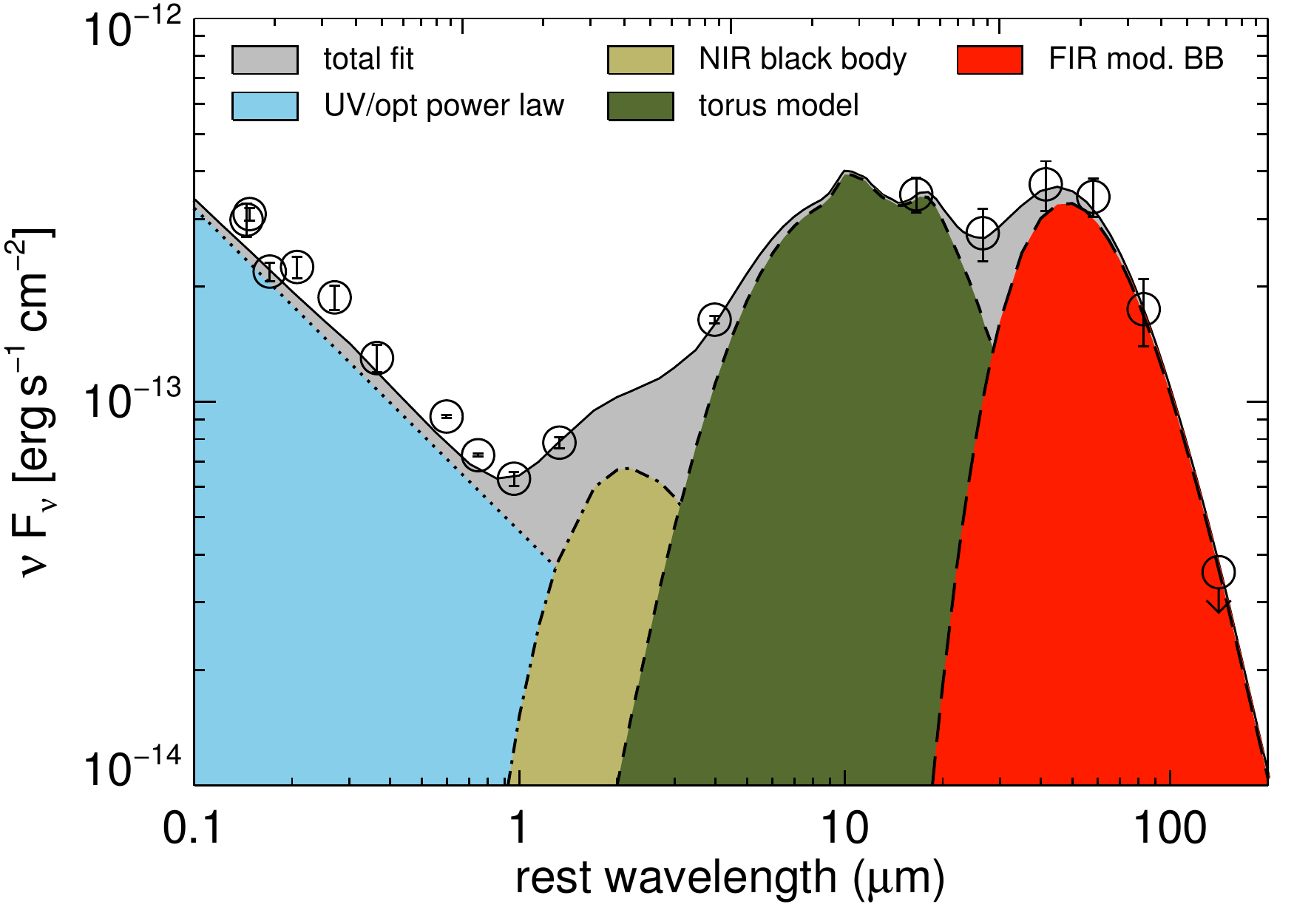}
\caption{Schematic representation of the components used for SED fitting. As an 
example we use the observed photometry of the $z=5.03$ QSO J1204$-$0021.\label{schematic_sed}}
\end{figure}

\begin{table*}[thp!]
\begin{center}
\caption{Results of the SED fitting.\label{fitting_results}}
\begin{tabular}{l r@{$\pm$}l r@{$\pm$}l r@{$\pm$}l r@{$\pm$}l r@{$\pm$}l}
\tableline\tableline
Source & \multicolumn{2}{c}{$L_{\rm UV/opt}$} & \multicolumn{2}{c}{$L_{\rm MIR}$} & \multicolumn{2}{c}{$T_{\rm FIR}$} & \multicolumn{2}{c}{$L_{\rm FIR}$}& \multicolumn{2}{c}{SFR}\\
       & \multicolumn{2}{c}{($10^{46}$\,erg\,s$^{-1}$)} & \multicolumn{2}{c}{($10^{46}$\,erg\,s$^{-1}$)} & \multicolumn{2}{c}{(K)} & \multicolumn{2}{c}{($10^{13}$\,$L_{\odot}$)}& \multicolumn{2}{c}{($10^3 M_{\odot}$\,yr$^{-1}$)}\\
(1)    & \multicolumn{2}{c}{(2)}     & \multicolumn{2}{c}{(3)}     & \multicolumn{2}{c}{(3)} & \multicolumn{2}{c}{(5)}& \multicolumn{2}{c}{(6)}\\
\tableline
J0002+2550   &  10.1&0.3     &  10.7&0.8                   & \multicolumn{2}{c}{47} & \multicolumn{2}{c}{$<$0.9} & \multicolumn{2}{c}{$<$1.5} \\
J0017$-$1000 &  10.2&0.2     &  10.5&0.3                   & \multicolumn{2}{c}{47} & \multicolumn{2}{c}{$<$0.7} & \multicolumn{2}{c}{$<$1.2} \\
J0338+0021   &   6.2&0.2     &  19.8&1.1                   & 51&6                   & 1.2&0.5                    &  2.1&0.6\\
J0756+4104   &   5.7&0.2     &  11.2&0.4                   & 40&2                   & 1.0&0.2                    &  1.7&0.3\\
J0842+1218   &   9.5&0.4     &  23.6&1.6                   & \multicolumn{2}{c}{47} & \multicolumn{2}{c}{$<$1.3} & \multicolumn{2}{c}{$<$2.2} \\
J0927+2001   &   5.8&0.2     &  \multicolumn{2}{c}{$<$9.9}  & 49&2                   & 1.1&0.2                   &  1.9&0.3\\
J0957+0610   &  10.4&0.2     &  13.4&0.7                   & \multicolumn{2}{c}{47} & \multicolumn{2}{c}{$<$0.7} & \multicolumn{2}{c}{$<$1.2} \\
J1044$-$0125 &   9.6&0.3     &  18.4&1.3                   & 53&3                   & 1.1&0.2                    &  1.9&0.3\\
J1048+4637   &  11.0&0.2     &  11.8&0.6                   & \multicolumn{2}{c}{47} & \multicolumn{2}{c}{$<$1.3} & \multicolumn{2}{c}{$<$2.2} \\
J1148+5251   &  14.5&0.2     &  18.2&0.8                   & 60&3                   & 3.5&0.5                    &  6.0&0.6\\
J1202+3235   &  16.9&0.3     &  20.9&0.7                   & \multicolumn{2}{c}{47} & \multicolumn{2}{c}{$<$0.9} & \multicolumn{2}{c}{$<$1.5} \\
J1204$-$0021 &   9.4&0.2     &  18.1&0.7                   & 51&5                   & 2.4&0.3                    &  4.1&0.5\\
J1334+1220   &   7.8&0.2     &  11.5&0.6                   & \multicolumn{2}{c}{47} & \multicolumn{2}{c}{$<$0.8} & \multicolumn{2}{c}{$<$1.3} \\
J1340+3926   &   8.8&0.2     &  20.3&0.8                   & \multicolumn{2}{c}{47} & \multicolumn{2}{c}{$<$0.7} & \multicolumn{2}{c}{$<$1.2} \\
J1340+2813   &   9.8&0.3     &  12.3&0.5                   & \multicolumn{2}{c}{47} & \multicolumn{2}{c}{$<$0.8} & \multicolumn{2}{c}{$<$1.3} \\
J1443+3623   &  13.1&0.4     &  27.2&1.0                   & \multicolumn{2}{c}{47} & \multicolumn{2}{c}{$<$0.8} & \multicolumn{2}{c}{$<$1.3} \\
J1602+4228   &   9.1&0.4     &  21.0&1.8                   & \multicolumn{2}{c}{47} & \multicolumn{2}{c}{$<$0.6} & \multicolumn{2}{c}{$<$1.0} \\
J1626+2751   &  23.2&0.4     &  22.9&1.5                   & \multicolumn{2}{c}{47} & 1.9&0.3                    &  3.2&0.5\\
J1659+2709   &  15.6&0.2     &  17.3&1.3                   & \multicolumn{2}{c}{47} & \multicolumn{2}{c}{$<$1.3} &  \multicolumn{2}{c}{$<$2.2}\\
\tableline
\end{tabular}
\tablecomments{(1) source name; (2) UV/optical luminosity determined by integrating the 
power-law component between 0.1\,$\mu$m and 1\,$\mu$m; (3) luminosity of the (presumably) 
AGN powered dust emission (NIR black body and torus model combined), integrated between 1.0\,$\mu$m and 1000\,$\mu$m; 
(4) temperature of the additional modified black body ($\beta$\,$=$\,1.6) in the FIR (temperature 
was held fixed in cases where no errors are given, see text for details); (5) luminosity of the 
additional FIR component, integrated between 8.0\,$\mu$m and 1000\,$\mu$m; (6) star-formation rate determined 
from the FIR luminosity under the assumption of pure starburst heating and using the relation in \citet{ken98}.}
\end{center}
\end{table*}

Data reduction was performed in HIPE (version 10) following
standard procedures as recommended by the SPIRE instrument
team. The SPIRE final maps are shown alongside the {\it Spitzer} and PACS images in 
the Appendix (Figure\,\ref{all_images}). The HIPE build-in source extractor  
'sourceExtractorSussextractor' \citep{sav07} was used to locate sources 
and determine source fluxes, including a pixelisation correction. Instead of using  global average 
confusion noise limits \citep{ngu10}, we estimated these uncertainties specifically 
for our target fields in the following manner \citep[see also,][]{elb11,pas11}: First, the source
extractor was run over the full calibrated maps. 
An artificial source image including all the sources 
found by the source extractor was created and subtracted from the observed map.
On this ``residual map'' we determined the pixel-to-pixel rms in a box
with a size of 8 times the FWHM (FWHM size: 18.2\arcsec, 24.9\arcsec,
and  36.3\arcsec~for default map pixel sizes of 6, 10, and 14\arcsec~at 
250, 350, and 500\,$\mu$m, respectively), centered
on the  nominal position of the QSO. The size  of this box was chosen
large enough to allow an appropriate sampling of the surroundings of
the source, but small enough to avoid including the lower coverage
areas at the edges of the map even for the longest wavelengths. In
addition,  the number of pixels per FWHM is approximately constant for the three wavelengths in the
final maps  ($2.5-3.0$\,px/FWHM) which translates into a similar number of
pixels used for determining the rms in the background box. 
The resulting estimates for the noise (limited by confusion) are 
comparable to the global average values given in 
\citet{ngu10}, but have a tendency to be slightly lower.  
Detections from the source extraction located within less than half
the FWHM from the nominal target position were tentatively considered
to belong to the quasar. The measured source flux was then compared to
the estimated confusion noise in the map. We also checked for
confusion with nearby FIR bright sources using our multi-wavelength
data to avoid mis-identifications. The final source photometry is
presented in Table\,\ref{tab_all}. Similar to PACS, the SPIRE errors in 
Table\,\ref{tab_all} do not include the $\sim$4\% uncertainty on the 
absolute flux calibration \citep{ben13}.

We note that a number of SPIRE
flux measurements in Table\,\ref{tab_all} are nominally below the estimated 3$\sigma$ 
value of the noise. In such cases, the images
reveal a clear excess of flux at the position of the quasar and
comparison with other wavelengths (e.g. {\it Spitzer}/24\,$\mu$m or PACS) 
shows no clear indication for possible confusion issues. 
The use of positional priors can reduce the effect of confusion noise 
by 20-30\,\% \citep{ros10}, and our data set provides accurate 
(relative and absolute) positional information as well as information on 
the SEDs of the quasar and potential confusing sources in the field. Therefore, 
we here include these flux measurements in our study, although they have to 
be treated with caution. Similarly, fluxes at 500\,$\mu$m should be 
considered tentative because at this wavelength the beam is large 
($\sim$\,36\,\arcsec~FWHM), the confusion noise is high, and the 
significance of the detections is often low.

\subsubsection{Millimeter regime}\label{sec:mm}

In total, 33 objects of our {\it Herschel} sample have published 
observations in the millimeter regime from the ground, typically at 
250\,GHz (see Tab.\,\ref{mmphot}). The 11 millimeter detections among 
those have been presented in detail in \citet{lei13}, but are also 
included here. The remaining 22 objects are undetected at 
millimeter wavelengths. Among those 22, five sources (J0002+2550, 
J0842+1218, J1048+4637, J1204$-$0021, J1602+4228) have {\it Herschel} 
detections in two or more bands, while the rest is also undetected with 
{\it Herschel}. A number of the millimeter observed 
objects have also been targeted in the sub-millimeter 
from the ground \citep{pri03,pri08,rob04,wan08a,wan10,bee06} and recently 
with ALMA \citep{wan13}.

% in 10^46 erg/s
%lfir_vec     = [-3.30,-2.62,4.60 ,3.95 ,-5.13,4.38 ,-2.87,4.40 ,-4.95,13.54,-3.54,9.38 ,-3.04,-2.75,-3.08,-3.02,-2.07,7.19, -5.17]
%lfir_err_vec = [0,0,1.44 ,0.41 ,0,0.63 ,0,0.61 ,0,1.27,0,1.21 ,0,0,0,0,0,1.20,0]

% in L_sun
%lfir_vec = [-8.6e+12,-6.8e+12, 1.2e+13, 1.0e+13,-1.3e+13, 1.1e+13,-7.5e+12, 1.1e+13,-1.3e+13, 3.5e+13,-9.2e+12, 2.4e+13,-7.9e+12,-7.2e+12,-8.0e+12,-7.9e+12,-5.4e+12, 1.9e+13,-1.3e+13]
%lfir_vec = [-9.0e+12,-7.0e+12, 1.2e+13, 1.0e+13,-1.3e+13, 1.1e+13,-7.0e+12, 1.1e+13,-1.3e+13, 3.5e+13,-9.0e+12, 2.4e+13,-8.0e+12,-7.0e+12,-8.0e+12,-8.0e+12,-6.0e+12, 1.9e+13,-1.3e+13]
%lfir_err_sun
%       0.0000000       0.0000000   3.7509770e+12   1.0679864e+12       0.0000000
%   1.6410523e+12       0.0000000   1.5889555e+12       0.0000000   3.3081531e+12
%       0.0000000   3.1518626e+12       0.0000000       0.0000000       0.0000000
%       0.0000000       0.0000000   3.1258141e+12       0.0000000

%sfr according to Kennicutt 1998
%      -1479.7712      -1174.8486       2062.7113       1771.2413      -2300.3716
%       1964.0600      -1286.9525       1973.0283      -2219.6567       6071.5460
%      -1587.3909       4206.1375      -1363.1831      -1233.1426      -1381.1197
%      -1354.2148      -928.22007       3224.1075      -2318.3082

\begin{figure}[t!]
%\centering
\includegraphics[angle=0,scale=.65]{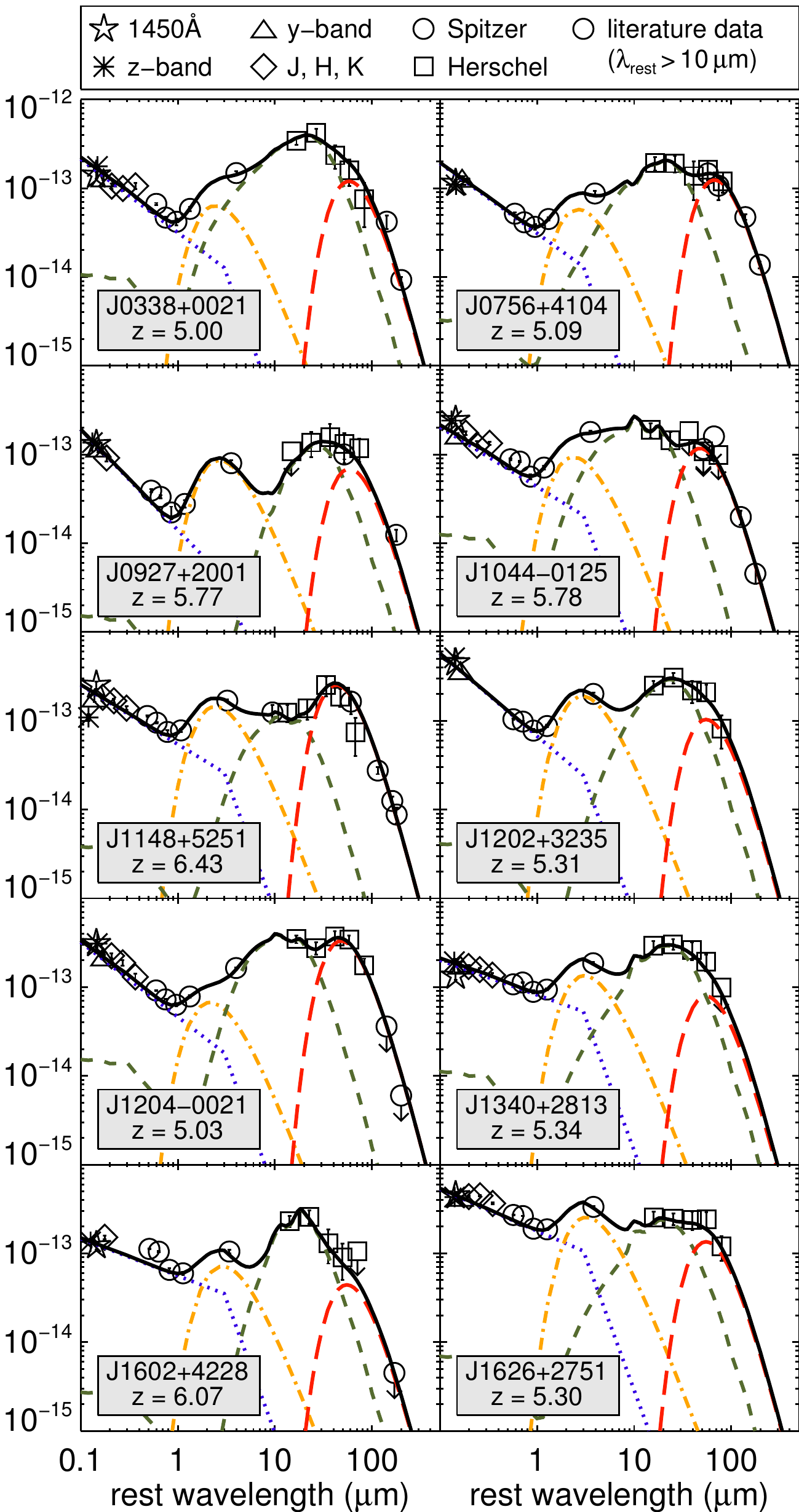}
\caption{The SEDs of the 10 quasars detected in at least four {\it
    Herschel} bands. The plots shows $\nu F_{\nu}$ in units of erg\,s$^{-1}$\,cm$^{-2}$ over 
  the rest frame wavelength. 
The colored lines indicate the results of a multi-component SED fit as described 
in Section\,\ref{sec:sed_fits}. They consist of a power-law (blue dotted), 
a black body of T\,$\sim$\,1200\,K (yellow dash-dotted), a torus model (green dashed), 
and a modified black body of $\sim$47\,K (see Table\,\ref{fitting_results}; red long dashed). The black solid line 
shows the total fit as the sum of the individual components. \label{seds_firdet}}
\end{figure}

\section{Detection rates}

Most of the objects in our sample had previously only been observed in
the optical or NIR. These data sample the rest frame UV/optical regime
and typically provide spatial resolution of $\sim$1\arcsec~or
higher. Therefore caution has to be exercised when matching such
objects with data in the FIR where often only (much) lower spatial
resolution is achievable and sources faint in the optical but bright
in the FIR could be mistaken as a counterpart. For a reliable source
matching, the multi-wavelength nature of our data set provided a 
powerful tool for
determining the exact position of the quasar in the {\it Herschel}
bands. In particular the {\it Spitzer} 24\,$\mu$m images were very
valuable in this regard. They provide spatial resolution bridging the
gap between the optical/NIR and FIR observations and strong detections 
for most quasars in our sample. In many cases we
can identify several sources per field that are visible both at {\it
Spitzer} and at {\it Herschel} wavelengths and the exact location of
the quasar in the {\it Herschel} maps can be determined from the relative
positional information.  With this procedure  we can robustly identify
faint {\it Herschel} detections with the quasars as well as avoid
mis-identifications due to nearby objects.  During this exercise we
observe absolute spatial offsets between {\it Spitzer} and {\it
Herschel} of typically $\lesssim$\,2\arcsec, in line with expectations
from the absolute pointing accuracies (S\'anchez-Portal et al. 2014, 
submitted).

\begin{figure}[t!]
\centering
\includegraphics[angle=0,scale=.45]{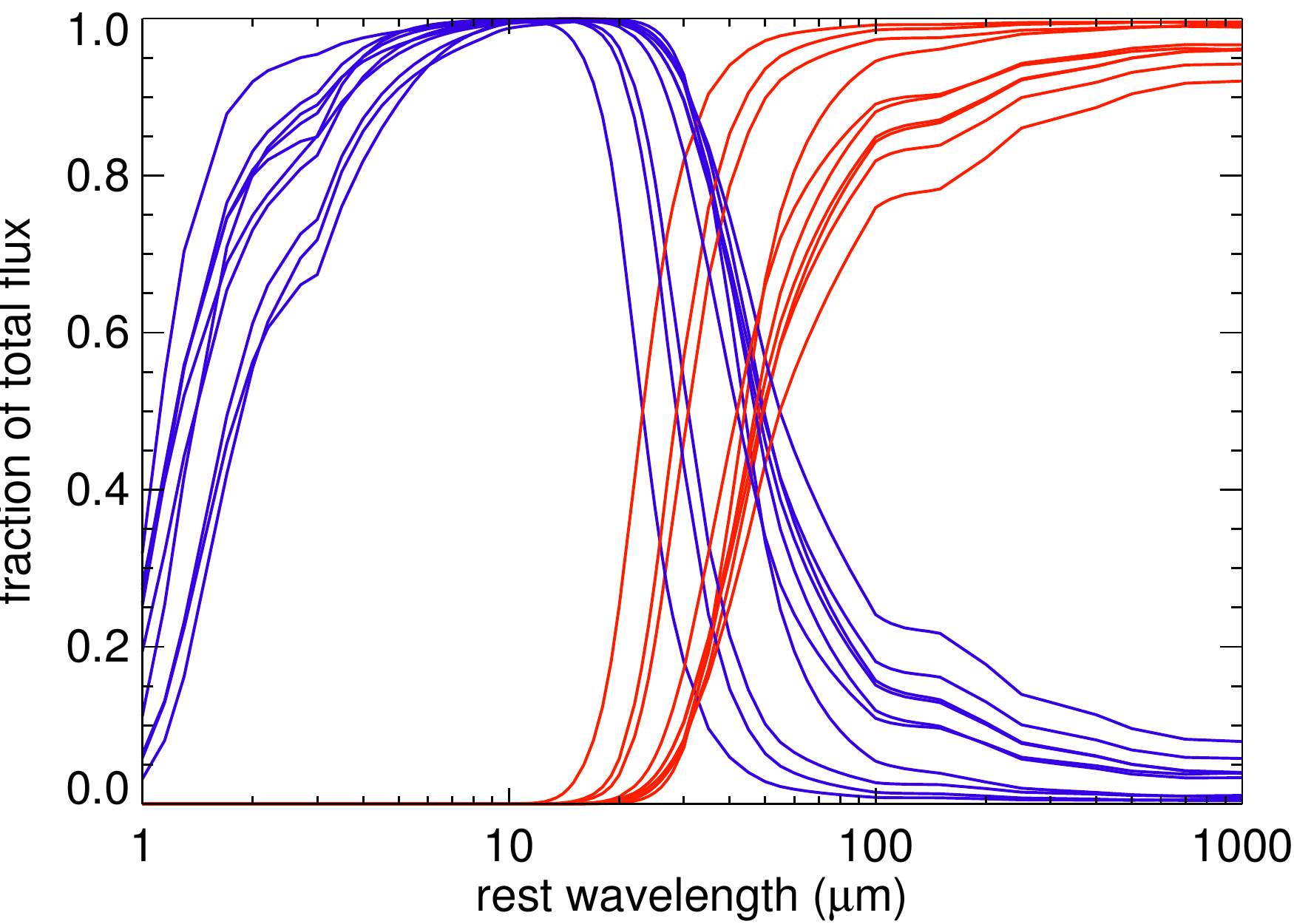}
\caption{For the ten objects where the FIR component could be well constrained due to 
additional millimeter data (see Figure\,\ref{seds_firdet}) we here show its relative
contributions (red) compared to the presumably AGN-heated dust (NIR black body 
plus torus model; blue) as a function of wavelength. For these FIR-bright sources, the 
FIR component dominates the total infrared emission at 
$\lambda_{\rm rest}$\,$\gtrsim$\,50\,$\mu$m. \label{flux_ratios}}
\end{figure}

In almost all cases we detect the observed quasars in the available 
{\it Spitzer}
bands at high significance (see Table\,\ref{tab_all}). One exception is
J1148+5253 which is neither detected with IRAC at 5.8 and 8\,$\mu$m nor
with MIPS at  24\,$\mu$m. However, this object is almost 3 magnitudes
fainter in z-band than the majority of the sample. The only other
exception is J1208+0010 which we do not detect in MIPS at
24\,$\mu$m. Our detections include those quasars which have previously
been dubbed 'dust-free' \citep{jia10}. In our analysis 
we see both sources (J0005$-$0006 and J0303$-$0019)
at all {\it Spitzer} wavelengths (Figure\,\ref{dust_free}). While
J0005$-$0006 is fairly isolated and can be identified readily, the
other object (J0303$-$0019) suffers from blending issues with a nearby
source. In the higher spatial resolution IRAC observations the two
objects can be well separated, but with IRS and MIPS the blending
becomes severe. In these cases we subtracted the  confusing source as
described in Section\,\ref{sec:spitzer}. In both bands we see significant 
residual flux at the
position of the quasar. The new detections of
these two objects, however, do not change the basic conclusion 
of \citet{jia10} that these quasars are clearly deficient of {\it hot} dust 
compared to the majority of the sample.

With PACS we only see 22 (100\,$\mu$m) and 19 (160\,$\mu$m) objects at
greater than 3$\sigma$ significance in our standard observations. In a number of objects the
exactly determined position (see above) was crucial to avoid
misidentifications. With SPIRE the detection rate is even lower and
we identify only 10 objects which are bright enough in the observed
FIR/sub-mm range to be detected systematically (i.e., at 250\,$\mu$m as well as at 
350\,$\mu$m) above the confusion noise.

The additional deep PACS observations for six objects undetected 
by our standard {\it Herschel} program result in two quasars detected in both bands and 
three sources detected only at 100\,$\mu$m at faint flux levels. One source remained 
undetected with an upper limit more than a factor of 2 below our standard limit.

%\begin{figure*}[t!]
%\centering
%\includegraphics[angle=0,scale=.33]{J0002+2550_seds_final_for_paper.pdf}
%\includegraphics[angle=0,scale=.33]{J0017-1000_seds_final_for_paper.pdf}
%\includegraphics[angle=0,scale=.33]{J0842+1218_seds_final_for_paper.pdf}
%\includegraphics[angle=0,scale=.33]{J0957+0610_seds_final_for_paper.pdf}
%\includegraphics[angle=0,scale=.33]{J1048+4637_seds_final_for_paper.pdf}
%\includegraphics[angle=0,scale=.33]{J1334+1220_seds_final_for_paper.pdf}
%\includegraphics[angle=0,scale=.33]{J1340+3926_seds_final_for_paper.pdf}
%\includegraphics[angle=0,scale=.33]{J1443+3623_seds_final_for_paper.pdf}
%\includegraphics[angle=0,scale=.33]{J1659+2709_seds_final_for_paper.pdf}
%\caption{The SEDs of the 7 quasars detected in 2 or 3 {\it Herschel} bands. 
%Scaling, symbols and colors as in Figure\,\ref{seds_firdet}. \label{seds_pacsdet}}
%\end{figure*}

\begin{figure}[t!]
%\centering
\includegraphics[angle=0,scale=.65]{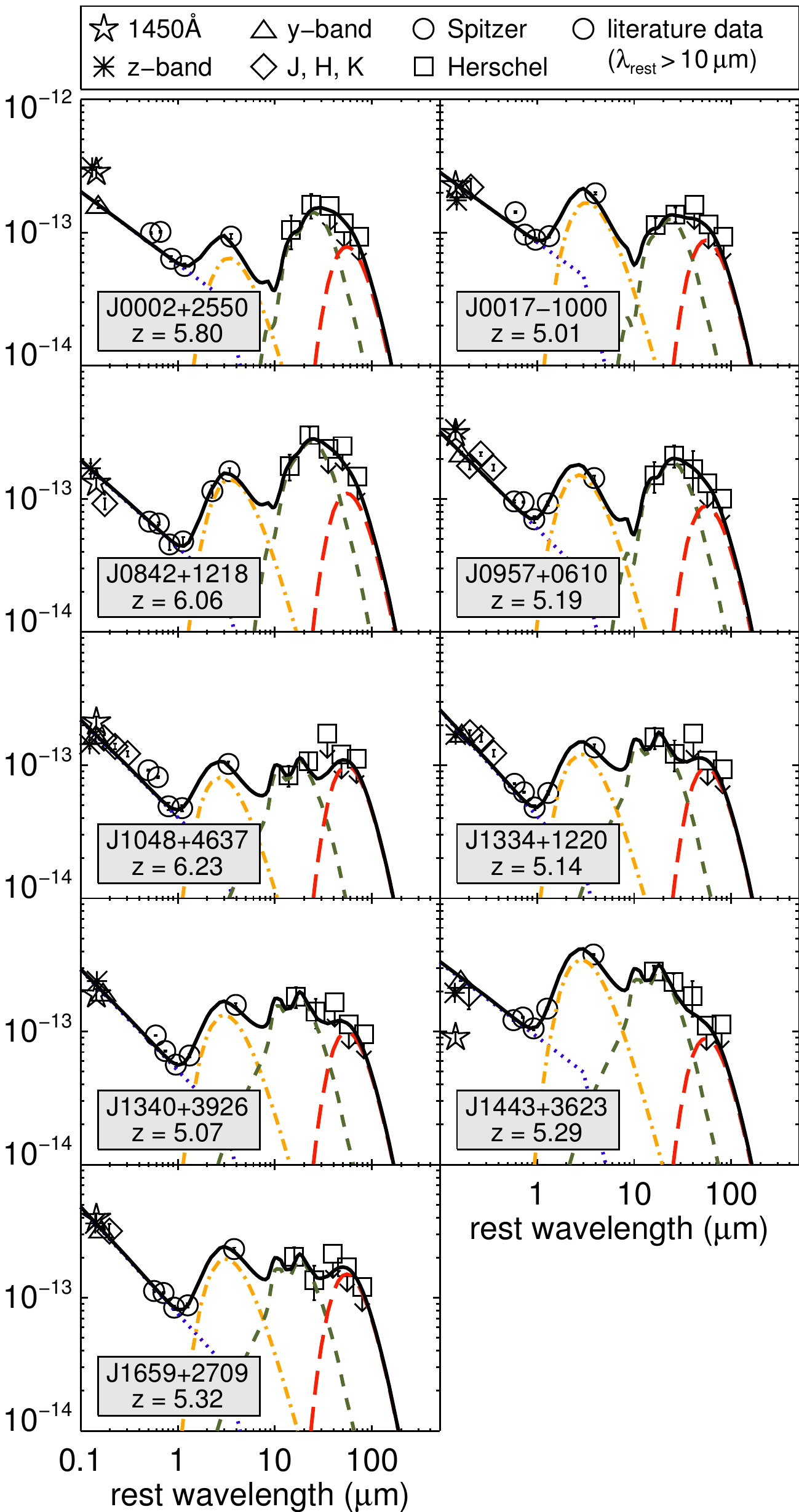}
\caption{The SEDs of the 9 quasars detected in 2 or 3 {\it Herschel} bands shown 
in $\nu F_{\nu}$ in units of erg\,s$^{-1}$\,cm$^{-2}$ over the rest frame wavelength.
Color coding of the fitted components as in Figure\,\ref{seds_firdet}. \label{seds_pacsdet}}
\end{figure}

\section{Analysis and dicussion}

\subsection{SED fitting}\label{sec:sed_fits}

Ten quasars in our sample have been detected in at least four of the
five {\it Herschel} bands (Table\,\ref{tab_all}). In combination with
their {\it Spitzer} fluxes and using supplemental NIR data, the
combined photometry  provides SEDs covering the rest frame wavelengths
from 0.1 to $\sim$80\,$\mu$m. To assess some basic physical properties
of these objects, we perform  SED fitting, following the approach
presented in \citet{lei13}. To  summarize briefly, the SEDs are fitted
with four components: a power-law in the UV/optical mainly representing
emission from the accretion disk, a black body from hot
($\sim$1200\,K) dust,  a torus model from the library of
\citet{hon10}, and an additional cool dust component in the  form of a
modified black body ($\beta$\,=\,1.6). We illustrate this approach and the 
arrangement of the fitted components schematically in Figure\,\ref{schematic_sed}.

In \citet{lei13} we already
presented five of the ten FIR detected quasars, all of which had
millimeter detections. The five additional sources presented here do not
have mm detections and sub-millimeter/millimeter upper limits exist
only for two of the five newly presented objects. In the case of J1204$-$0021
\citep{car01,pri03} those data points can be used to provide
additional constraints on  the temperature which is consequently
treated as a free parameter. For  J1602+4228 \citep{wan08b} the
250\,GHz upper limit does not strongly  constrain the temperature of
the fitted FIR component and we fix  $T_{\rm FIR}$ to a value of 47\,K
for this object \citep{bee06,wan07,lei13}. For the remaining three objects, the temperature of
the FIR component was also fixed to 47\,K. Due to the lack of mm data 
which would help to anchor the Rayleigh-Jeans tail of the FIR component, 
the fits would otherwise predict artificially increased dust
temperatures \citep{lei13}.

\begin{figure}[t!]
%\centering
\includegraphics[angle=0,scale=.5]{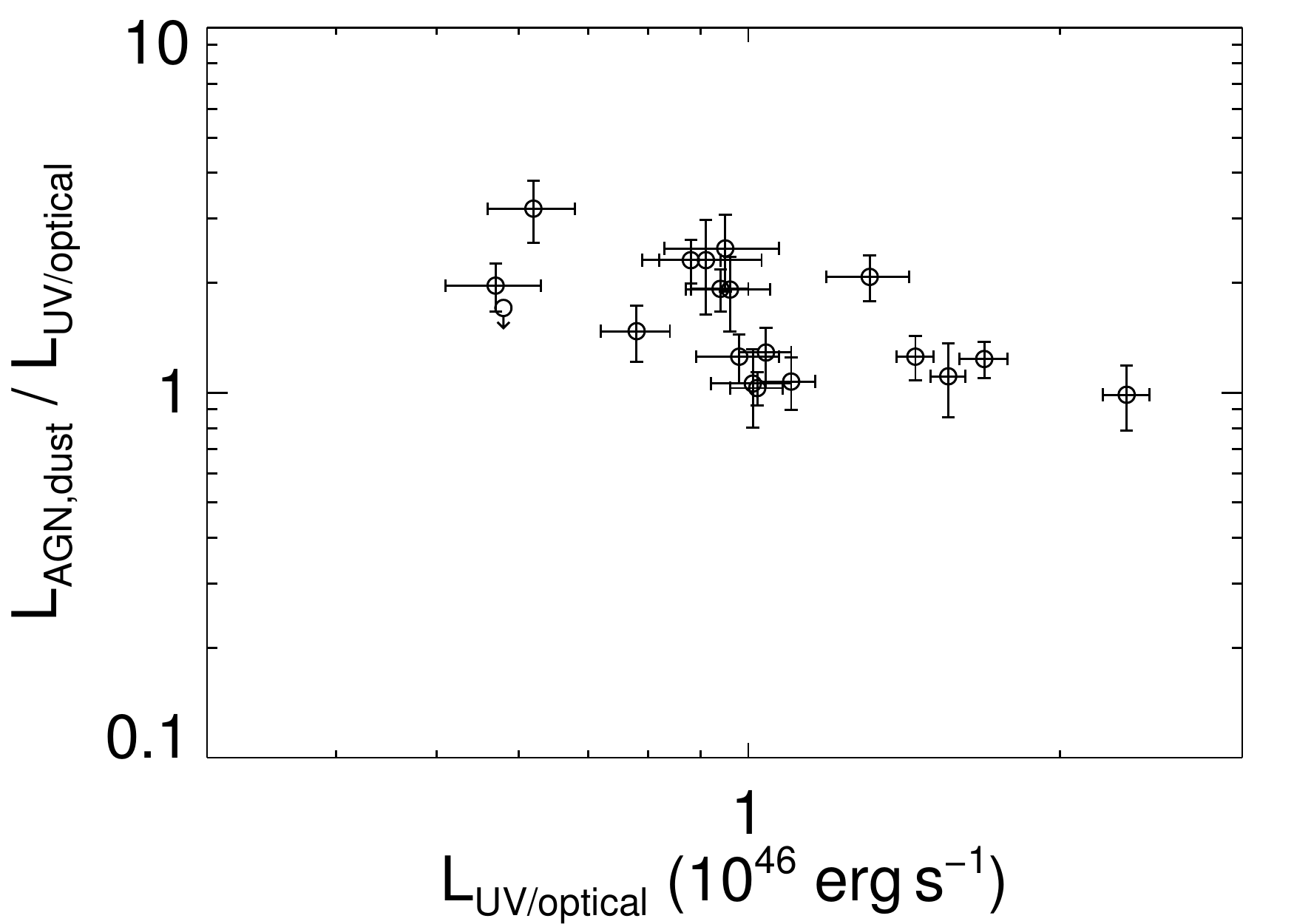}
\caption{The ratio of the AGN-dominated dust-to-accretion disk emission decreases 
with UV/optical luminosity. The data are taken from Table\,\ref{fitting_results} and 
correspond to the UV/optical luminosity between 0.1\,$\mu$m and 1\,$\mu$m and the 
AGN-heated dust emission between 1\,$\mu$m and 1000\,$\mu$m (NIR bump plus torus, but 
excluding the additional FIR component). All wavelengths refer to the rest frame of the 
source. Errorbars correspond to $\pm$3\,$\sigma$. \label{lumratrios_fit}}
\end{figure}

\begin{figure*}[t!]
\centering
\includegraphics[angle=0,scale=.35]{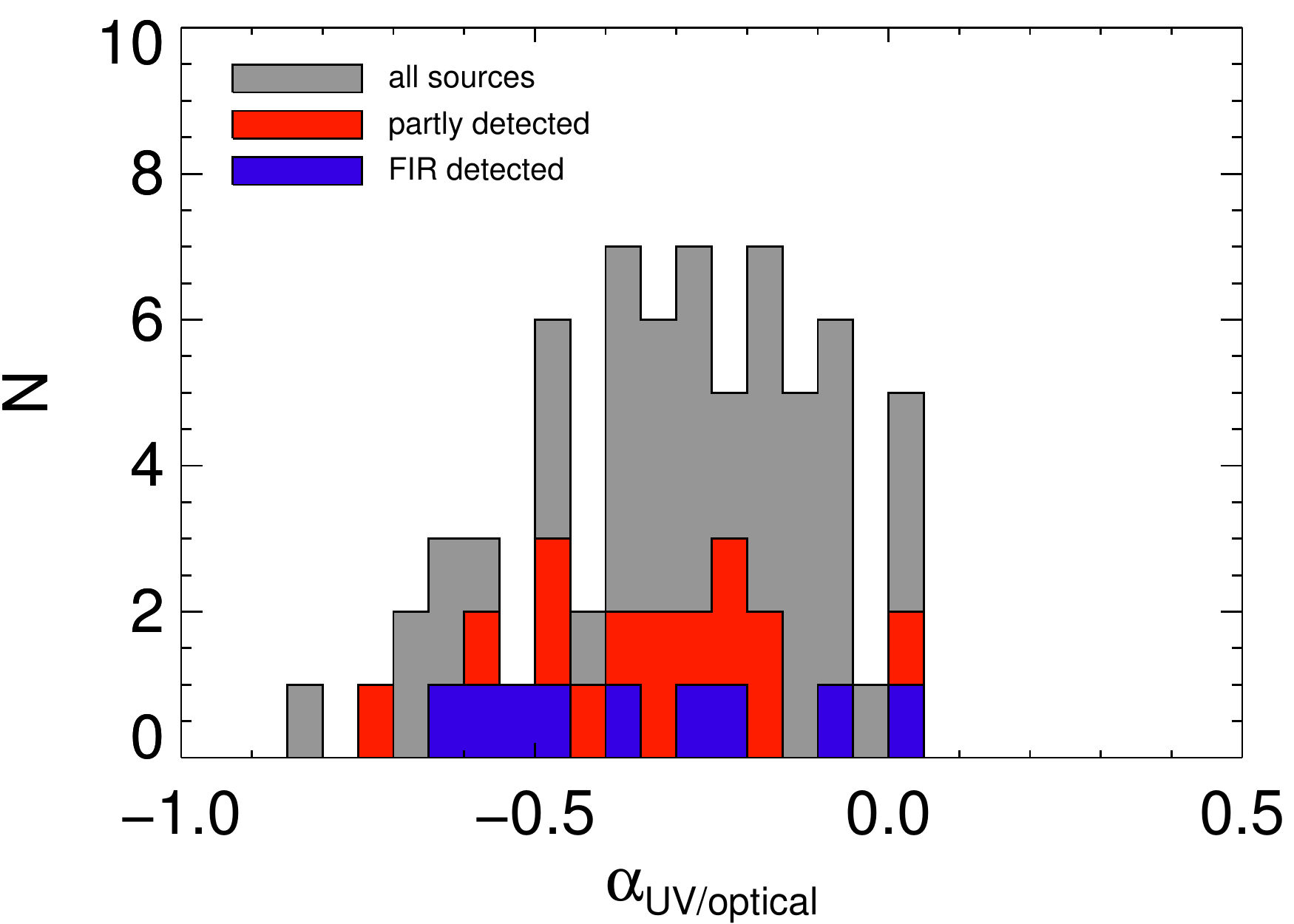}
\includegraphics[angle=0,scale=.35]{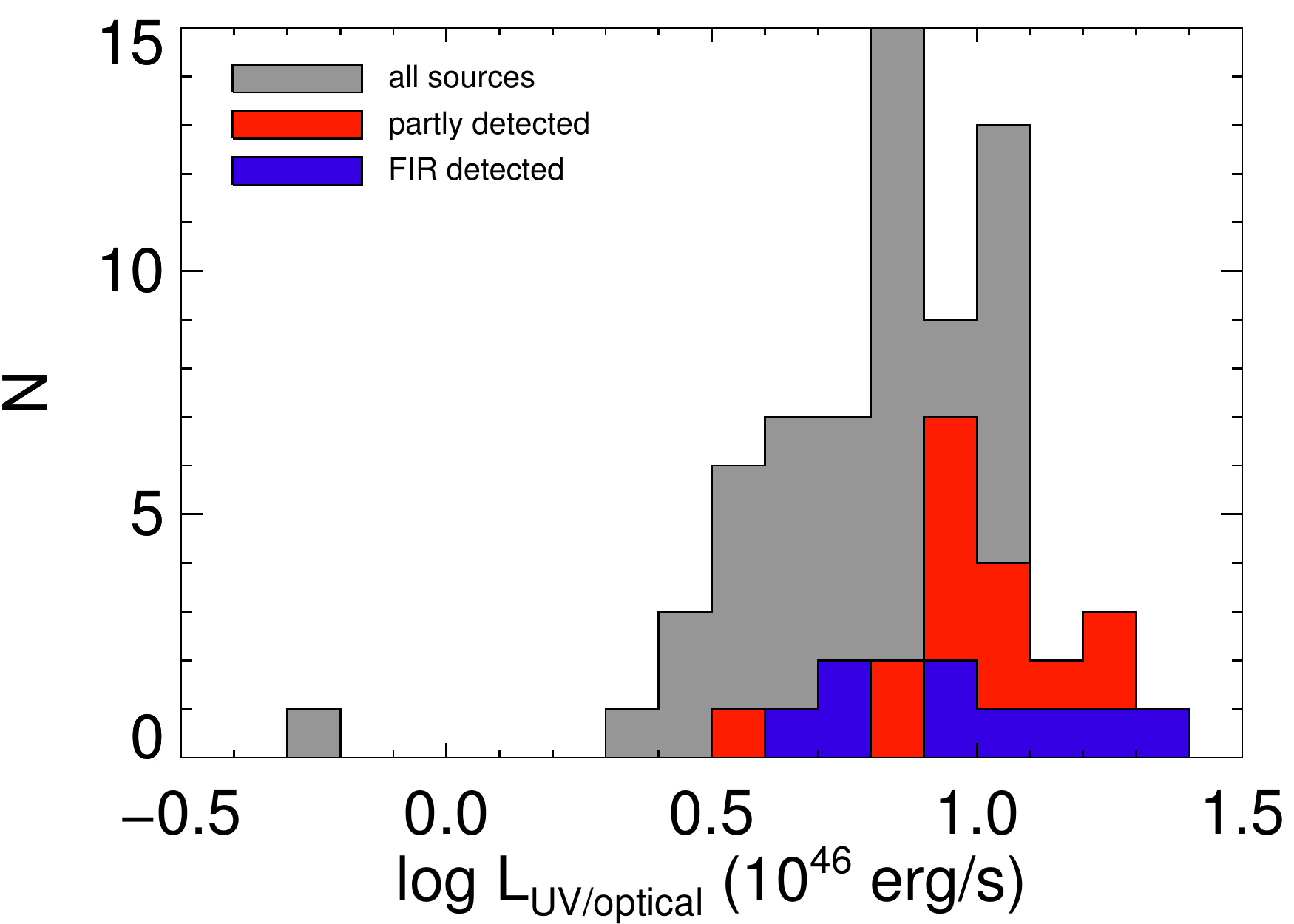}
\includegraphics[angle=0,scale=.35]{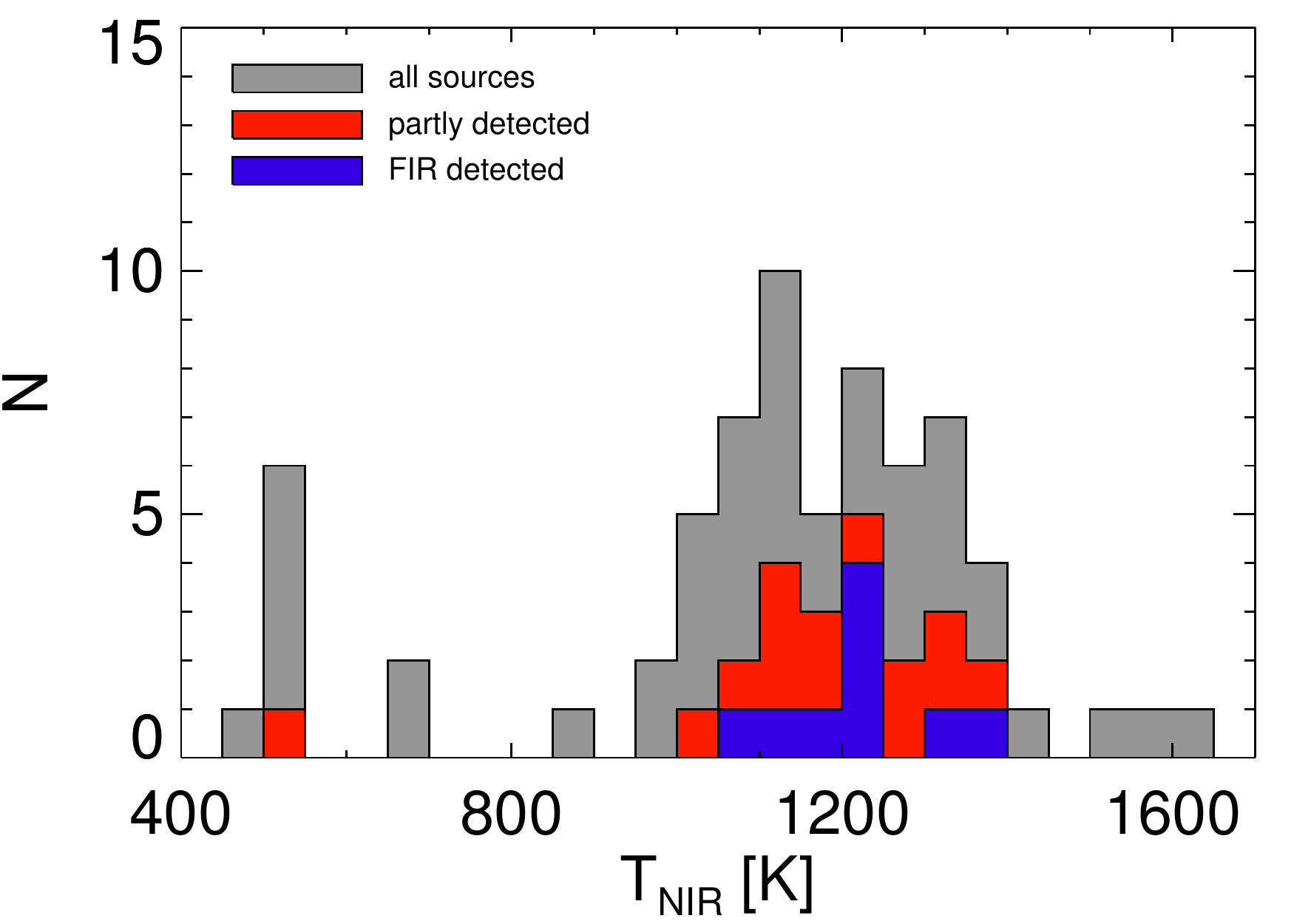}
\includegraphics[angle=0,scale=.35]{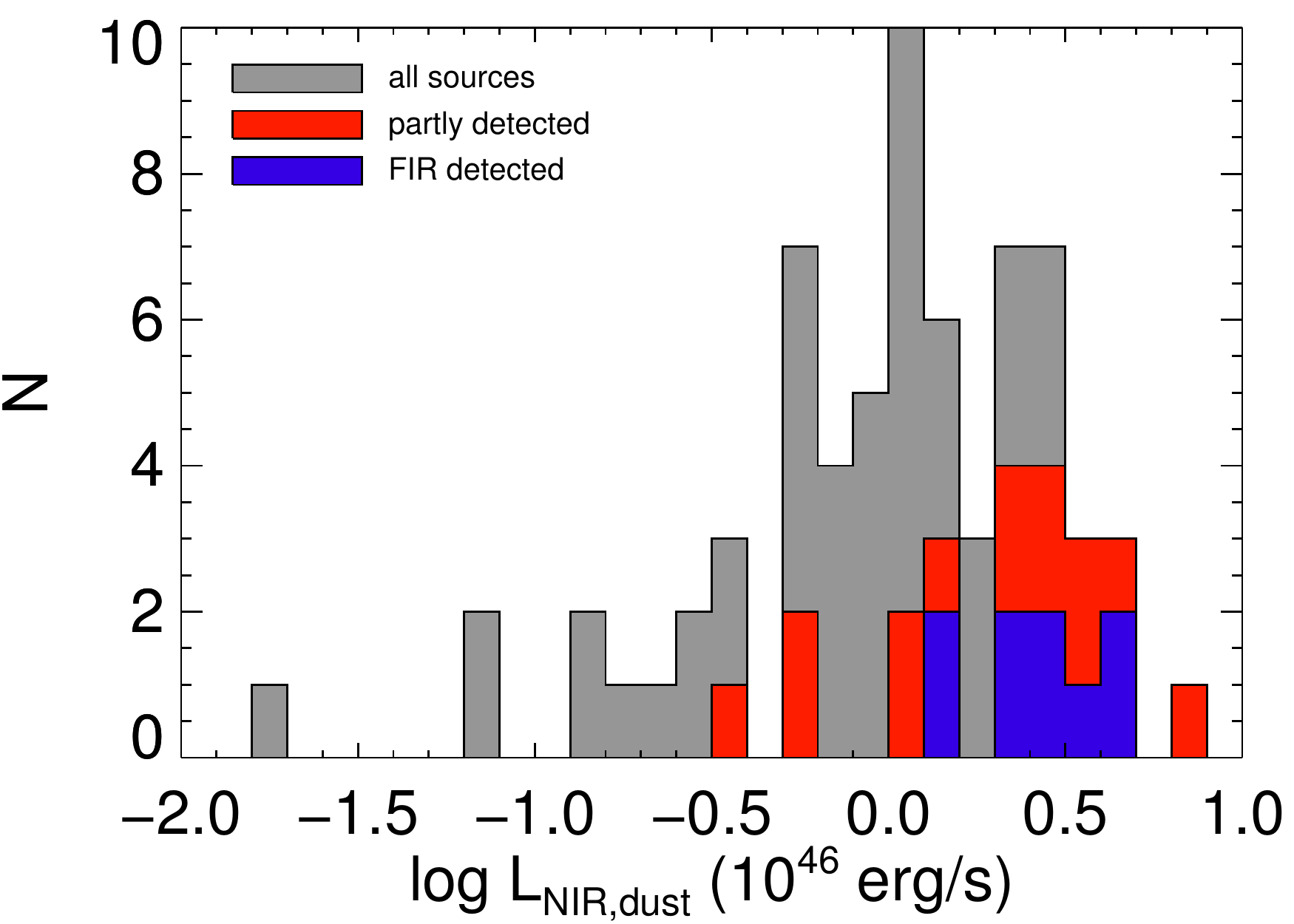}
\caption{Parameter value distributions from the UV/optical plus NIR fits 
(Section\,\ref{sec:alpha_fits}). {\it Top, left}: UV/optical power-law 
index $\alpha$ ($F_{\nu}$\,$\propto$\,${\nu}^{\alpha}$). {\it Top, right}: 
luminosity of the power-law component integrated between 0.1\,$\mu$m and 1\,$\mu$m. 
{\it Bottom, left}: temperature of the hot dust component.
{\it Bottom, right}: luminosity of the hot dust component integrated between 
1\,$\mu$m and 3\,$\mu$m. The {\it Herschel} detected objects (blue and red histograms) 
are preferentially found at the high-luminosity end in $L_{\rm UV/opt}$ as well as in 
$L_{\rm NIR}$  but show no particular trends in $\alpha$ or  $T_{\rm NIR}$. \label{uvseds}}
\end{figure*}

The rest frame UV/optical and infrared SEDs of these ten objects can
be  fitted well with a combination of these four components. The best
fitting model  combinations are shown in Figure\,\ref{seds_firdet} and  
Table\,\ref{fitting_results} summarizes some basic properties determined 
from the fitting. Using these  fits we also determine the
relative contributions of the different components to the  total
infrared SED. For this we combine the dust component in the NIR
and the  torus model, both of which are likely to be powered by the
AGN. We compare this AGN  related emission to the additional FIR
component and show  their relative
contributions to the total infrared emission as a function of
wavelength in Figure\,\ref{flux_ratios}. We see that in the presence 
of luminous FIR emission ($L_{\rm FIR}$\,$\sim$\,10$^{13}$\,$L_{\odot}$), 
this component dominates the total infrared SED at rest frame wavelengths above  
$\sim$50\,$\mu$m for all ten
objects. This means that in such cases of strong FIR/sub-millimeter 
emission, rest frame wavelengths
$\gtrsim$50\,$\mu$m isolate the additional FIR component without the
need for full SED fits (at least in our modeling approach). 
The possible heating source for the additional FIR component 
(AGN versus star formation) is further discussed in section \ref{sec:stacking}.

We also extend a similar SED fitting approach to objects with fewer
{\it Herschel} detections. In  cases where two PACS detections are
available (9 sources), these data provide sufficient constraints  for
the torus model,  while the upper limits in the SPIRE bands (and in the 
millimeter where avilable, se Tab.\,\ref{mmphot}) limit the
contribution of the additional FIR component  (fixed to a temperature of
47\,K). These fits are presented in Figure\,\ref{seds_pacsdet} and
some basic properties derived from the fitted components are presented
in Table\,\ref{fitting_results}. From this table we use the 
UV/optical luminosity and the AGN-dominated dust luminosity to show 
that the ratio of the AGN-dominated dust-to-accretion disk emission 
decreases with increasing  UV/optical luminosity 
(Figure\,\ref{lumratrios_fit}). This behavior may reflect 
the increase of the dust sublimation radius for more luminous UV/optical 
continuum emitters \citep[e.g.,][]{bar87} which, under the assumption of 
a constant scale height, is often explained in terms of a decreasing 
dust covering factor with increasing luminosity in the context of the 
so-called receding torus model \citep{law91}.

The measured FIR fluxes for our 10 FIR detected objects fall only
moderatly above the 3$\sigma$ confusion noise limit (Table\,\ref{tab_all}). 
Thus, the photometric upper 
limits for the 9 FIR non-detections (i.e. only detected in PACS) yield
upper limits on $L_{\rm FIR}$ that do not differ significantly from 
the detection on an individual basis (Table\,\ref{fitting_results}). 
Further constraints on the average FIR
properties of the PACS-only sources are provided by a stacking
analysis as presented in Section \ref{sec:stacking}.

%{schematic_sed}

\subsection{The SEDs at $\lambda_{\rm rest}$\,$<$\,4\,$\mu$m}\label{sec:alpha_fits}

For two thirds of the sample, the upper limits in the {\it Herschel}
observations do not provide strong constraints to MIR or FIR
components to allow full SED fitting. We therefore chose to limit the
fitting to rest frame wavelengths corresponding  to the MIPS
24\,$\mu$m band ($\sim$3-4\,$\mu$m rest frame) and shorter where the
majority of the sources is well detected. For these  data we fit a
combination of a power-law in the UV/optical and a hot black body in
the NIR. To minimize the influence from emission lines (e.g., 
Ly$\alpha$, H$\alpha$) and the small
blue bump on the fitted power-law slope, we limit the data points to
{\it Spitzer} bands at $\lambda_{\rm obs}$\,$\geq$\,5.8\,$\mu$m  and
only using the y-band photometry in the rest frame UV. In those cases
where no y-band photometry is available (5 objects), we use the z-band
instead.  For selected sources the UV part of the rest frame SEDs was
excluded from the  fitting to avoid broad absorption line features
(e.g. J0203+0012, J1427+3312).  The resulting fits are shown in
Figure\,\ref{seds}.

We derive UV/optical luminosities for the quasars by integrating the 
fitted  power-law
between 0.1\,$\mu$m and 1\,$\mu$m. Similarly, for the hot dust
component, the NIR dust luminosity is provided by integrating the
fitted  black body between 1\,$\mu$m and 3\,$\mu$m. The fitted values
for $\alpha_{\rm UV/opt}$  ($F_{\nu}$\,$\propto$\,${\nu}^{\alpha}$)
and $T_{\rm NIR}$, as well as the  integrated luminosities for the two
components, are given in Table\,\ref{tab_uvopt_ew}  and their
distributions are shown in Figure\,\ref{uvseds}.\footnote{We estimated  
uncertainties on these values and tested for the influence of possible 
variability within our non-simultaneous data set by creating 1000 random, 
normally distributed magnitude offsets ($\sigma = \pm 0.1$\,mag), applied 
each of these to the y-band flux and re-fitted the photometry. The width 
of the the resulting distributions in the four parameters 
($\alpha_{\rm UV/opt}$, $L_{\rm UV/opt}$, $L_{\rm NIR,dust}$, $T_{\rm NIR}$)
was taken as their uncertainties (Table\,\ref{tab_uvopt_ew}).}

\begin{figure*}%[t!]
\centering
\includegraphics[angle=0,scale=.4]{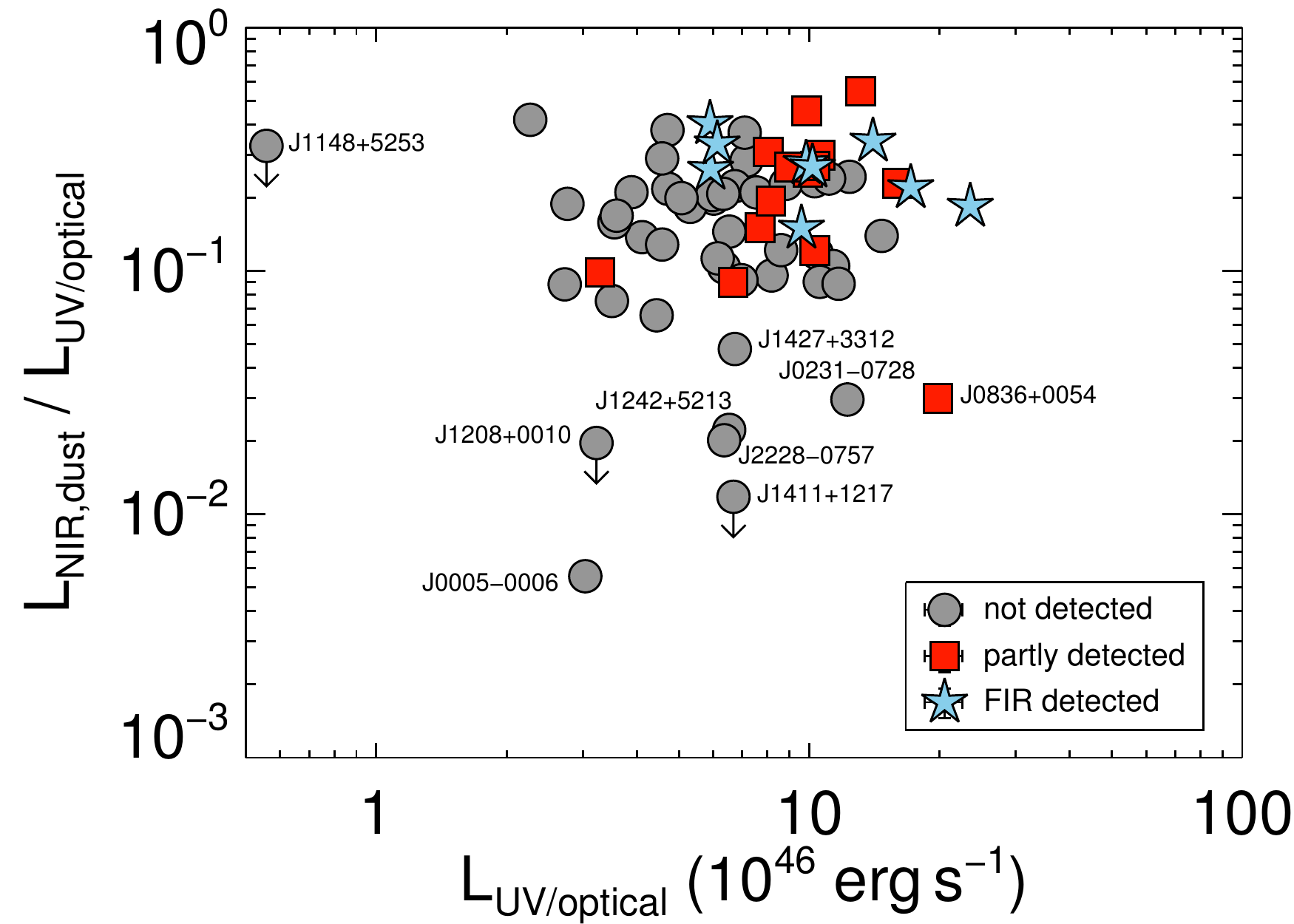}
\includegraphics[angle=0,scale=.4]{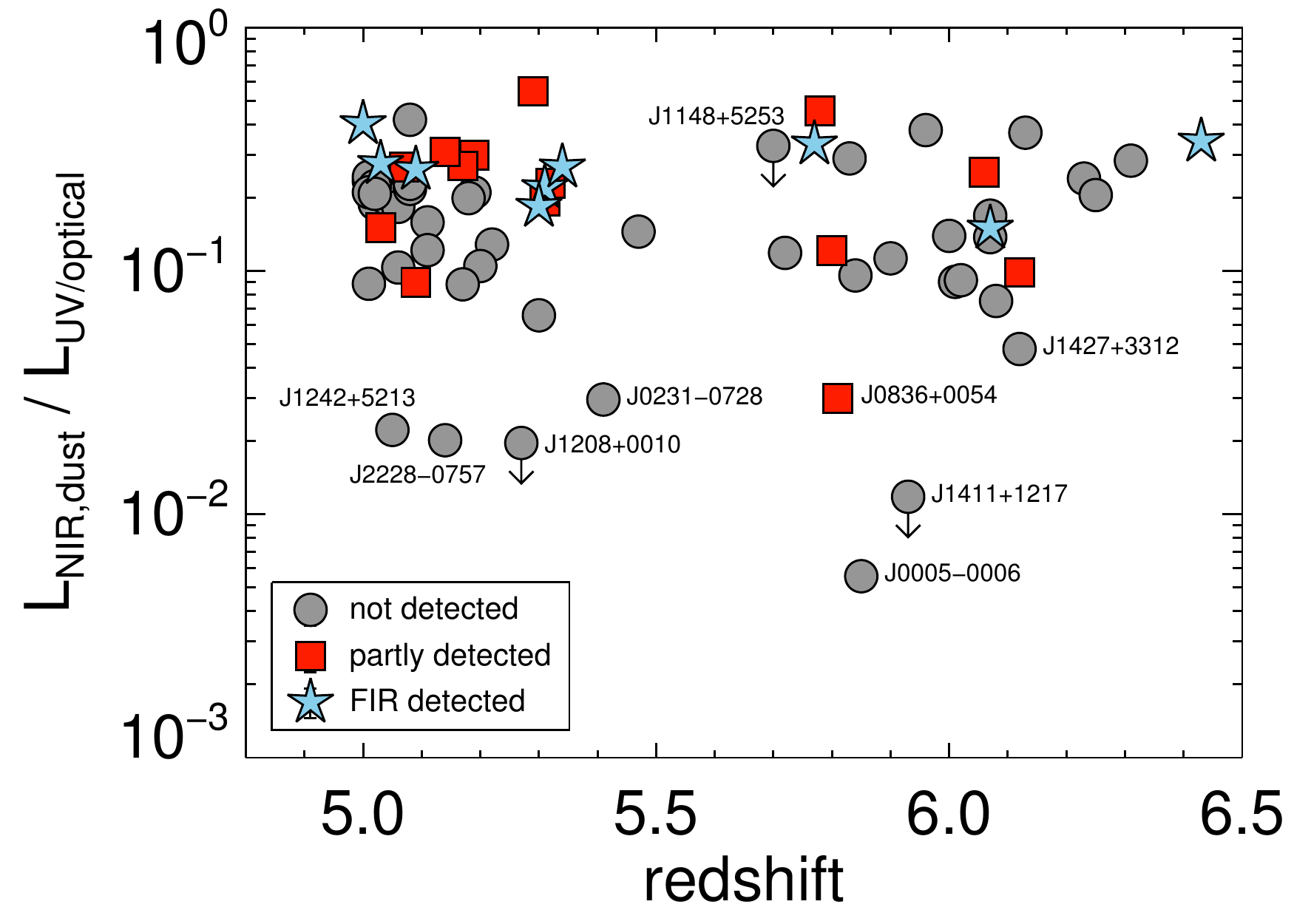}\\
\includegraphics[angle=0,scale=.4]{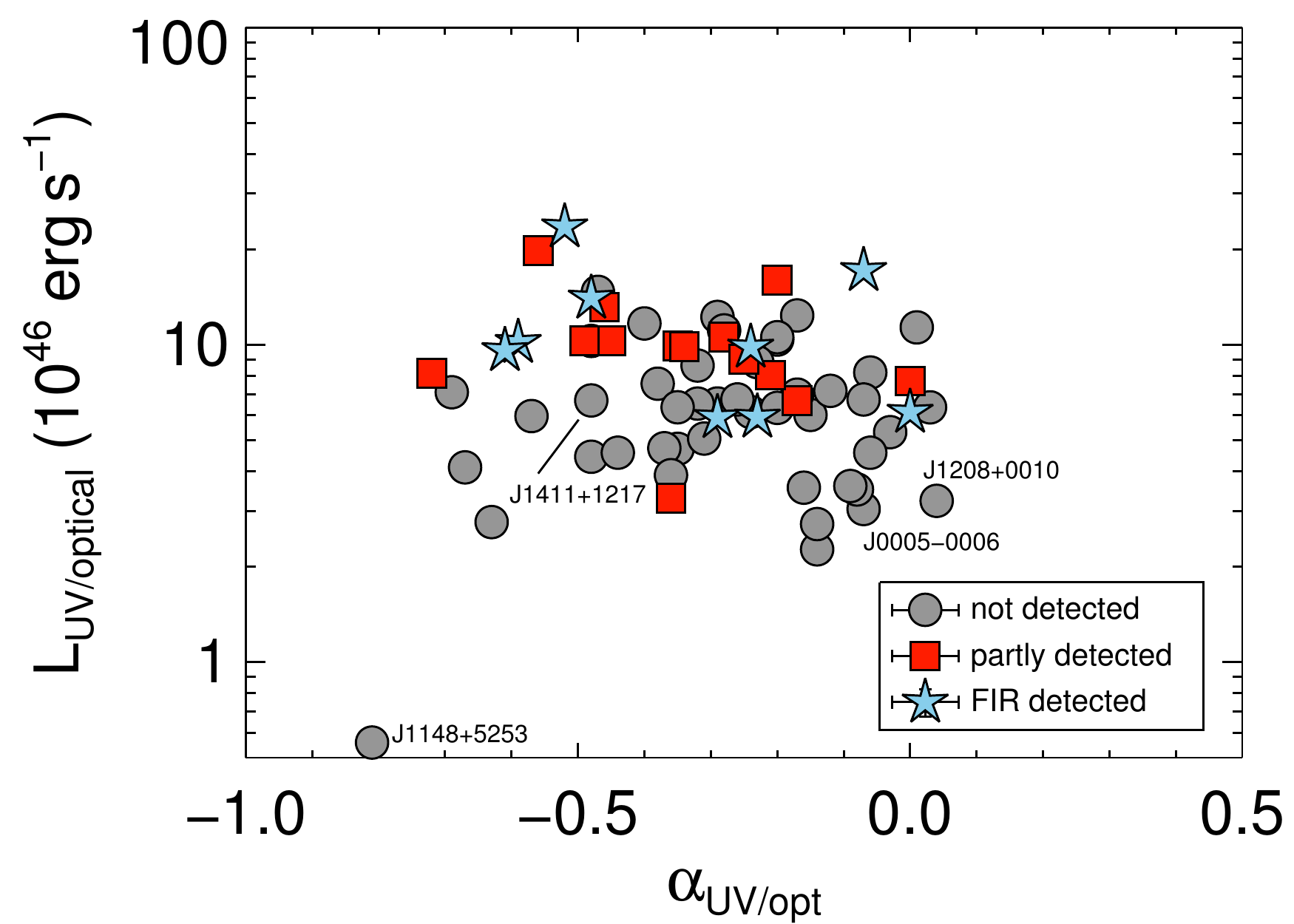}
\includegraphics[angle=0,scale=.4]{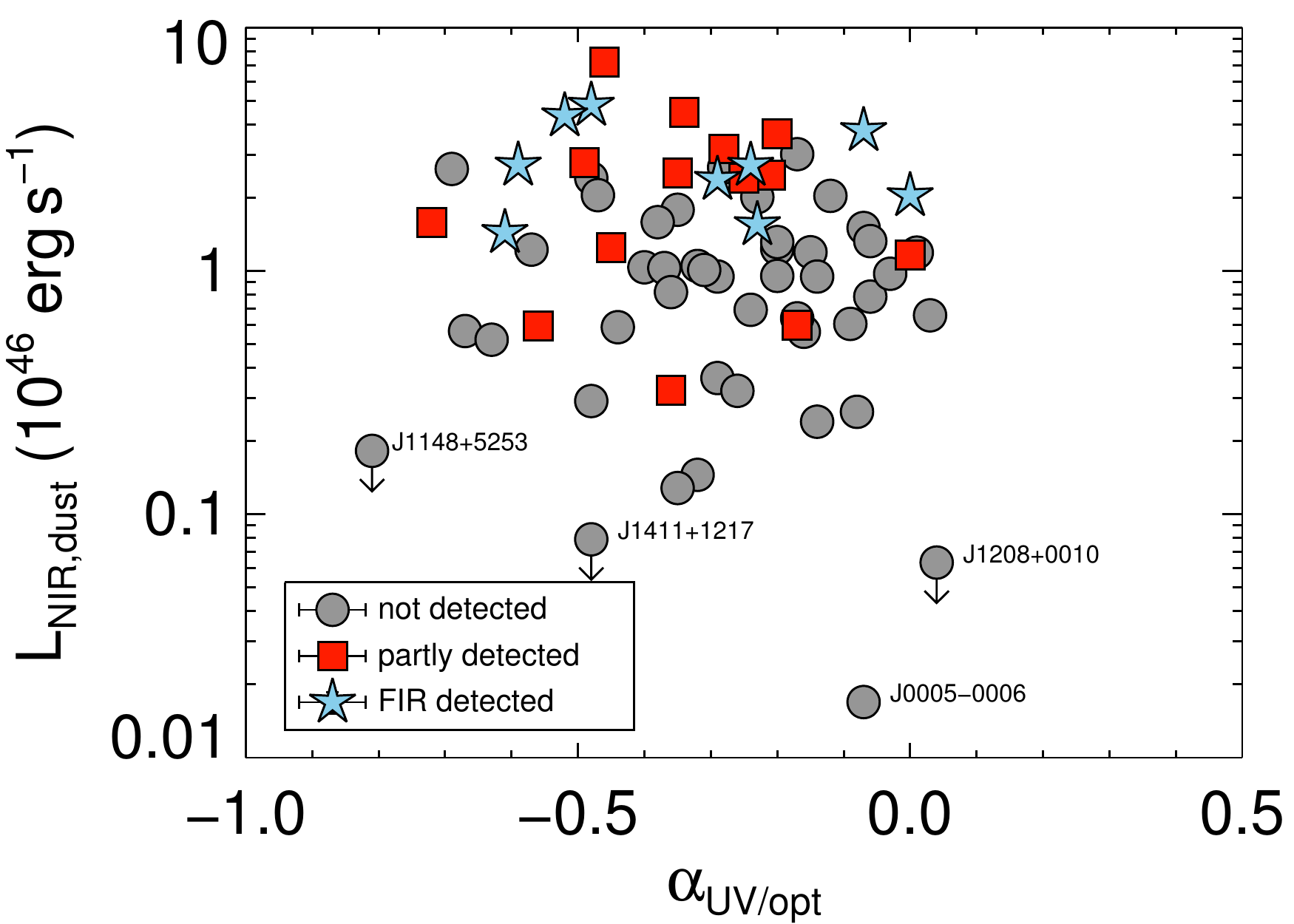}
\caption{Results from the UV/optical plus NIR fits (Section\,\ref{sec:alpha_fits}). Error bars 
in the legend indicate typical errors in the respective panels.
{\it Top, left:} UV/optical to NIR luminosity ratio as a function of UV/optical luminosity. 
The luminosity ratio does not show any obvious trend over the UV/optical luminosity range we 
sample, indicating that both luminosity measures are correlated. There is also no trend 
with redshift seen ({\it top right}) which could indicate either redshift evolution or 
possible artifacts from sampling SEDs at different redshifts with fairly broad filters. 
Both panels show a group of objects which have low luminosity ratio compared to the majority 
of sources. {\it Bottom row:} Both luminosity measures do not show any obvious trends with the
 UV/optical power-law index $\alpha$ ($F_{\nu}$\,$\propto$\,${\nu}^{\alpha}$). \label{uvopt}}
\end{figure*}

In the distributions in Figure\,\ref{uvseds} we also indicate the 
{\it Herschel} FIR
detected objects (blue) and the partly detected objects (red; see
Section \ref{sec:stacking} for the definition of these samples). While
no  specific trends can be identified for $\alpha_{\rm UV/opt}$ or
$T_{\rm NIR}$,  Figure\,\ref{uvseds} reveals that {\it Herschel}
detections are  preferentially found at the high end of the UV/optical
luminosity distribution \citep[see also,][]{net13}. This is even more
pronounced  for the NIR luminosity $L_{\rm NIR,dust}$
(Figure\,\ref{uvseds}, bottom right).

\begin{figure}[t!]
\centering
\includegraphics[angle=0,scale=.45]{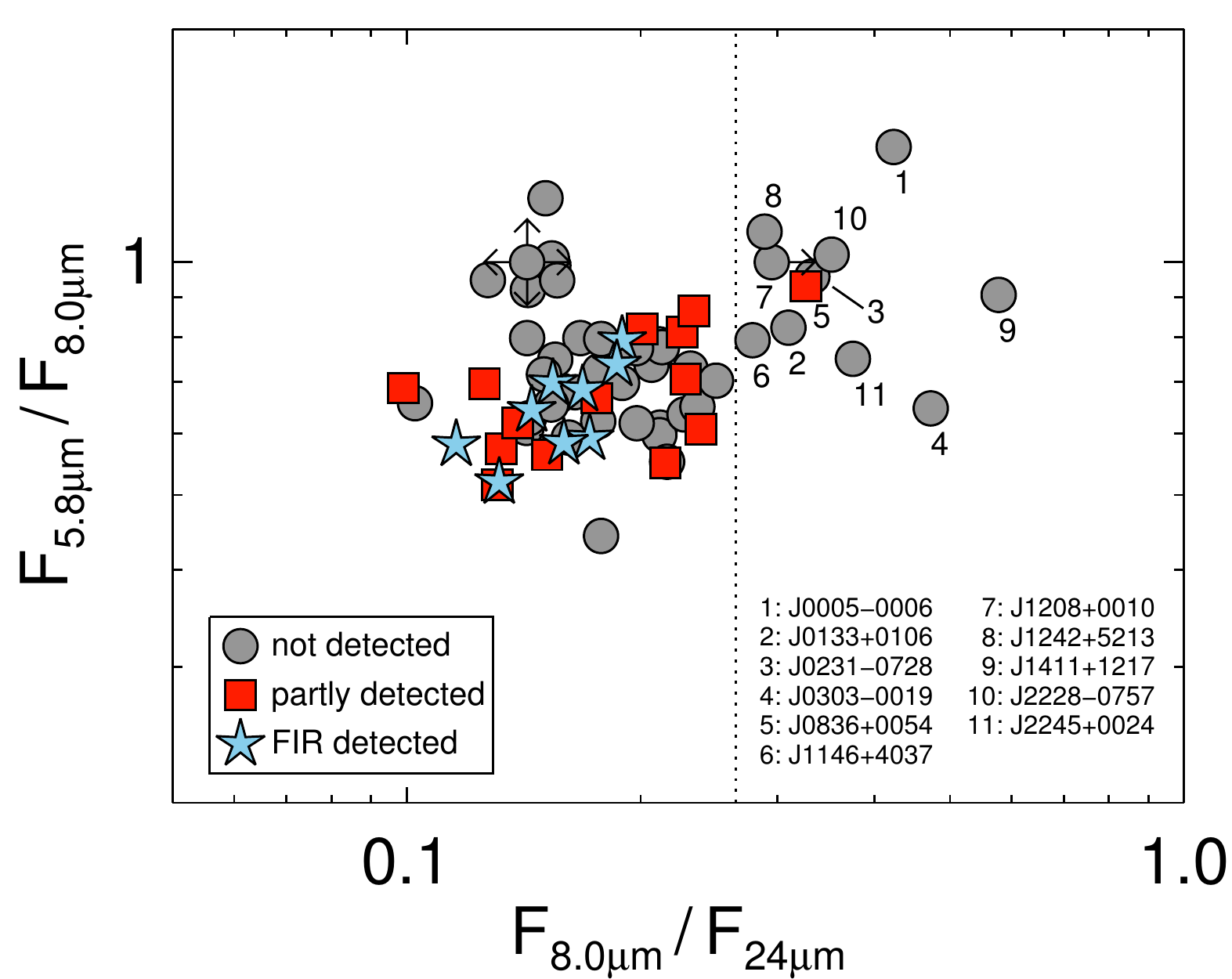}
\caption{Observed flux ratios for three {\it Spitzer} bands demonstrating the selection of 
sources deficient in hot dust emission on observational grounds alone. Most of the QSOs in 
our sample that fall to the right of the dotted line (which indicates $F_{8\mu{\rm m}}/F_{24\mu{\rm m}} = 0.25$) 
also have low $L_{\rm NIR,dust}/L_{\rm UV/opt}$\,$<$\,0.05 (Figure\,\ref{uvopt} and 
Table\,\ref{tab_uvopt_ew}).\label{fnu_ratios}}
\end{figure}

We also find a group of objects which have very low
temperatures of the hot dust component in our fitting approach
(Figure\,\ref{uvseds}, bottom left). We caution that the actual
temperature values provided by our fitting are not well defined in
these cases because the data only poorly constrain  $T_{\rm
NIR}$. Nevertheless, the SEDs clearly demonstrate that these objects
have a dearth of very hot dust compared to their UV/optical luminosity,
and  compared to the remainder of the sample. The reduced
contributions from hot dust  to the SEDs is also reflected in their
lower values for  $L_{\rm NIR,dust}$.\footnote{Instead of integrating
under the poorly constrained  black body, we here follow a different
approach to determine $L_{\rm NIR,dust}$.  First, the observed
photometry is interpolated linearly in log $\nu F_{\nu}$.  Then we
determine $L_{\rm NIR,dust}$ as the  excess emission of the
interpolated photometry over the fitted power-law  contributions
between 1\,$\mu$m and 3\,$\mu$m.}  In the individual SED plots
(Figure\,\ref{seds}) these objects can be identified from their
shallow  rise in flux between the observed bands at 8\,$\mu$m and
24\,$\mu$m (Figure\,\ref{fnu_ratios}). Often, the observed IRAC
8\,$\mu$m data point is still  dominated by the power law and not
by the onset of the hot  dust emission as traced by the
MIPS 24\,$\mu$m photometry. In such cases the 24\,$\mu$m  photometry 
itself only very moderately exceeds the predictions from the power law.

Altogether we find  that $\sim$12--16\% of the sample have NIR to 
UV/optical properties that are quite different from the rest of
the sample.\footnote{The exact number depends on the method used to
identify  the objects, e.g., $L_{\rm NIR,dust}/L_{\rm
UV/opt}$\,$<$\,0.05 in Figure\,\ref{uvopt}, or  $F_{8\mu{\rm
m}}/F_{24\mu{\rm m}}$\,$\gtrsim$\,0.25 in Figure\,\ref{fnu_ratios}.}
Such sources have been found in similar proportions in other samples 
\citep[e.g.,][]{jia10,hao11,mor12}. \citet{jun13} show that 
the fraction of dust-poor quasars increases with optical luminosity and 
redshift, and our numbers are consistent with their trends. These authors 
suggest that the dust-poor phase is a transient phenomenon during the 
evolution of the quasar \citep[e.g.,][]{jia10}, rather than a distinct 
population of quasars with low covering factors \citep[e.g.,][]{hao11}.

The rest of our objects has  $L_{\rm
NIR,dust}$/$L_{\rm UV/opt}$ between $\sim$0.08\,$-$\,0.5 and  we see
no trends in this ratio with redshift or $L_{\rm UV/opt}$ (Figure\,\ref{uvopt}).
The latter implies that the luminosities in the UV/optical and NIR
are well correlated for most objects \citep[see also,][]{mor12}. This
is not surprising considering that the accretion disk emission, here
traced by the UV/optical luminosity, is expected to be 
the  primary heating source of the hot dust. Neither $L_{\rm UV/opt}$
nor $L_{\rm NIR}$ show any trend with $\alpha_{\rm UV/opt}$ 
(Figure\,\ref{uvopt}).
Similarly, $T_{\rm NIR}$ shows no obvious trends with $\alpha_{\rm
UV/opt}$ or  $L_{\rm UV/opt}$, while $\alpha_{\rm UV/opt}$ is
uncorrelated with redshift.

\input{latex_table_optical_nir_results_new_plus_ew.tex}

\begin{figure}[th!]
\centering
\includegraphics[angle=0,scale=.50]{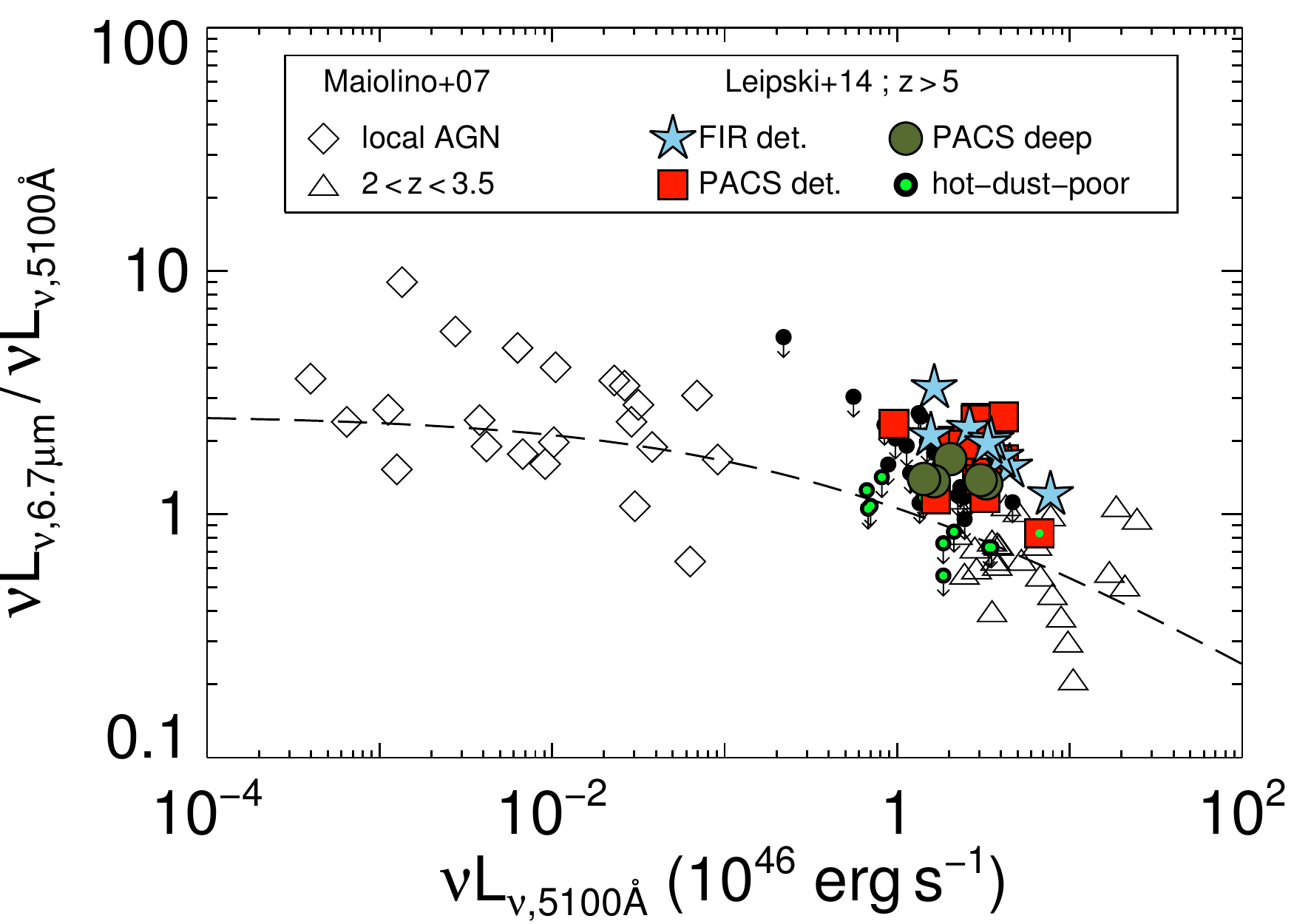}
\caption{The ratio of the MIR luminosity at 6.7\,$\mu$m and the optical 
luminosity at 5100\AA~(all in the rest frame) as a proxy for the dust 
covering factor plotted over the optical luminosity. Open symbols as well 
as the dashed trend line are taken from \citet{mai07}. Filled symbols 
refer to all 68 high-redshift QSOs from this work for which the relevant 
luminosities could be determined. 
The dark green circles show the 
five quasars for which the deep re-observations resulted 
in a 100\,$\mu$m detection. Data points marked with bright green dots 
represent objects with a dearth of hot dust as determined from 
Figure\,\ref{flux_ratios}. Many of the sources with 3$\sigma$ 
upper limits in the plotted luminosity ratio populate the same area 
as (and are concealed by) the {\it Herschel} detections (i.e., 
by the filled colored symbols).
\label{covering_factor} }
\end{figure}

By including the PACS 100\,$\mu$m band, we extend the analysis to 
slightly longer infrared wavelengths and  determine the flux at a
rest frame wavelength of 6.7\,$\mu$m through interpolation of the observed
photometry at 24\,$\mu$m and 100\,$\mu$m using a power
law in $\nu F_{\nu}$. In combination with the monochromatic luminosity at
5100\AA~(rest frame) as provided by our UV/optical power-law fits, we
can study the monochromatic ratio of MIR-to-optical emission as a function of
(monochromatic) optical luminosity (Figure\,\ref{covering_factor}).
Under the assumption that the infrared-to-optical luminosity ratio is a
proxy of the dust covering factor in (type 1) AGN, a plot as in
Figure\,\ref{covering_factor} has been used by \citet{mai07} to
identify a trend where the dust covering factor decreases with
increasing optical luminosity. Such a  general behavior of the dust
covering factor (or obscured fraction of AGN) has been detected for
many different samples and using various techniques \citep[e.g.][and
references therein]{tre08,has08,lus13} and is also seen in our sample 
for the {\it Herschel} detected objects (Figure\,\ref{lumratrios_fit}). 
However, the question whether the
covering factor also  changes with redshift remains controversial,
with claims for \citep{tre06,has08} and against \citep{ued03,lus13}
significant  redshift evolution.

In this context, our high-redshift
QSOs show a systematic, albeit very moderate offset in the MIR-to-optical 
luminosity ratio with respect to $2.0
\lesssim z \lesssim 3.5$ QSOs (Figure\,\ref{covering_factor}). 
Much of the observed offset is currently driven by the {\it Herschel}
detections where the 6.7\,$\mu$m flux is determined as the interpolation
between two significantly detected data points at $\lambda_{\rm obs}$ of 
24\,$\mu$m and 100\,$\mu$m. 
However, about 60\% of the 
$z>5$ objects have only upper limits on the MIR-to-optical ratio, mostly 
due to non-detections in the 100\,$\mu$m band. While in 
Figure\,\ref{covering_factor} these objects 
currently populate the same area as the 
{\it Herschel} 100\,$\mu$m detected objects (colored symbols), 
their effect on the observed trends remain unclear. This is in particular 
emphasized when considering the wide range of intrinsic SED shapes that 
may be present among the {\it Herschel} non-detected sources 
(see Section\,\ref{sec:stacking}), which could potentially result in 
a wider range of luminosity ratios than seen currently for the {\it Herschel} 
detected objects. For example, if the $z>5$ objects intrinsically showed the 
same spread in the MIR-to-optical ratio as the $2.0 \lesssim z \lesssim 3.5$ 
sample (almost 1 dex in Figure\,\ref{covering_factor}) then the resulting 
distribution would be roughly consistent with the observed trends at 
lower redshift. 

On the other hand Figure\,\ref{covering_factor} reveals 
that the data points and upper 
limits with the lowest MIR-to-optical ratio among our 
sample in Figure\,\ref{covering_factor} almost 
exclusively belong to the group of objects where the
SEDs indicate a dearth of hot dust. These objects may be of different 
nature or reside in a different evolutionary state 
\citep[e.g.,][]{jia10,hao11,mor11,jun13} and possibly cannot be 
directly compared with the other QSOs from either sample.

\subsection{H$\alpha$ equivalent widths}\label{sec:halpha}

H$\alpha$ is one of the most prominent emission lines in the
UV/optical spectra of common quasars  \citep[e.g.,][]{van01}. For our
high-redshift objects ($z>5$), this emission line is  redshifted into
the observed mid infrared, largely precluding direct spectroscopic
observations with current facilities.   However, we can use our 
high signal-to-noise {\it Spitzer} photometry to estimate H$\alpha$ 
fluxes. For redshifts greater than
$z\sim5.2$, the influence of
this line can be seen in the individual SEDs (Figure\,\ref{seds}) 
where H$\alpha$ emission boosts the flux in the 4.5\,$\mu$m IRAC
band compared to a power-law continuum (e.g., J0840+5624). At lower redshift ($z<5.2$), H$\alpha$
falls onto the flanks of the filter transmission or largely into the
small gap between the 3.6\,$\mu$m and 4.5\,$\mu$m IRAC filter. This
makes it difficult to extract reliable emission-line flux estimates
from the observed photometry. At
$z\gtrsim5.2$ and up to our maximum redshift ($z=6.42$), the H$\alpha$
emission line is  fully covered by the filter transmission window and
peaks within the plateau region of the 4.5\,$\mu$m filter.\footnote{We  here
assume a rest frame line width for H$\alpha$ as determined from the
SDSS composite spectrum  \citep{van01}.}

\begin{figure*}[t!]
\centering
\includegraphics[angle=0,scale=.4]{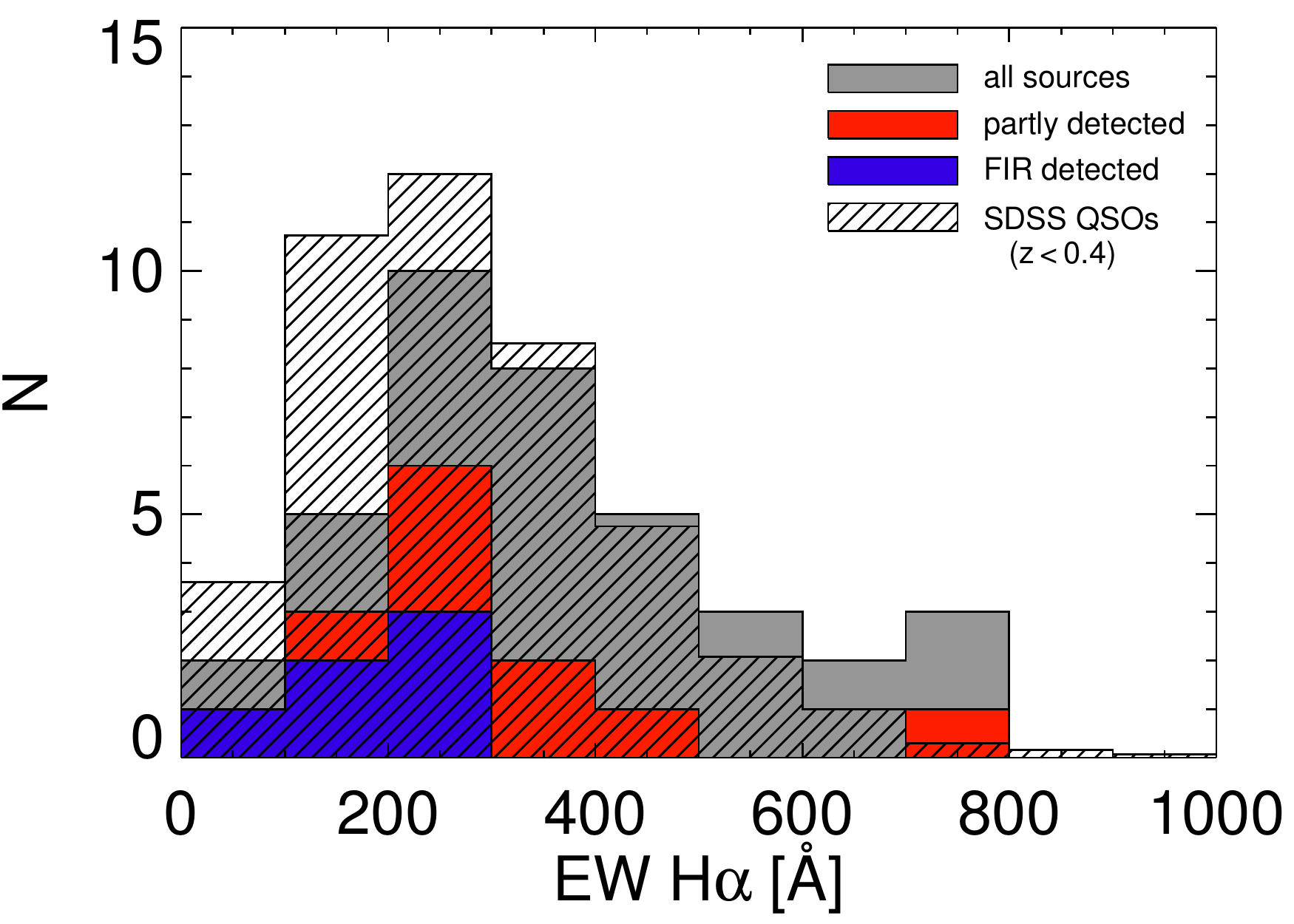}
\includegraphics[angle=0,scale=.4]{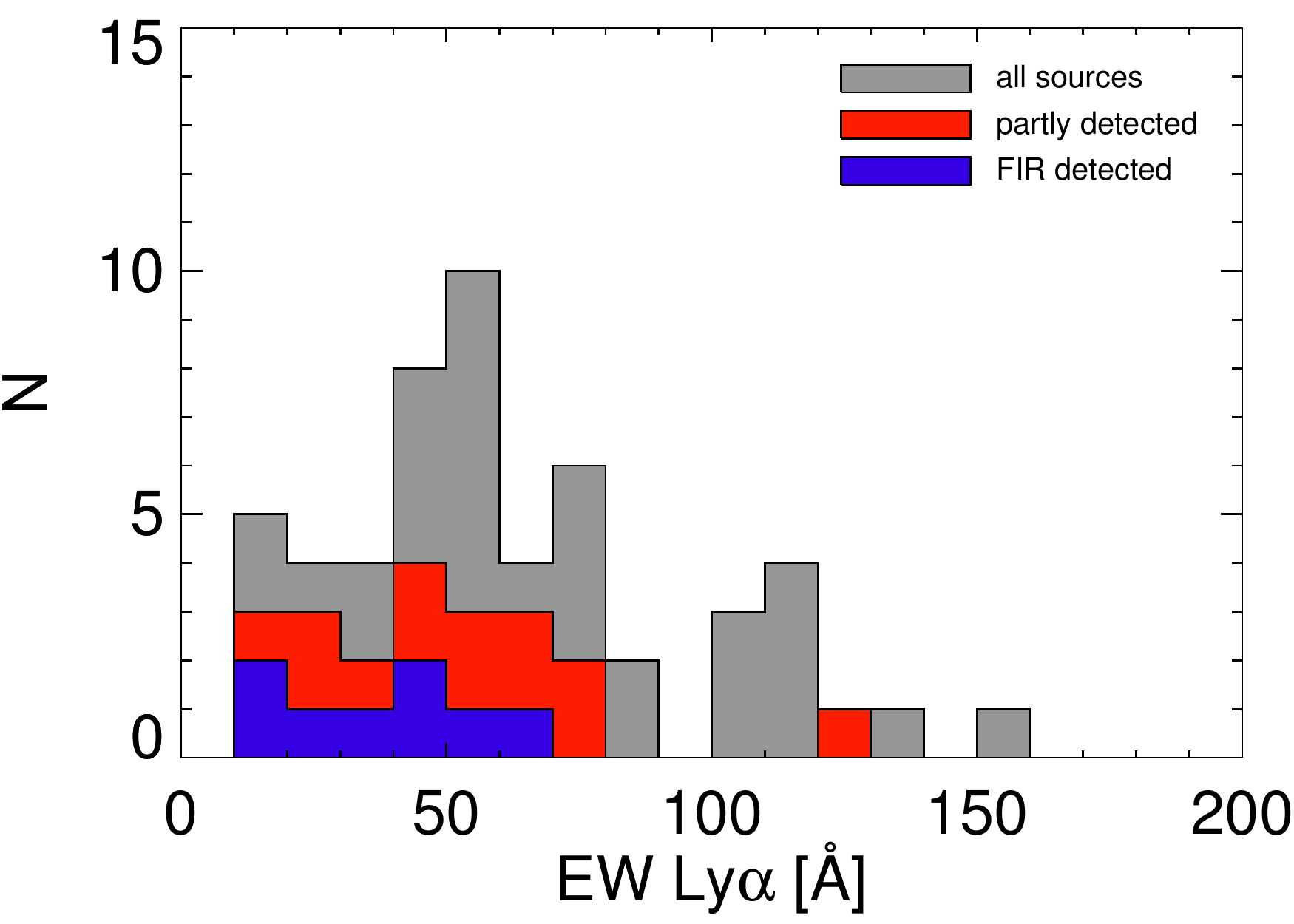}
\caption{{\it Left}: Distribution of H$\alpha$ equivalent widths in the sample 
(grey filled area, estimated as outlined in Section\,\ref{sec:halpha}. As a 
hashed region we show the histogram of the H$\alpha$ equivalent widths of 
$\sim$4800 SDSS quasars taken from \citet{she11}. {\it Right}: Distribution of 
Ly$\alpha$ equivalent widths for our high-$z$ sample, taken from the 
literature.\label{halpha}}
\end{figure*}

For the purpose of estimating H$\alpha$ line fluxes we fit the SEDs
slightly  differently as compared to Section\,\ref{sec:alpha_fits} or
as shown in Figure\,\ref{seds}. Instead of  considering the full
UV/optical continuum for a power-law fit we now limit the fit to the
neighboring  photometric points in order to isolate the local
continuum. This means that for an H$\alpha$ line  falling into the
4.5\,$\mu$m band we fit the power law to the 3.6\,$\mu$m and
5.8\,$\mu$m bands only. From the offset of the  measured flux in the
4.5\,$\mu$m filter compared to the local estimate of the power-law
continuum we then calculate the H$\alpha$ emission-line flux and
equivalent width (EW).\footnote{For redshifts  $z\gtrsim6.0$ the
H$\beta$ emission line enters the 3.6\,$\mu$m band, thus potentially
increasing the  flux in this filter compared to the underlying
continuum. In our approach this would result in slightly
underestimated H$\alpha$ fluxes due to a steeper  
fitted continuum (in $\nu F_{\nu}$). However, H$\beta$ is expected to be a factor  of
$\sim$3 fainter than H$\alpha$ and its effect on the H$\alpha$ EWs is considered negligible
here.}  We show  the distribution of the estimated H$\alpha$ EWs in
Figure\,\ref{halpha} and the derived values are also provided in
Table\,\ref{tab_uvopt_ew}. When comparing our high-$z$  results to
spectroscopic H$\alpha$ EWs from low redshift ($z \lesssim 0.4$) SDSS quasars
\citep{she11}, we see that the two distributions are quite
similar in width and shape. This similarity between low and
high-redshift quasars indicates a lack of redshift evolution in the
H$\alpha$ EWs, which agrees with similar results for rest-frame 
UV emission lines \citep{iwa04,jua09,der11}.

From Figure\,\ref{halpha} ({\it left}) we can also see that the  {\it
Herschel} detected objects, and in particular the FIR detected
objects, have preferentially low H$\alpha$ EWs. This trend is also
seen in the Ly$\alpha$ EWs (Figure\,\ref{halpha}, top right) as taken
from the literature (Table\,\ref{tab_uvopt_ew}). Such a prevalence of FIR
bright objects among sources with low Ly$\alpha$ EW has previously
been indicated for mm-detected high-redshift quasars
\citep[e.g.,][]{omo96,ber03,wan08b}.  \citet{wan08b} speculated that
in these objects a special dust geometry that only affects the broad
emission line clouds (and not the continuum) could in principle lead
to such an effect.   Because the impact of dust obscuration on
H$\alpha$ would be much smaller than for  Ly$\alpha$, the persisting
trend of {\it Herschel} FIR detected objects to be found  at low H$\alpha$
EW values questions such a scenario. However, while the effects of
obscuration are indeed reduced for H$\alpha$ compared to Ly$\alpha$,
they can still be  non-negligible. A more definite answer requires
higher precision direct spectroscopic  measurements, preferably of
(rest frame) NIR emission lines, to further reduce  the
effect of possible dust obscuration.

\begin{figure*}[t!]
\centering
\includegraphics[angle=0,scale=.18]{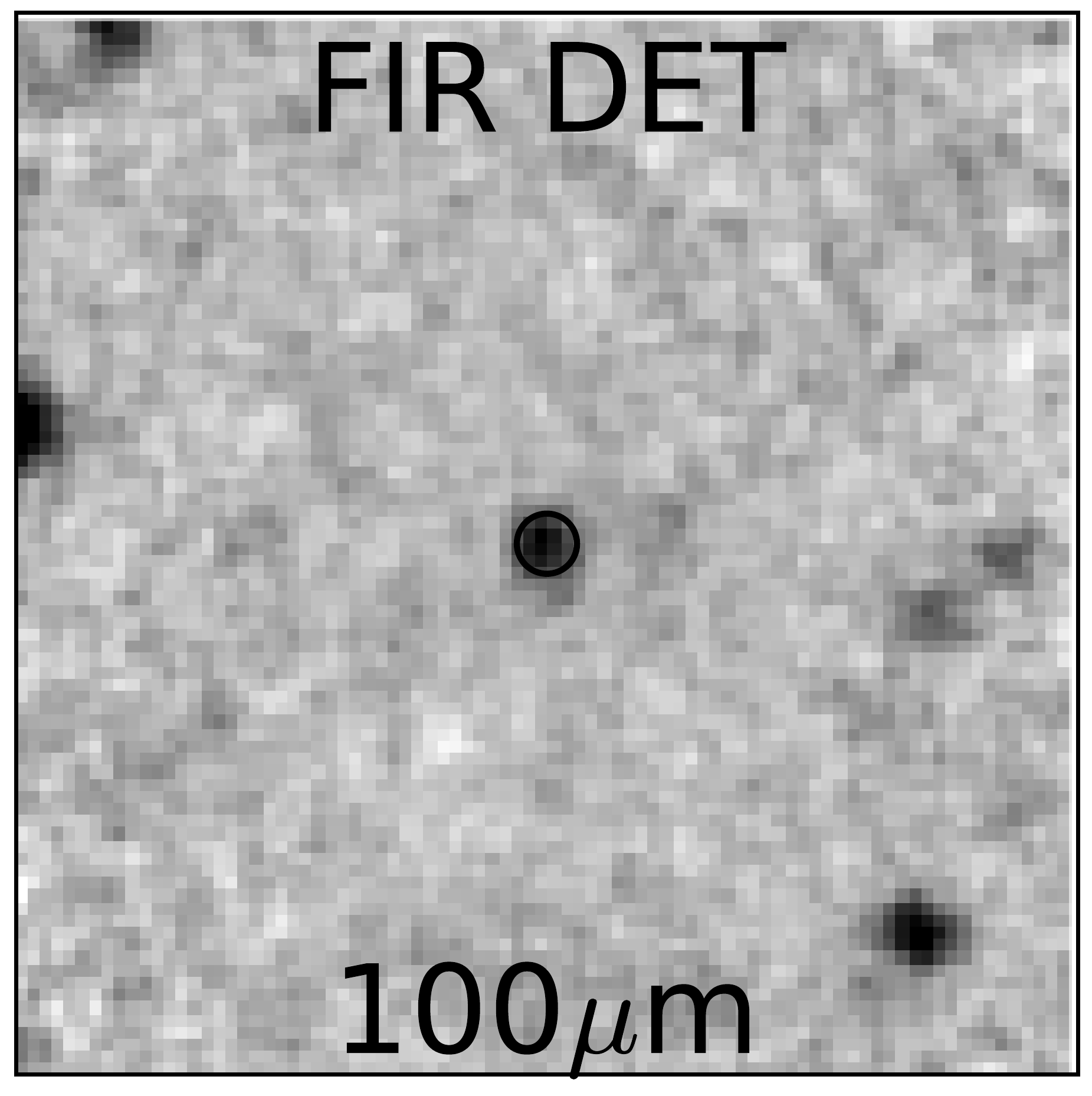}
\includegraphics[angle=0,scale=.18]{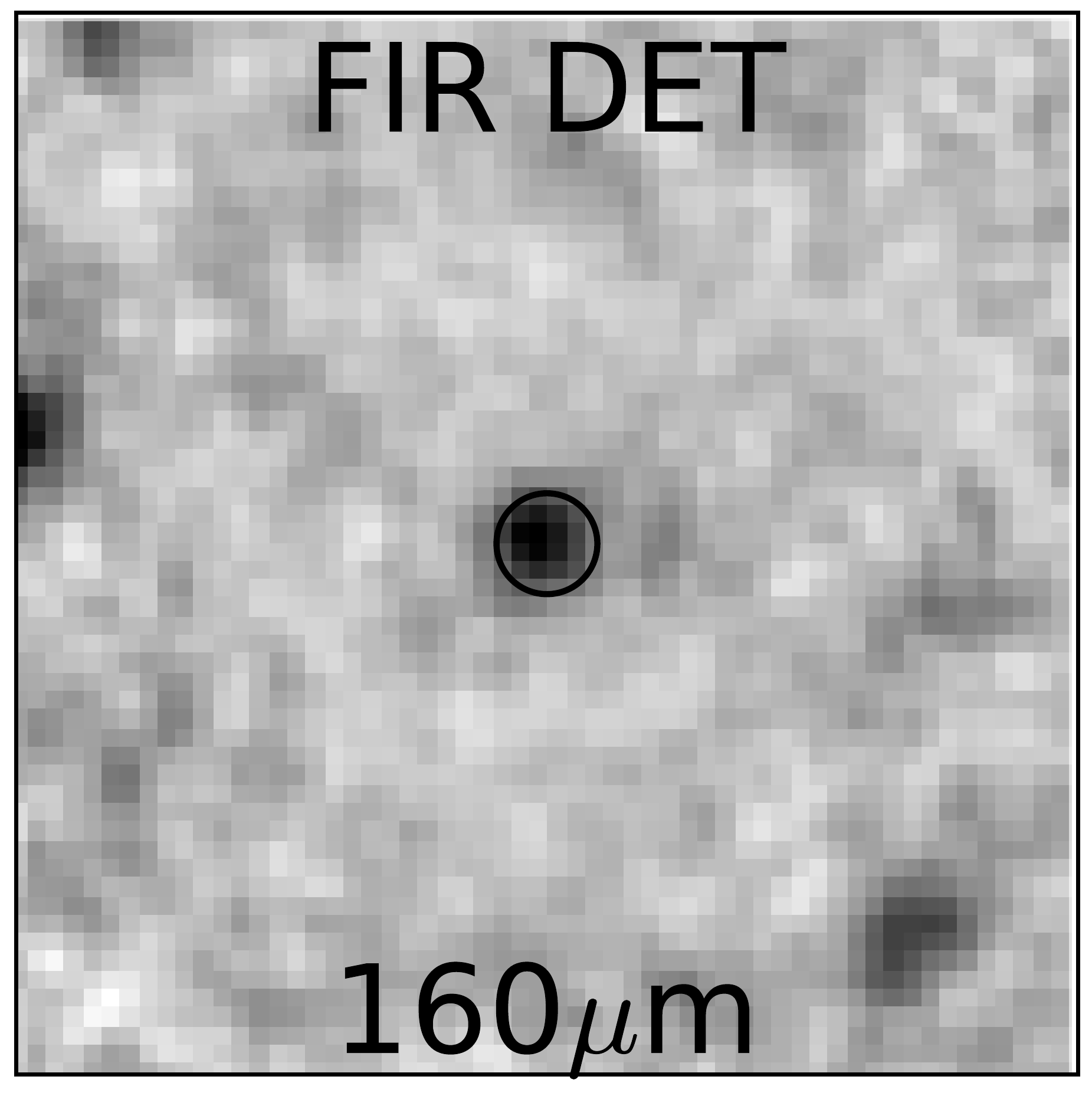}
\includegraphics[angle=0,scale=.18]{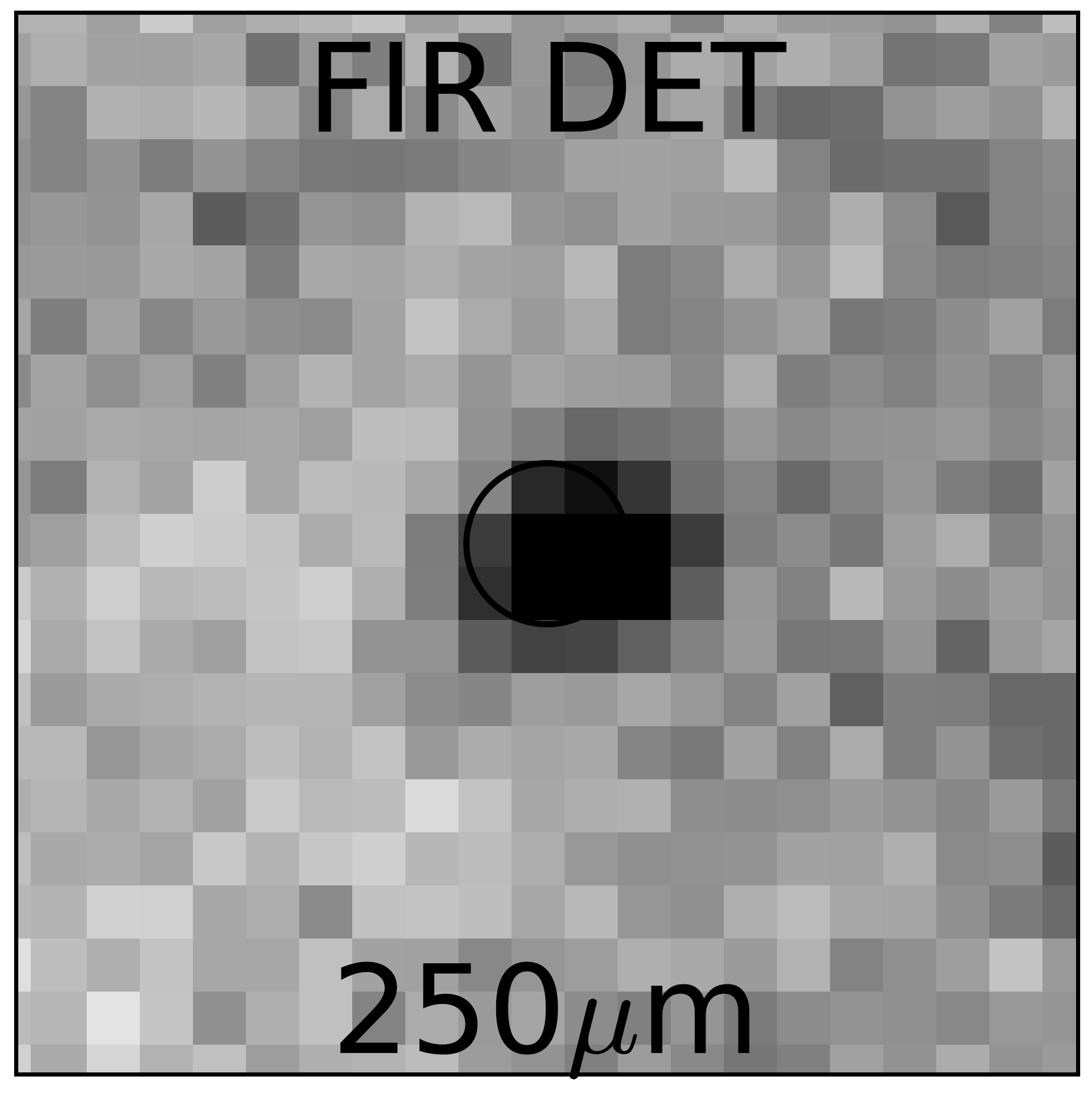}
\includegraphics[angle=0,scale=.18]{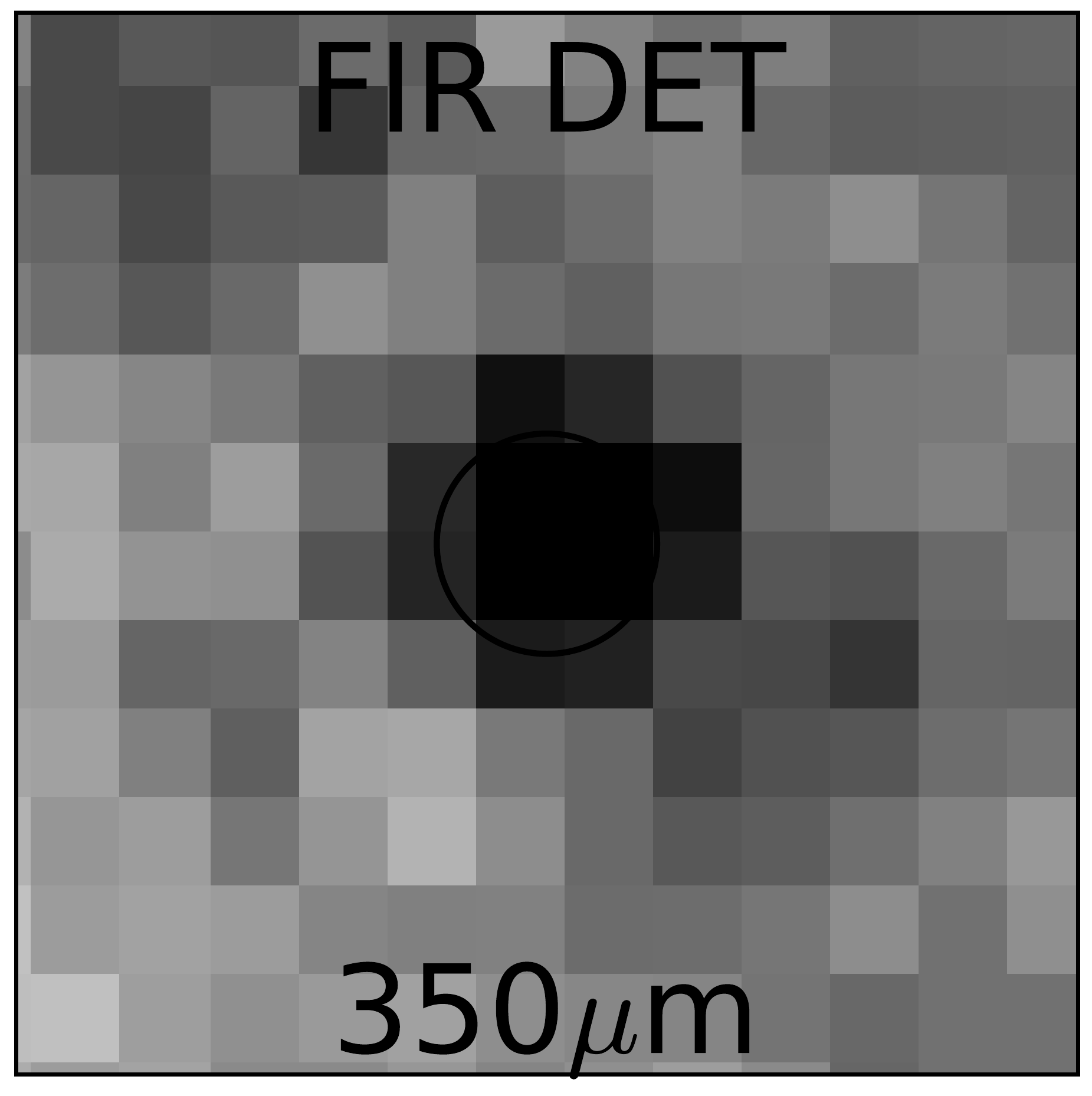}
\includegraphics[angle=0,scale=.18]{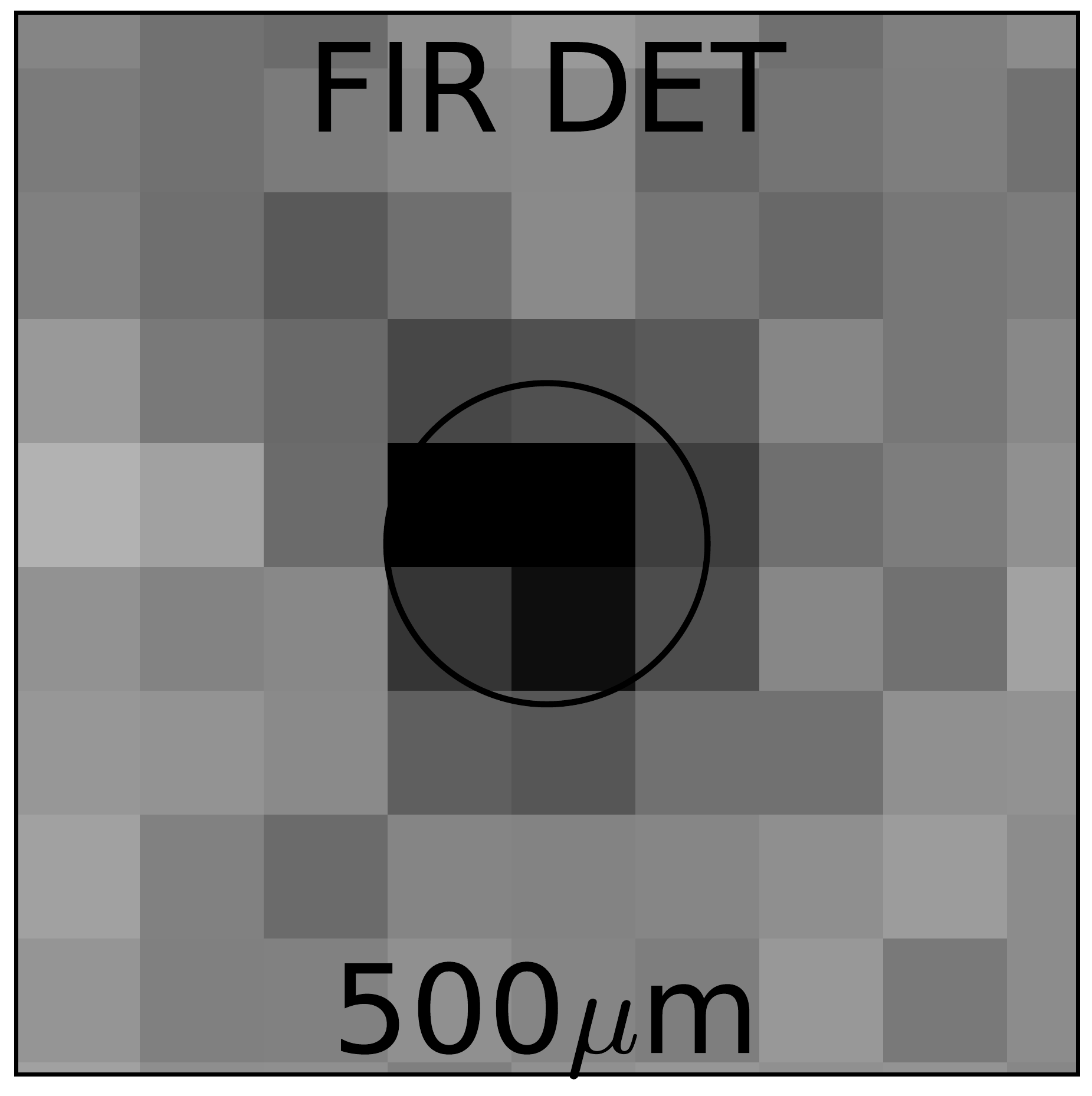}\\
\includegraphics[angle=0,scale=.18]{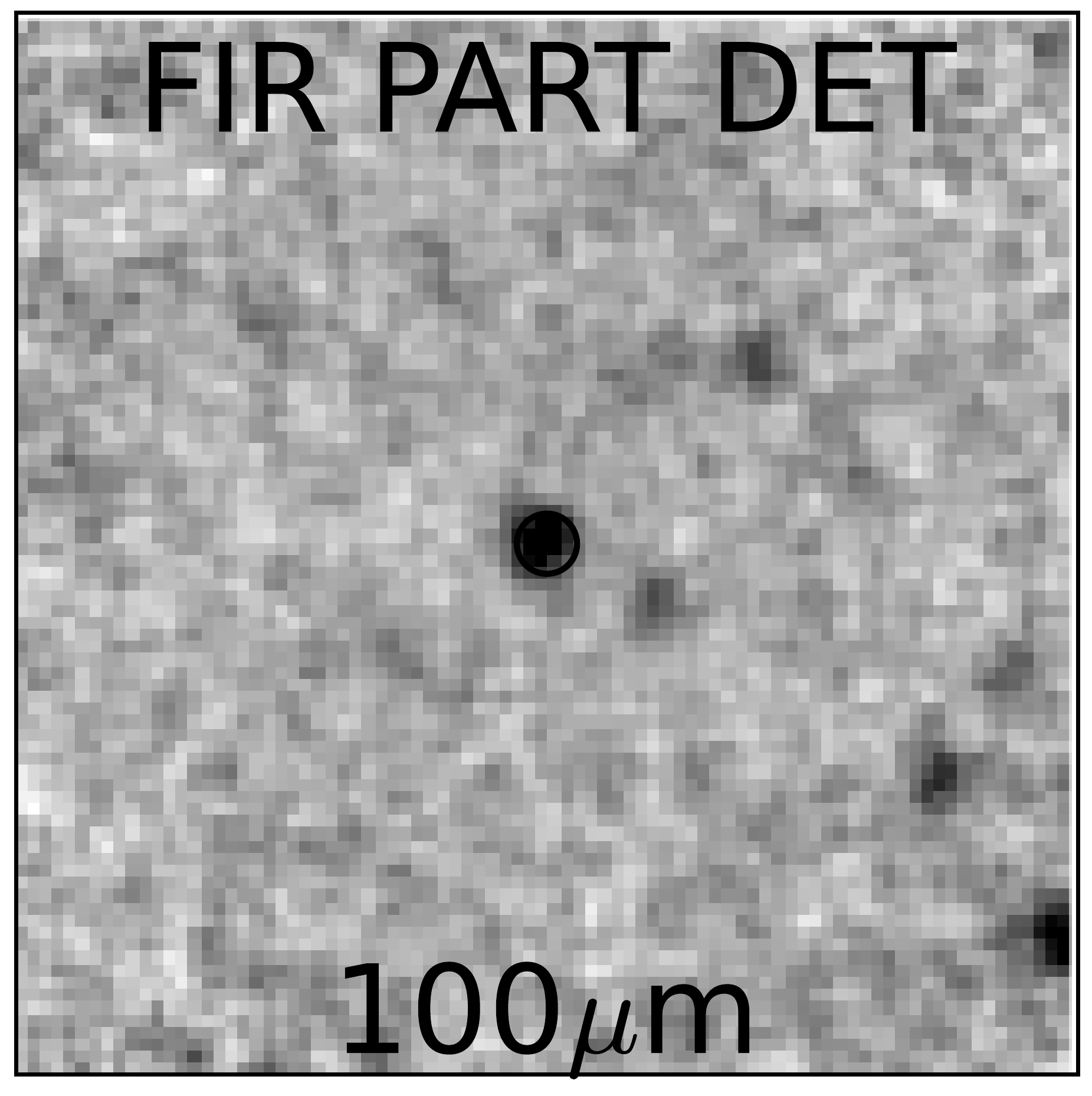}
\includegraphics[angle=0,scale=.18]{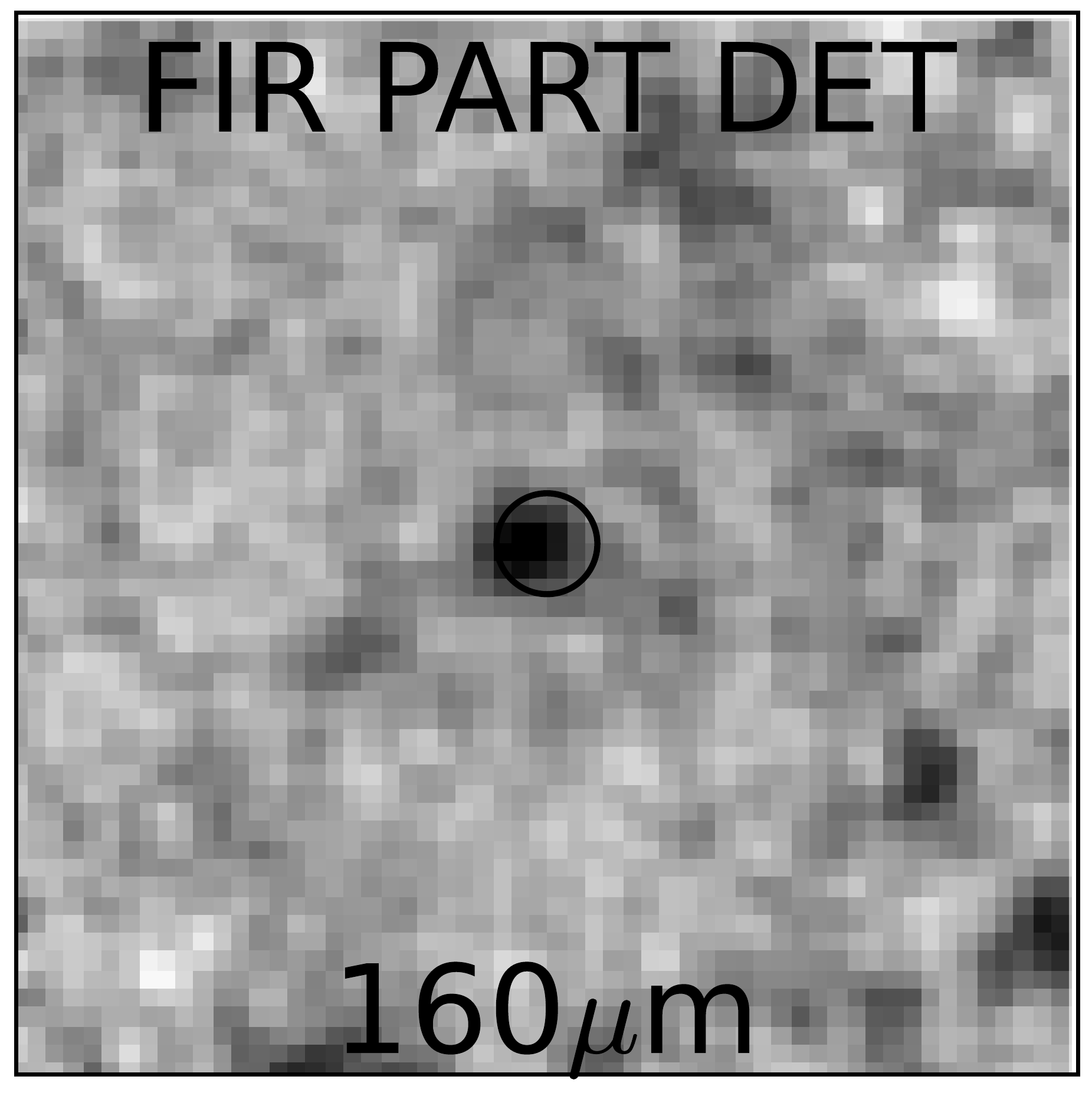}
\includegraphics[angle=0,scale=.18]{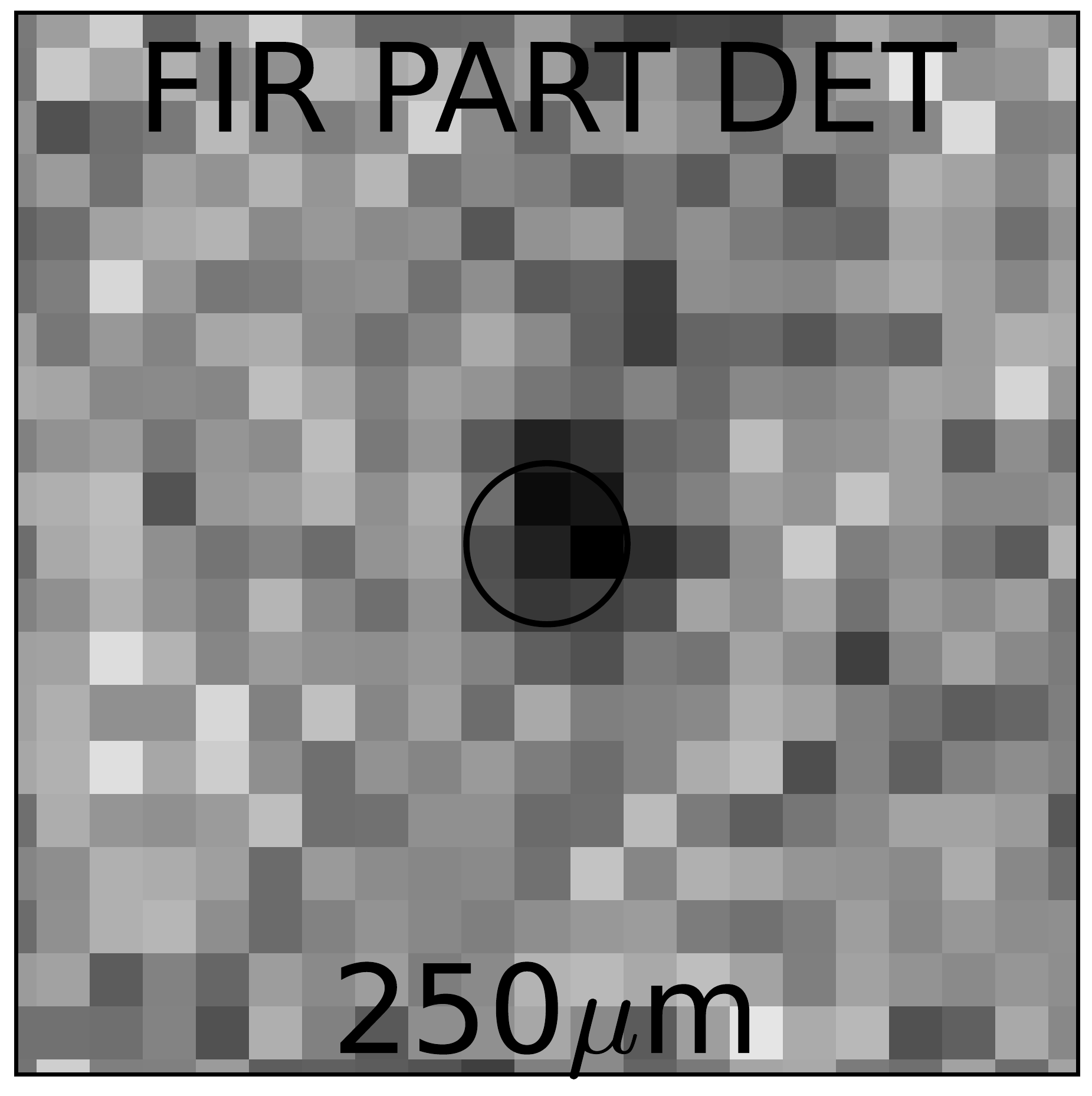}
\includegraphics[angle=0,scale=.18]{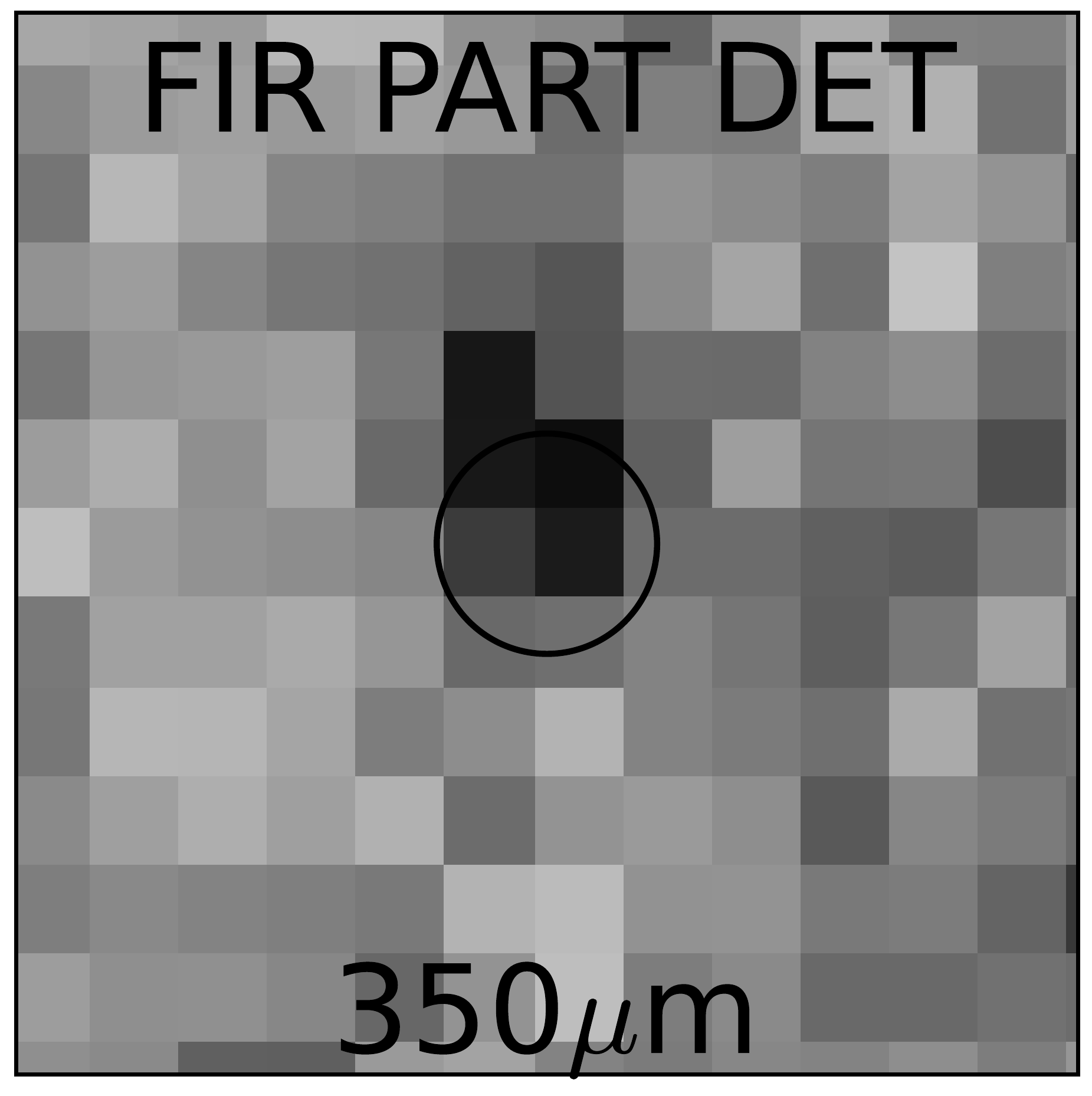}
\includegraphics[angle=0,scale=.18]{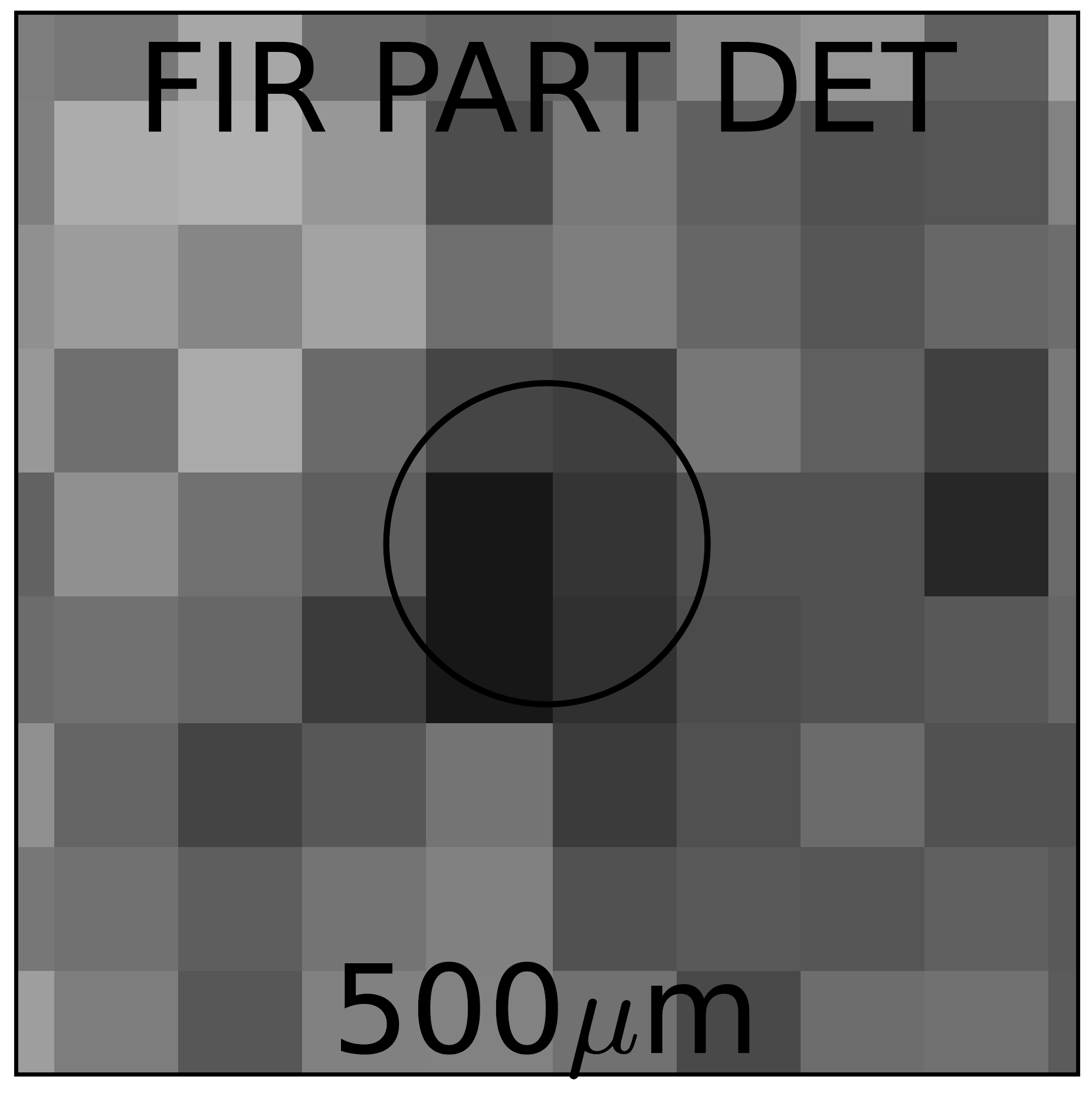}\\
\includegraphics[angle=0,scale=.18]{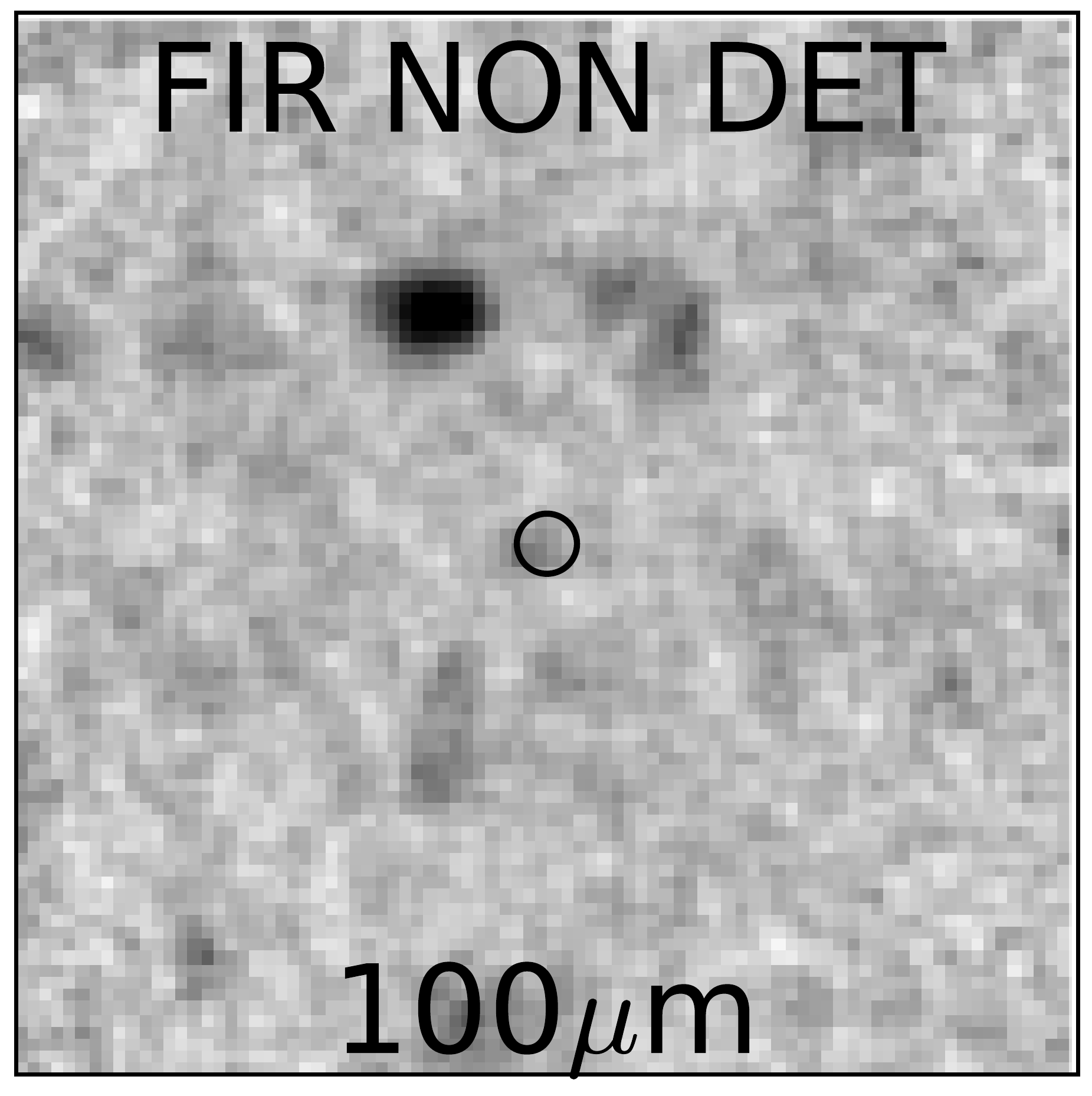}
\includegraphics[angle=0,scale=.18]{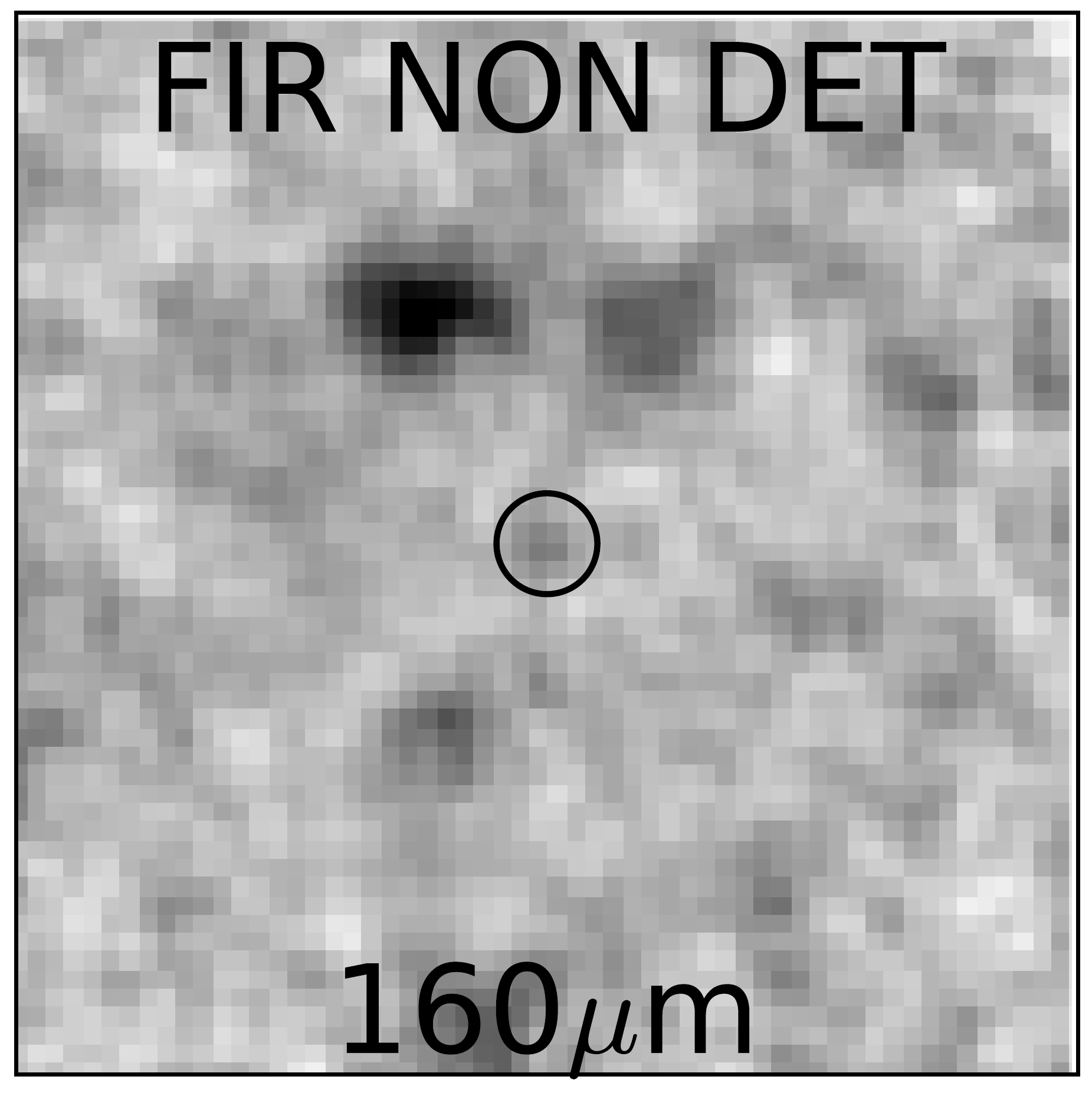}
\includegraphics[angle=0,scale=.18]{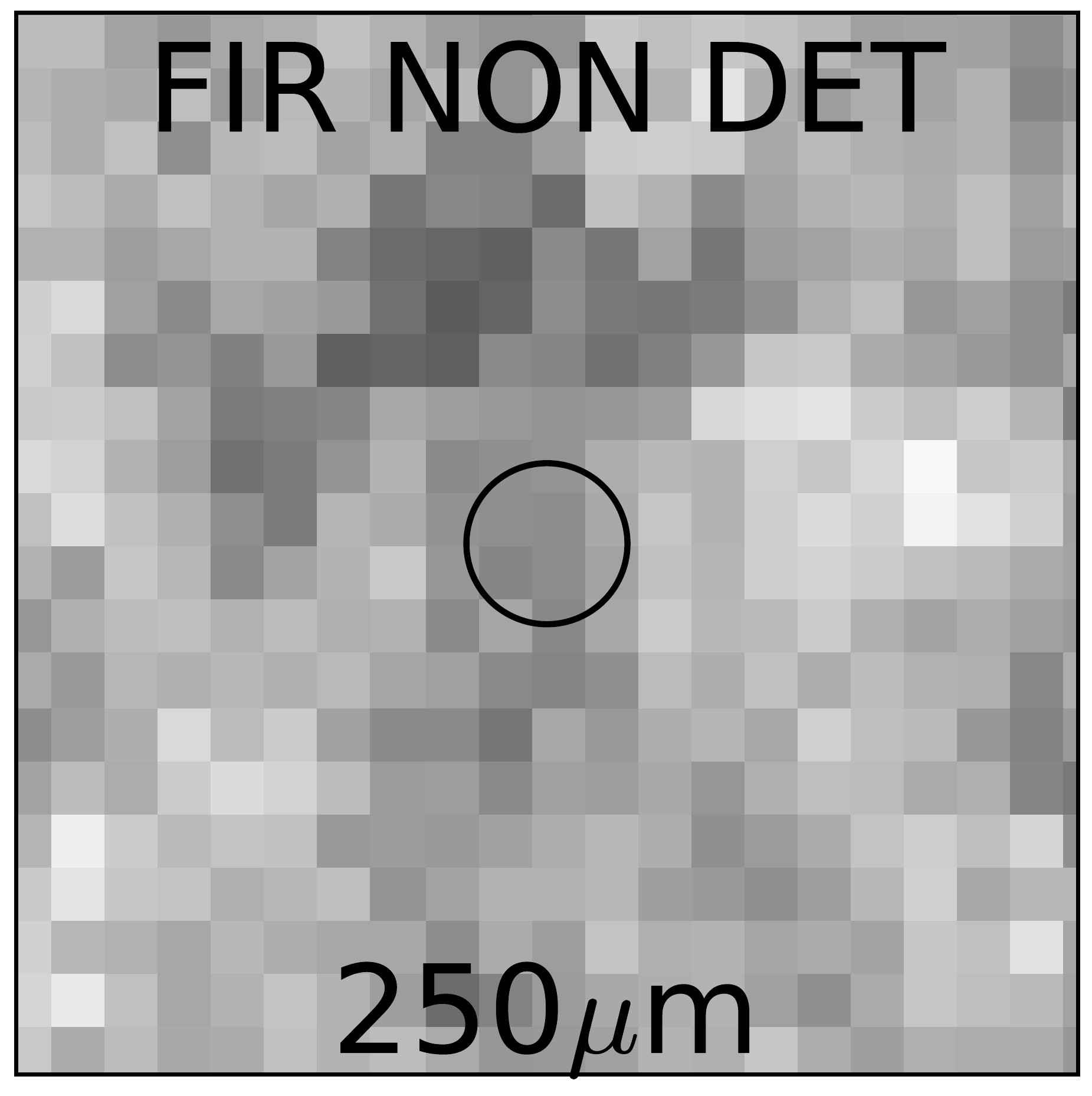}
\includegraphics[angle=0,scale=.18]{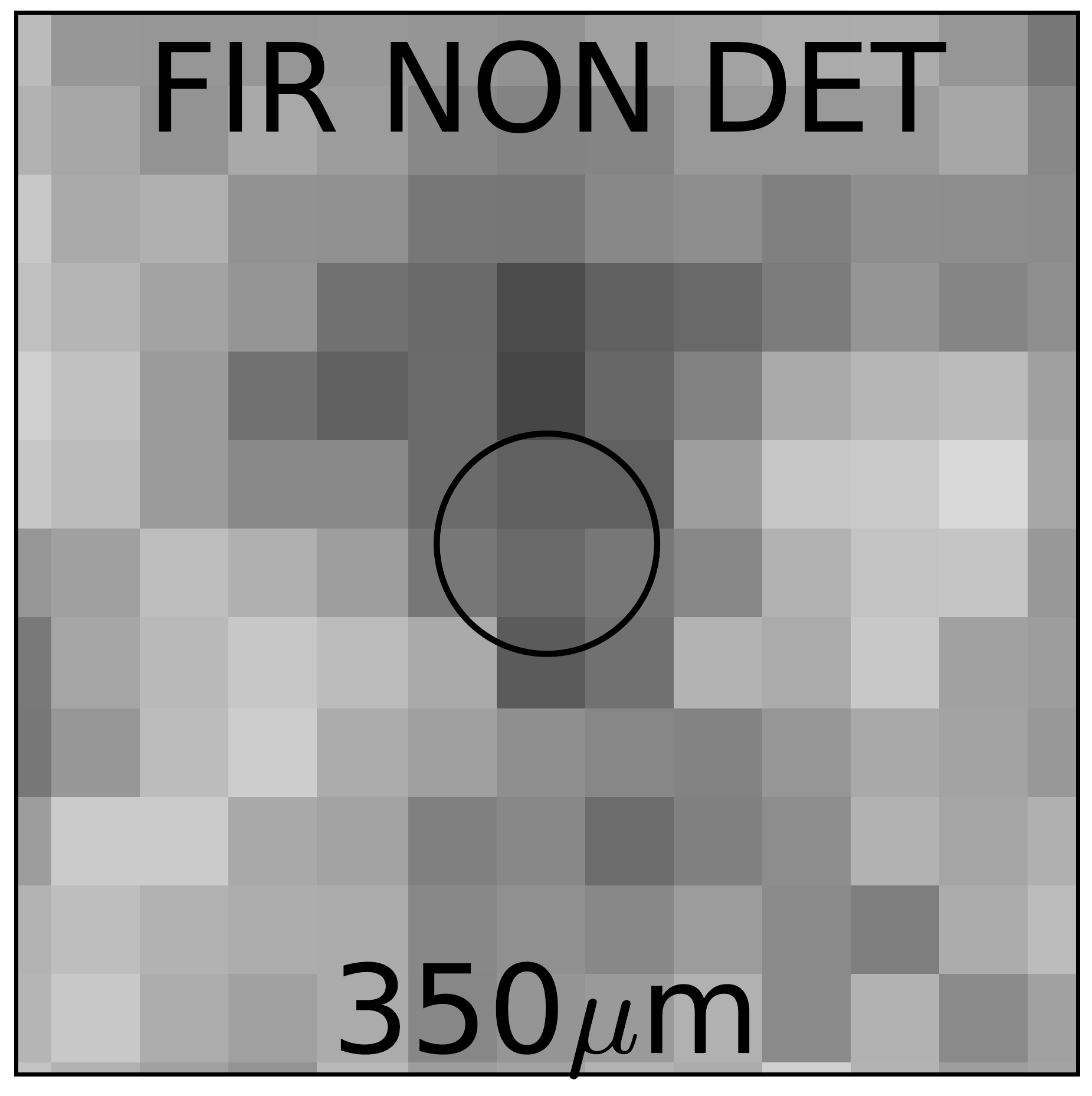}
\includegraphics[angle=0,scale=.18]{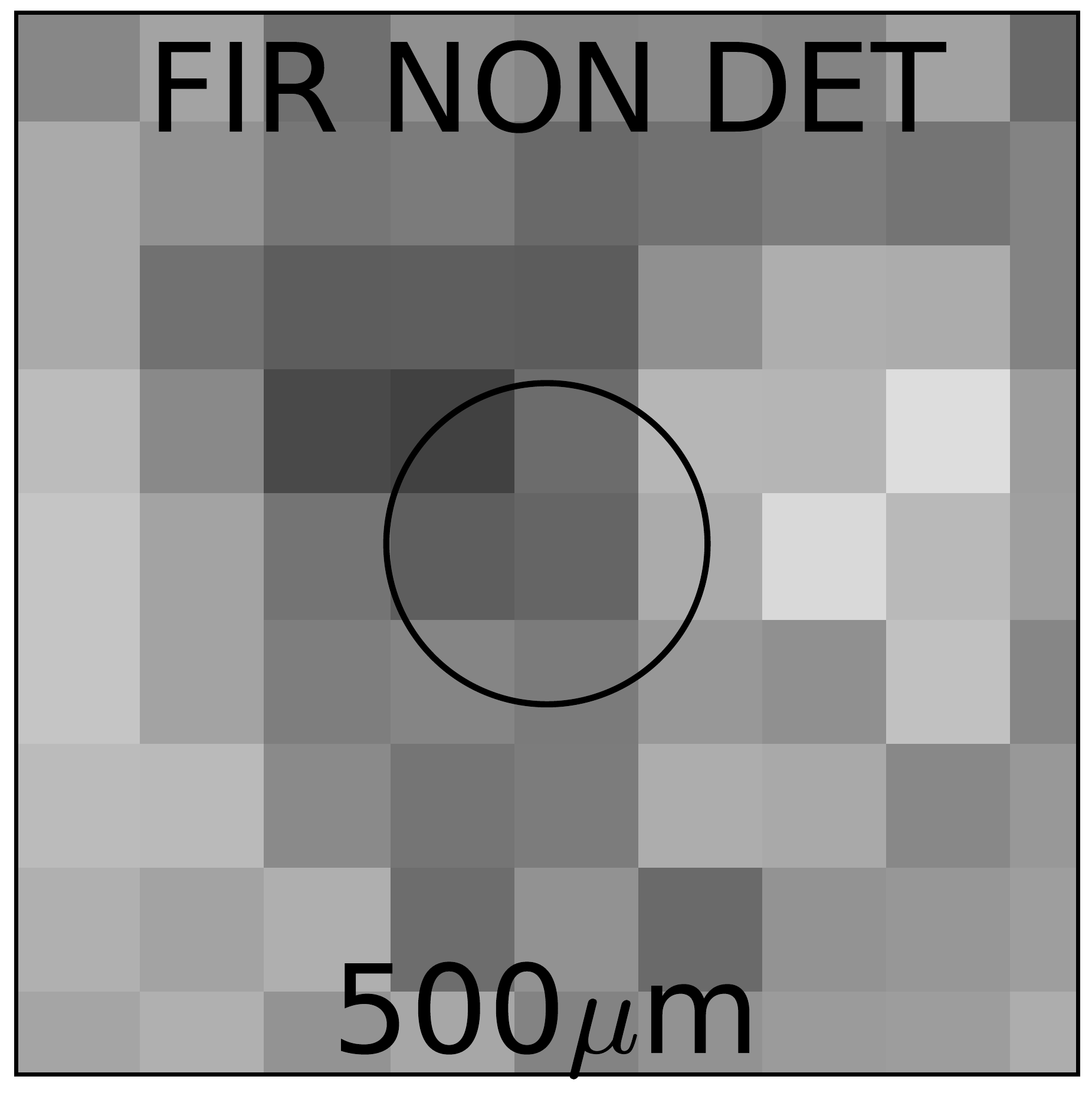}
\caption{Stacked images in the {\it Herschel} bands for the three subsamples outlined in 
section \ref{sec:stacking}. The images have a size of 2\arcmin on a side, the circle 
indicating the central position has a diameter corresponding to the FWHM at the 
respective wavelength.\label{stacked_images}}
\end{figure*}

\subsection{Stacking}\label{sec:stacking}

Due to the large number of {\it Herschel} non-detections in our sample,
we have used a stacking  approach to study the average infrared
properties of the high-redshift quasars. For this purpose  we divided
the full sample into three subsamples:\footnote{We
here use the standard  {\it Herschel} data and do not include the
additional deep photometry available for six objects.}

\begin{enumerate}
\item 10 FIR-detected objects with detections in at least three 
  {\it Herschel} bands (160\,$\mu$m, 250\,$\mu$m, and 
  350\,$\mu$m).\footnote{Except for J0927+2001 all these objects are 
  also detected at 100\,$\mu$m.}
\item 14 partly ({\it Herschel}) detected objects with significant 
  PACS 100\,$\mu$m and/or 160\,$\mu$m flux. We refer to this subsample 
  also as the PACS-only objects.\footnote{Although two of 
    these, J0957+0610 and J1443+3623, are also seen at the 
    $\lesssim$3$\sigma$ level at 250\,$\mu$m.}
\item 33 ({\it Herschel}) non-detections.
\end{enumerate}

The remaining objects have been excluded from the stacking analysis on
various grounds: 10 objects  suffer from confusion with nearby FIR
bright sources which would influence the stacked fluxes.  The science
targets on these images are not detected with {\it Herschel} individually. Two
additional sources, J1148+5253 and  J2245+0024, have been excluded
because  they are significantly fainter in the optical/UV than the
rest of the sample. Both are also {\it
Herschel} non-detections.

The stacking was performed in flux in the observed 
frame and the resulting mean SEDs were shifted into rest frame 
using the median redshift in the respective subsamples. In the y-band
and in the {\it Spitzer} bands (where virtually all objects in
all three subsamples are individually detected) we used the observed
photometry as input. For {\it Herschel} the final images where stacked pixel by
pixel centered on the position of the
quasar. Photometry on the resulting stacks was performed as described for
the individual images (Sections\,\ref{sec:pacs} and \ref{sec:spire}).
The mean stacks in the {\it Herschel} bands for the three subsamples
are  presented in Figure\,\ref{stacked_images}. To estimate the
variation present  within these subsamples we followed a
bootstrapping approach. For a given subsample we randomly selected
as many objects as there are members in that subsample, allowing for
replacements, created a new stack and performed photometry. This was 
done for 1000 random combinations of objects in each
subsample.  The centroid of the distribution of these 1000 individual
stacked photometry values was then taken as the final average flux of the
subsample. We use the standard deviation of this distribution, which
can be considered a measure for the variety of intrinsic SED shapes 
present in the subsample, as the uncertainty on the average flux.
The overall significance of the final stacked mean
value in the {\it Herschel} bands was determined as follows: we stacked the images at
random positions on the background, following a similar procedure as
for the quasar positions. If
the mean value of the source stack distribution is larger than
three times the mean value of the background stack distribution we consider
the stacked quasar signal to be significant.

\begin{figure*}[t!]
\centering
\includegraphics[angle=0,scale=.50]{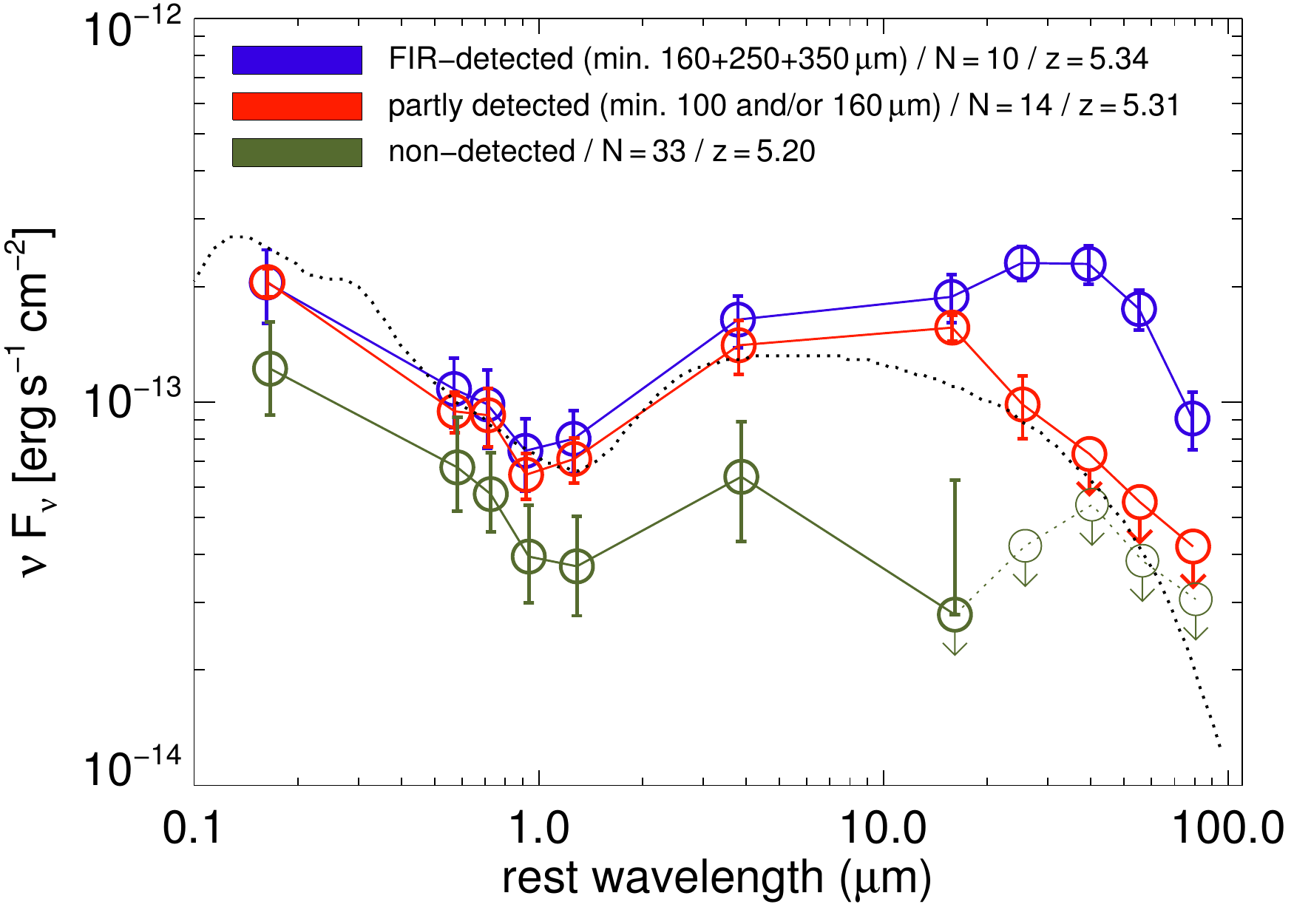}
\includegraphics[angle=0,scale=.50]{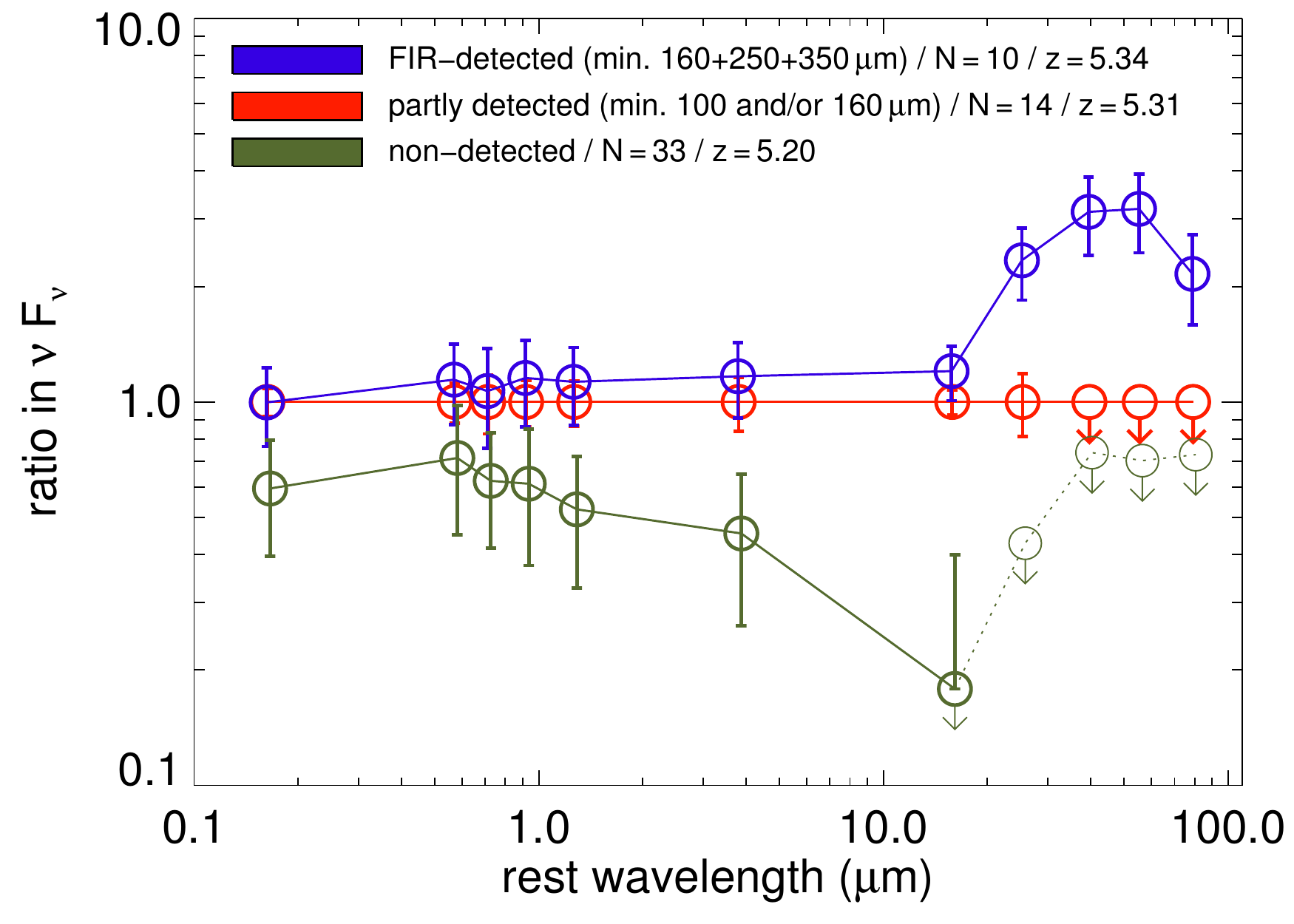}
\caption{{\it Left:} Average SEDs for three subsamples as defined in
  Section \ref{sec:stacking}. The dashed line is the SDSS quasar template \citep{ric06}. 
{\it Right}: The average SEDs divided by the SED of the partly {\it Herschel} detected 
sample (red). This emphasizes the differences in the SED shape between the three samples. 
\label{stacked_seds}}
\end{figure*}

In Figure\,\ref{stacked_seds} ({\it left}) we compare the average SEDs
of the  FIR-detected objects (blue SED) with that of the partly
detected objects (red SED). The SEDs are very similar in absolute
scaling and spectral shape up to and including the observed
100\,$\mu$m band ($\sim$15\,$\mu$m rest frame). At longer wavelengths,
however, the SEDs are very different. In the $\nu F_{\nu}$
representation of Figure\,\ref{stacked_seds} ({\it left}), the
PACS-only objects show a steep drop above $\sim$20\,$\mu$m rest frame
while the FIR  detected objects display an additional component
towards the FIR. This behavior is emphasized in
Figure\,\ref{stacked_seds} ({\it right}) where we show the average
SEDs normalized by the shape of the mean SED of the partly-detected
objects.

The partly {\it Herschel} detected sources (red SED in
Figure\,\ref{stacked_seds}) are  optically luminous AGN with powerful
NIR and MIR emission, but without exceptional FIR brightness, at least
on average.  The shape of the SED is very similar to the average SDSS
quasar SED and beyond $\sim$20\,$\mu$m broadly resembles the shape of
typical torus models \citep[e.g.,][]{sch08,nen08,hon10,sta12}. In
these cases the AGN is likely contributing significantly or even
dominantly to the FIR  emission
\citep[][]{net07,lut08,wan08b}. However, the upper limits in the SPIRE
bands are not very stringent and would still be consistent with a FIR
component of $\sim$10$^{12}$\,$L_{\odot}$ (assuming a modified black
body of T\,=\,47\,K and $\beta$\,=\,1.6). Therefore, it cannot be
ruled out that star formation contributes FIR emission on levels of a
few tens to a few hundred solar masses per year as found for other
high-redshift QSOs \citep{wan08b,ven12,wil13,net13}. We note that for
some combinations of objects the bootstrapping indeed reveals
significant detections in the SPIRE bands, indicating that some
sources in this subsample were just below the individual detection
limit. In the global mean, however, the partly {\it Herschel} detected
subsample only reaches $\sim$2$\sigma$  significance in the stacked
values at 250\,$\mu$m and 350\,$\mu$m.

The comparison of the average SEDs of the two {\it Herschel}-detected
subsamples emphasizes that AGN with strong emission in the
UV/optical (from the accretion disk) throughout the NIR and MIR (from
AGN-powered hot and warm dust) do not necessarily show considerable FIR 
emission
as well (both panels in Figure\,\ref{stacked_seds};  see also
\citealt{dai12}).  The fact that some optical and MIR luminous QSOs
show strong FIR emission and others do not may indicate that  
star formation is the dominant driver for the additional FIR component
observed in the {\it Herschel} FIR detected QSOs.  This is consistent
with the results of \citet{lut08} who show that for a sample of
millimeter bright QSOs at $z\sim2$ the PAH and FIR luminosities correlate, 
which also
supports star formation  as the source of the FIR emission in
powerful {\it FIR bright} AGN (these objects have optical luminosities
comparable to our sources).  For some high-redshift millimeter bright QSOs 
it has been shown
that the potentially star-formation
dominated FIR continuum and line emission (e.g. of [CII]) is concentrated 
in the innermost
kiloparsecs \citep[e.g.,][]{wal09,wan13}. At the highest luminosities,
the close proximity to the AGN may thus still lead to significant 
contributions from the AGN to the FIR emission
(e.g., \citealt{dai12}, see also \citealt{val11}). For such sources
future observation at high resolution may provide
additional clues on the relative AGN-to-SF contributions to the dust heating 
from the spatial distribution of the cold dust emission.

%In this case the 
%presence of strong star formation is not ubiquitous in high redshift
%quasars considering our detection rate of $\sim$15\%\footnote{This detection rate is 
%slightly higher than for the sample of \citet{dai12} (lower redshift and AGN luminosity) 
%and slightly lower compared to \citet{net13} ($z\sim4.8$, AGN luminosity towards our 
%high-luminosity end).}

% Dai+12:  10\% detection rate in FIR for flux-limited (MIPS24) sample

% Maiolino+07, Treister+08, Lutz+08: MIR-to-optical flux ratio changes 
% with luminosity -> smaller dust covering factor
% Determine L_5100 from PL and 6.7 from 24-100mu connection
% 
% Also try: NIR lum vs. 24/100 ratio

The average SED of the {\it Herschel} non-detections (green SEDs in
Figure\,\ref{stacked_seds})  differs from the 
SEDs of the {\it Herschel} detected sources in several aspects.  Even
at 100\,$\mu$m (observed) we only find a barely significant detection in the
maximum flux case (as provided  by the bootstraping), and
non-detections for most realizations as well as for the mean of this 
subsample. No detection was obtained for
any combination of objects at longer wavelengths. A significant
difference between the average SED of this subsample and those of
objects with individual {\it Herschel} detections is that the UV/optical and
NIR/MIR  fluxes are systematically smaller, and accordingly
  also the luminosities because the median redshifts of the three
subsamples are very similar. This has already been
indicated by the UV/optical and NIR luminosity distributions of the
full sample (Figure\,\ref{uvseds}) where the {\it Herschel} detected
sources are found to be clustered at the high luminosity end. 
This flux difference in the mean
increases towards longer wavelengths (factor of 1.4 at
$\sim$0.5\,$\mu$m rest frame and factor of 5 at $\sim$15\,$\mu$m rest
frame; see also \citealt{bla13}). 
%We have already noted  in Section\,\ref{sec:alpha_fits} that
%the {\it Herschel} non-detections are  preferentially those with
%fainter UV/optical or NIR luminosities
%(Figure\,\ref{uvseds}). 
From Figure\,\ref{stacked_seds} it appears that on average the shape of
the infrared SED is changing for fainter QSOs, even though taking into 
account the full errors bars shown in that figure could reduce the trend 
seen for the mean.\footnote{We have
here included the objects with low $L_{\rm NIR}$ in the average
SED. Excluding these sources from the stack reduces the number of
objects to 26, which slightly lowers the differences in scaling
compared to the other two average SEDs, but the main trends of the SED
shape remain.}  This behavior is in principle supported by the
individual objects that have  deeper re-observations. These were drawn
from the {\it Herschel} non-detected subsample, and by selection  
correspond to sources with high observed 24\,$\mu$m fluxes in this
sample (above the average for all but one of the six sources). 
Individually they also show a decline in $\nu F_{\nu}$
between 24\,$\mu$m and 100\,$\mu$m, but shallower than the average
SED.

The flux levels of the {\it Herschel}/PACS 100\,$\mu$m 
detections for these deeper data show that they were not far 
below the sensitivity limit of our standard observations.
 Even though a number of objects apparently only barely avoided detection 
in our standard data, the final average SED of the Herschel 
non-detections (which includes these objects just below the detection limit) 
shows steeper 24\,$\mu$m to 100\,$\mu$m slopes and remains 
{\it Herschel} PACS non-detected in most cases. This supports the 
idea that the {\it Herschel} non-detected subsample includes 
objects with a wide range of intrinsic SED shapes and that many 
of the sources in this sub sample are much fainter in the 
100\,$\mu$m band than our detection limit.

We note that four objects in this {\it Herschel} non-detected 
subsample have 
individual millimeter detections \citep[][]{ber03,wan07,wan11} and
were discussed in \citet{lei13}. Considering the shape of the SED in the 
(AGN dominated) MIR, we can speculate that the 
cold dust emission responsible for the millimeter flux (typically 
$\sim$$2-3$\,mJy at 1.2\,mm observed) is probably powered by star formation 
in these cases \citep[see also][]{wan08b,lei13}. 

% sources in FIRNONDET stack
% SDSSJ001714.67-100055.4
% SDSSJ013326.84+010637.7
% SDSSJ020332.35+001228.6 -> mm det
% SDSSJ0231-0728
% SDSSJ035349.72+010404.4
% SDSSJ073103.12+445949.4
% SDSSJ083317.66+272629.0
% SDSSJ084035.09+562419.9 -> mm det
% SDSSJ084119.52+290504.4
% SDSSJ090245.77+085115.8
% SDSSJ0913+5919
% SDSSJ091543.64+492416.7
% SDSSJ101336.33+424026.5
% SDSSJ1030+0524
% SDSSJ1048+4637          -> mm det
% SDSSJ111920.64+345248.2
% SDSSJ114657.79+403708.7
% SDSSJ115424.74+134145.8
% SDSSpJ1208+0010
% SDSSJ122146.42+444528.0
% SDSSJ124247.91+521306.8
% SDSSJ1306+0356
% SDSSJ133550.81+353315.8 -> mm det
% SDSSJ134141.46+461110.3
% SDSSJ1411+1217
% FIRSTJ142738.5+331241
% SDSSJ151035.29+514841.0
% SDSSJ152404.10+081639.3
% SDSSJ162629.19+285857.6
% SDSSJ1630+4012
% SDSSJ211928.32+102906.6
% SDSSJ222845.14-075755.2

%\addtocounter{figure}{-1}
%\begin{figure*}[t!]
%\centering
%\includegraphics[angle=0,scale=.23]{J0840+5624_mips.pdf}
%\caption{{\it continued}}
%\end{figure*}

\begin{figure*}[t!]
\centering
\includegraphics[angle=0,scale=.42]{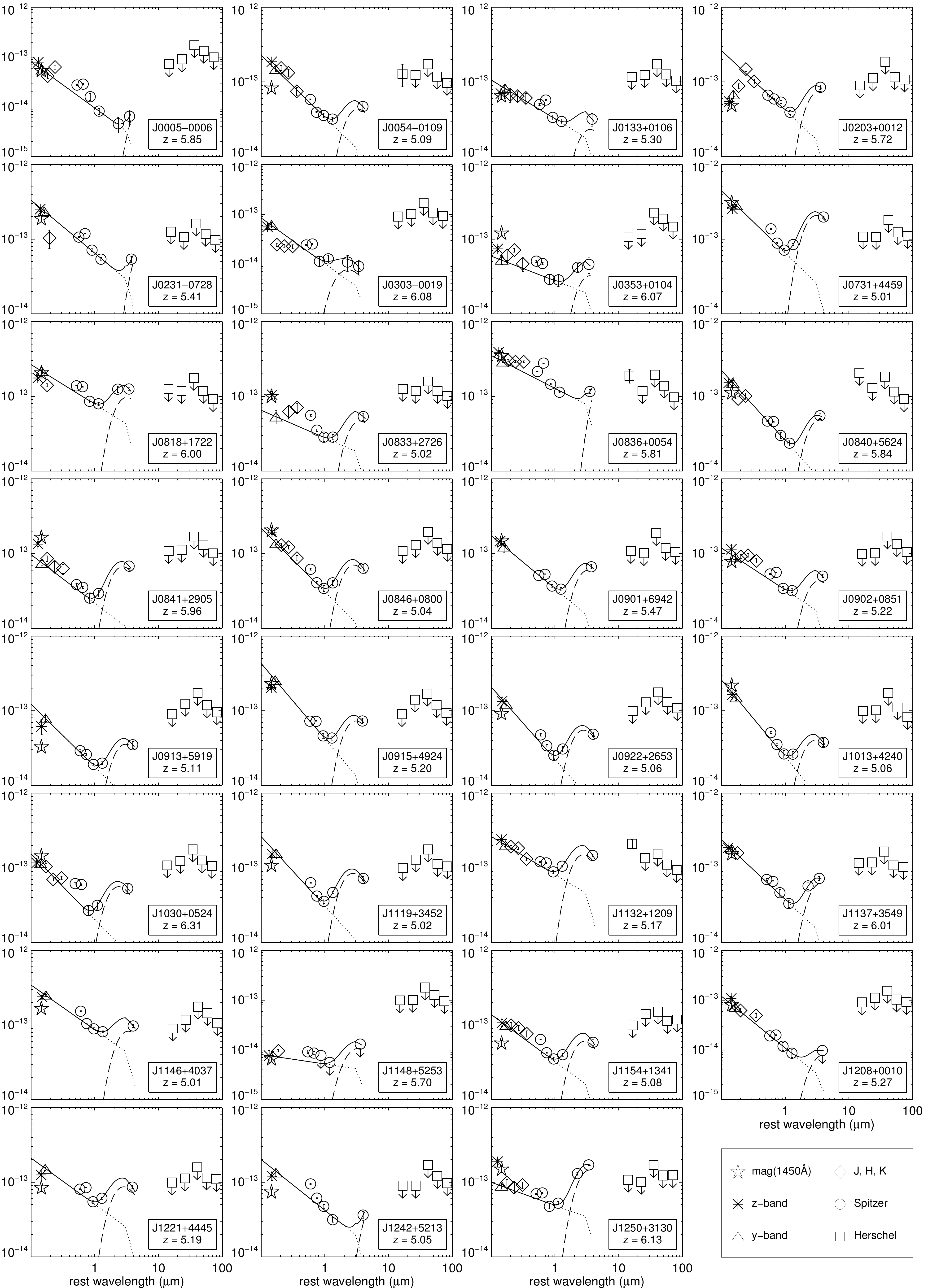}

\caption{The SEDs of the 52 quasars with only one (6 sources) 
or no (45 sources) {\it Herschel} detection, or where no SED fitting 
could be performed (J2054$-$0005) due to missing {\it Spitzer} data, 
shown in $\nu F_{\nu}$ in units of erg\,s$^{-1}$\,cm$^{-2}$ over the rest frame wavelength.
Photometry symbols as in Figures\,\ref{seds_firdet} and \ref{seds_pacsdet}. The solid line 
shows a combined power-law (dotted line) and black body (dashed line) 
fit as outlined in section \ref{sec:alpha_fits}. \label{seds}}
\end{figure*}

\addtocounter{figure}{-1}
\begin{figure*}[t!]
\centering
\includegraphics[angle=0,scale=.42]{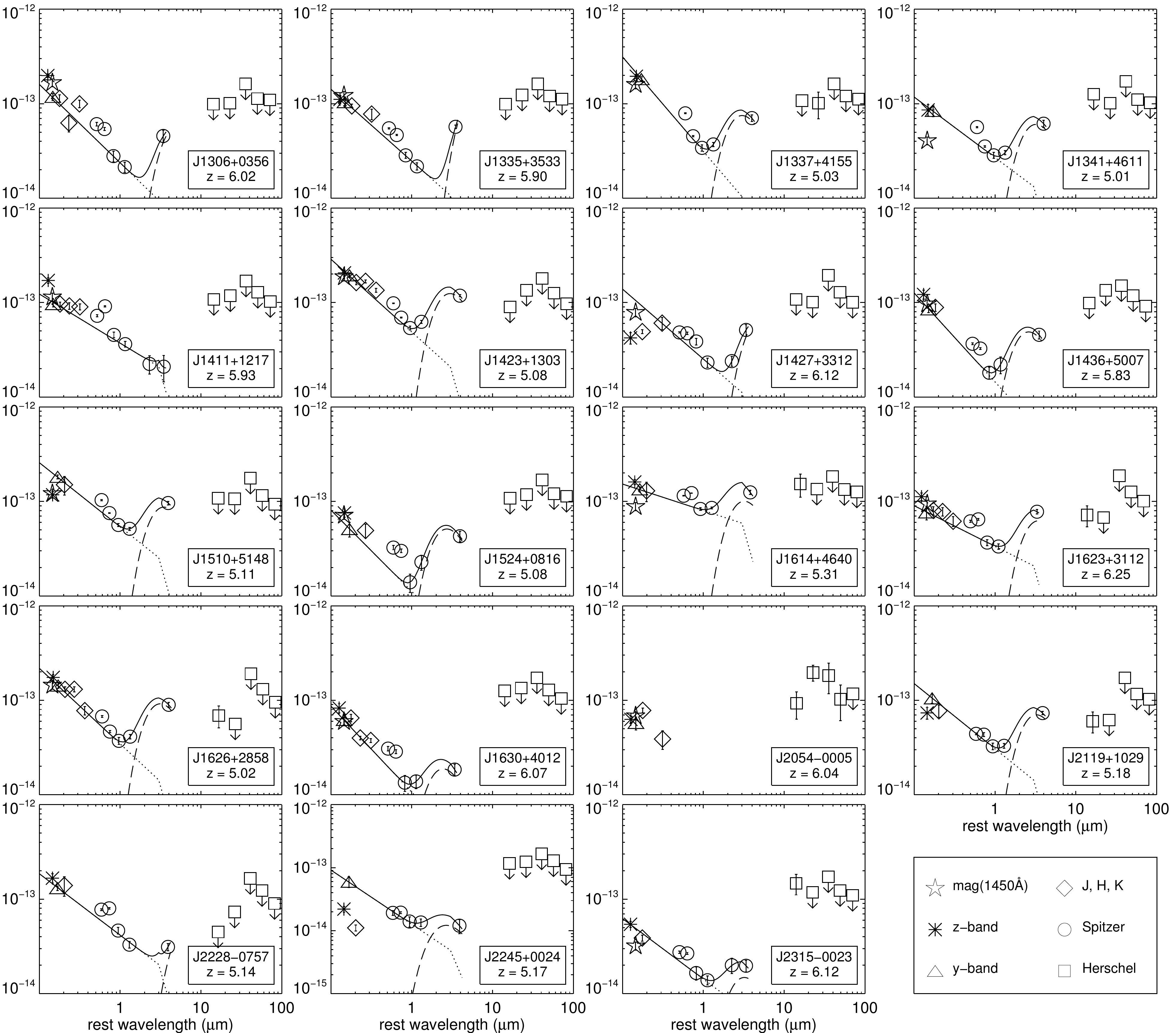}
\caption{{\it continued}}
\end{figure*}

\section{Summary and Conclusions}

We studied the spectral energy distributions of 69 QSOs at redshift $z>5$, covering 
rest frame wavelengths from  0.1 to $\sim$80\,$\mu$m. For this purpose we presented 
new {\it Herschel} observations in five bands between 
100\,$\mu$m and 500\,$\mu$m  which we combined with mostly unpublished {\it Spitzer} 
data, also in five bands (3.6\,$\mu$m to 24\,$\mu$m). Photometry from
the literature and large area surveys in the optical 
and near infrared completed the wavelength coverage. Our main results are:

\begin{enumerate}

\item The detection rate with {\it Spitzer} is very high, with only
two objects lacking detections  in the longest bands. The detection
rate decreases towards the (observed) far infrared  where {\it
Herschel} detected $\sim$30\% in PACS (100/160\,$\mu$m) and
$\sim$15\% in SPIRE (250/350\,$\mu$m).
	
\item All objects with a sufficient number of {\it Herschel}
detections (typically in at least two bands) were  subject to
multi-component SED fitting using the full wavelength range. All of
them required a hot dust  component (T\,$\sim$\,1200\,K) in addition
to an AGN torus model to fit the near and mid-infrared emission.  The
objects with rest frame FIR detections (i.e. in the SPIRE bands)  also
needed an additional cold (T\,$\sim$\,50\,K)  component with $L_{\rm
FIR}$  on the order of 10$^{13}$\,$L_{\odot}$.

\item At shorter wavelengths ($\lambda_{\rm
obs}$\,$\leq$\,24\,$\mu$m, $\lambda_{\rm
rest}$\,$\lesssim$\,4\,$\mu$m, ), the high detection rate facilitated the
study of  the UV/optical and NIR properties of most objects in our
sample. For this purpose we fitted a power-law in the rest frame UV/optical  in
combination with a black body in the NIR to the observed photometry. The distribution of the
resulting parameters shows that  the {\it Herschel} detected objects
are preferentially found at the high luminosity end of our sample (for
$L_{\rm UV/opt}$  and in particular for $L_{\rm NIR}$). No such trends
are seen for the UV/optical power-law index or the  temperature of the
NIR black body.  $L_{\rm UV/opt}$ and $L_{\rm NIR}$ are correlated and
their luminosity ratio does not  show significant trends with optical
luminosity or redshift. We identify a group of objects corresponding
to $\sim$15\% of the  full sample that shows low $L_{\rm NIR} / L_{\rm
UV/opt}$ ratios.  Such objects seem to be deficient in hot dust
compared to most  of the other quasars.

\item We determined the monochromatic luminosities at a rest frame
wavelength of 6.7\,$\mu$m (from the observed photometry in the
24\,$\mu$m and 100\,$\mu$m bands) and at 5100\AA~(from the UV/optical
power-law fit). The resulting MIR-to-optical luminosity  ratio tends to be 
higher at $z>5$ than for redshift $2-3$ QSOs of comparable
optical luminosity,  at least for the objects with {\it Herschel} 
detections at 100\,$\mu$m. However, about 60\% of the $z>5$ sample 
have only upper limits on the MIR-to-optical luminosity ratio. Depending 
on the intrinsic SED shape of these {\it Herschel} non-detected objects, 
the high-$z$ sample could still be consistent with the trends observed at 
lower redshift.
%If this ratio can be used as a proxy for the dust
%covering factor, then the $z>5$ QSOs might show larger dust covering 
%factors than the QSOs at intermediate redshift.

\item At $z>5.2$ we derived the equivalent width of the H$\alpha$
emission line from the {\it Spitzer} photometry using  the offset of the
4.5\,$\mu$m band compared to  a continuum fit using the 3.6\,$\mu$m and 
5.8\,$\mu$m bands. The distribution
of the EWs is similar  to that of local ($z<0.4$) SDSS QSOs,
suggesting little evolution over cosmic time, as previously seen for
rest frame UV emission lines.  Among the full sample, the {\it Herschel} detected objects
(and in particular the FIR detected objects) show low EWs in H$\alpha$
as well as in Ly$\alpha$.  
%The  latter has previously been seen for
%millimeter bright QSOs at intermediate to high redshifts.

\item We studied the average SEDs by stacking the observed data in the
{\it Spitzer} and {\it Herschel} bands. This was performed  for three
subsamples: objects detected in the FIR with {\it Herschel}, objects
only detected in the shorter {\it Herschel} bands, and  those not
detected with {\it Herschel}. The strong similarity in the optical and
MIR for the two samples with {\it Herschel} detections is taken as an
indication that star formation powers the additional FIR component in
the FIR-detected subsample. 		  The average SED of the {\it
Herschel} non-detections is fainter (factor $\sim$1.5) in the rest
frame optical than the  {\it Herschel} detected SEDs, and this discrepancy
increases towards the MIR (factor $\sim$5 at 15\,$\mu$m, rest frame). 
  This possibly indicates that these objects on average have a stronger 
emphasis on hotter dust, i.e. higher NIR-to-MIR luminosity ratios in their 
rest frame SEDs, when compared to the (optically slightly more luminous) 
average SEDs of the {\it Herschel} detected objects.

\end{enumerate}

\acknowledgments

CL acknowledges funding through DFG grant LE 3042/1-1. XF acknowledges support 
from NSF grants AST 08-06861 and 11-07682. MH is supported by the 
Nordrhein-Westf\"alische Akademie der Wissenschaften 
und der K\"unste.  We thank the referee for constructive 
comments which helped to improve the paper. This work is based in part on data obtained from 
the UKIRT Infrared Deep Sky Survey (UKIDSS). The Pan-STARRS1 Surveys (PS1) 
have been made possible through
contributions of the Institute for Astronomy, the University of
Hawaii, the Pan-STARRS Project Office, the Max-Planck Society
and its participating institutes, the Max Planck Institute for
Astronomy, Heidelberg and the Max Planck Institute for
Extraterrestrial Physics, Garching, The Johns Hopkins
University, Durham University, the University of Edinburgh,
Queen's University Belfast, the Harvard-Smithsonian Center for
Astrophysics, the Las Cumbres Observatory Global Telescope
Network Incorporated, the National Central University of
Taiwan, the Space Telescope Science Institute, the National
Aeronautics and Space Administration under Grant No. NNX08AR22G
issued through the Planetary Science Division of the NASA
Science Mission Directorate, the National Science Foundation
under Grant No. AST-1238877, the University of Maryland, and
Eotvos Lorand University (ELTE). We thank the PS1
Builders and PS1 operations staff for construction and
operation of the PS1 system and access to the data products
provided.

{\it Facilities:} \facility{Herschel}, \facility{Spitzer}.

%\appendix
\begin{appendix}

\section{Appendix material}
\setcounter{figure}{0} 
\renewcommand{\thefigure}{A.\arabic{figure}}
\begin{figure*}[t!]
\centering
\includegraphics[angle=90,width=1.0\textwidth]{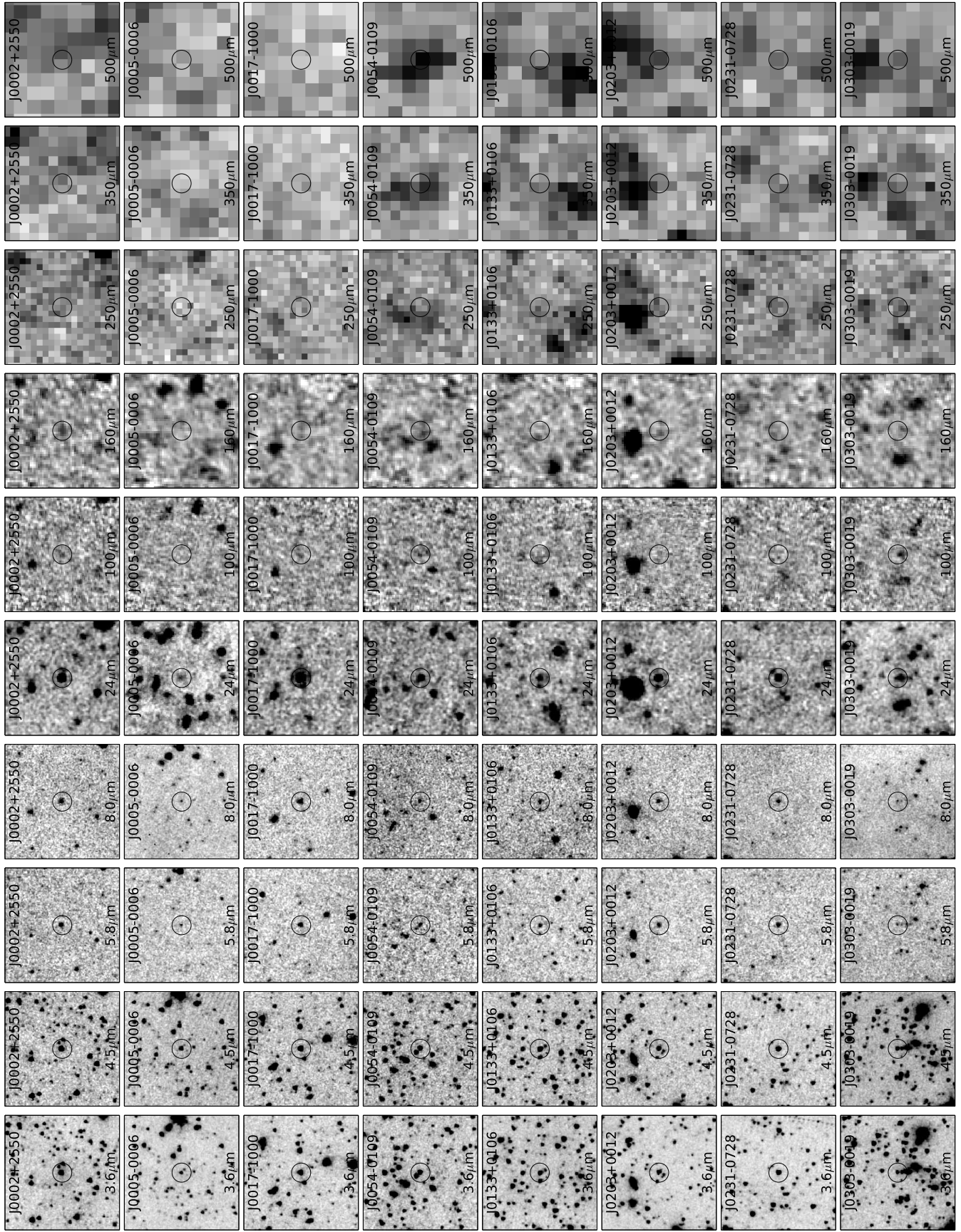}
\caption{The {\it Spitzer} and {\it Herschel} images for all objects in this 
paper. Depicted are from left to right: 3.6\,$\mu$m, 4.5\,$\mu$m, 5.8\,$\mu$m, 
8\,$\mu$m, 24\,$\mu$m, 100\,$\mu$m, 160\,$\mu$m, 250\,$\mu$m, 350\,$\mu$m, and 
500\,$\mu$m. The individual panels show an area of 2\arcmin\,$\times$\,2\arcmin~and 
the circle indicating the quasar position has a diameter of 20\arcsec. {\it This 
excerpt is shown for guidance. See the online version of the journal for the 
remaining objects.}\label{all_images}}
\end{figure*}

%\addtocounter{figure}{-1}
%\begin{figure*}[t!]
%\centering
%\includegraphics[angle=90,width=1.0\textwidth]{plot_of_all_images_2.pdf}
%\caption{{\it continued}}
%\end{figure*}
%\addtocounter{figure}{-1}
%\begin{figure*}[t!]
%\centering
%\includegraphics[angle=90,width=1.0\textwidth]{plot_of_all_images_3.pdf}
%\caption{{\it continued}}
%\end{figure*}
%\addtocounter{figure}{-1}
%\begin{figure*}[t!]
%\centering
%\includegraphics[angle=90,width=1.0\textwidth]{plot_of_all_images_4.pdf}
%\caption{{\it continued}}
%\end{figure*}
%\addtocounter{figure}{-1}
%\begin{figure*}[t!]
%\centering
%\includegraphics[angle=90,width=1.0\textwidth]{plot_of_all_images_5.pdf}
%\caption{{\it continued}}
%\end{figure*}
%\addtocounter{figure}{-1}
%\begin{figure*}[t!]
%\centering
%\includegraphics[angle=90,width=1.0\textwidth]{plot_of_all_images_6.pdf}
%\caption{{\it continued}}
%\end{figure*}
%\addtocounter{figure}{-1}
%\begin{figure*}[t!]
%\centering
%\includegraphics[angle=90,width=1.0\textwidth]{plot_of_all_images_7.pdf}
%\caption{{\it continued}}
%\end{figure*}
%\addtocounter{figure}{-1}
%\begin{figure*}[t!]
%\centering
%\includegraphics[angle=90,width=1.0\textwidth]{plot_of_all_images_8.pdf}
%\caption{{\it continued}}
%\end{figure*}
%\addtocounter{figure}{-1}
%\begin{figure*}[t!]
%\centering
%\includegraphics[angle=90,width=1.0\textwidth]{plot_of_all_images_9_add2054.pdf}
%\caption{{\it continued}}
%\end{figure*}

\end{appendix}

%\clearpage

%\LongTables
%\begin{landscape}
%\input{latex_table_obslog_astroph.tex}
%\clearpage
%\end{landscape}
%\clearpage

%\LongTables
%\input{latex_table_optical_nir_results_new_plus_ew_astroph.tex}

%\clearpage

%\clearpage
%\LongTables
%\begin{landscape}
%\input{latex_table_fluxes.tex}
%\clearpage
%\end{landscape}

%\clearpage
%\LongTables
%\begin{landscape}
%\input{latex_table_nir_magnitudes.tex}
%\clearpage
%\end{landscape}

%\input{latex_table_fluxes.tex}

%\input{latex_table_nir_magnitudes.tex}

%\input{latex_table_optical_nir_results.tex}
%\input{latex_table_ew_data.tex}

\end{document}

%% file: latex_table_obslog_astroph.tex
\begin{table*}[t!]
\scriptsize
\begin{center}
\caption{Sample and observation log.\label{obslog}}
\begin{tabular}{l c c c c c c c}
\tableline \tableline
Source &
redshift &
m$_{1450{\rm \AA}}$ &
ref &
\multicolumn{2}{c}{PACS} &
\multicolumn{2}{c}{SPIRE} \\
 &
 &
(mag) &
 &
OD &
OBSIDs &
OD &
OBSID \\
(1) &
(2) &
(3) &
(4) &
(5) &
(6) &
(7) &
(8) \\
\tableline 
SDSSJ000239.39+255034.8                    &  5.80  &  19.0  &   7 & 262 & 1342189945/1342189946  &    424 & 1342201376 \\
SDSSJ000552.34$-$000655.8                  &  5.85  &  20.8  &  10 & 615 & 1342213123/1342213124  &    411 & 1342199391 \\
SDSSJ001714.67$-$100055.4\tablenotemark{a} &  5.01  &  19.4  &  14 & 418 & 1342199873/1342199874  &    411 & 1342199382 \\
SDSSJ005421.42$-$010921.6                  &  5.09  &  20.5  &  14 & 615 & 1342213061/1342213062  &    424 & 1342201381 \\
SDSSJ013326.84+010637.7                    &  5.30  &  20.7  &  15 & 627 & 1342213530/1342213531  &    439 & 1342201322 \\
SDSSJ020332.35+001228.6                    &  5.72  &  20.9  &  10 & 636 & 1342213950/1342213951  &    439 & 1342201319 \\
SDSSJ023137.65$-$072854.5                  &  5.41  &  19.5  &  14 & 636 & 1342213965/1342213966  &    626 & 1342213482 \\
SDSSJ030331.40$-$001912.9                  &  6.08  &  21.3  &  10 & 787 & 1342223852/1342223853  &    808 & 1342224969 \\
SDSSJ033829.31+002156.3                    &  5.00  &  20.0  &   1 & 661 & 1342216135/1342216136  &    648 & 1342214565 \\
SDSSJ035349.72+010404.4                    &  6.07  &  20.2  &  10 & 668 & 1342215978/1342215979  &    467 & 1342203626 \\
SDSSJ073103.12+445949.4                    &  5.01  &  19.1  &  14 & 516 & 1342206338/1342206339  &    495 & 1342204959 \\
SDSSJ075618.14+410408.6                    &  5.09  &  20.1  &  11 & 539 & 1342208981/1342208982  &    495 & 1342204966 \\
SDSSJ081827.40+172251.8                    &  6.00  &  19.3  &   8 & 513 & 1342206072/1342206073  &    515 & 1342206224 \\
SDSSJ083317.66+272629.0                    &  5.02  &  20.3  &  15 & 539 & 1342208985/1342208986  &    515 & 1342206173 \\
SDSSJ083643.85+005453.3                    &  5.81  &  18.8  &   4 & 545 & 1342208480/1342208481  &    515 & 1342206212 \\
SDSSJ084035.09+562419.9                    &  5.84  &  20.0  &   8 & 545 & 1342208512/1342208513  &    495 & 1342204960 \\
SDSSJ084119.52+290504.4                    &  5.96  &  19.6  &   9 & 513 & 1342206070/1342206071  &    515 & 1342206172 \\
SDSSJ084229.23+121848.2                    &  6.06  &  19.9  &  13 & 545 & 1342208494/1342208495  &    515 & 1342206222 \\
SDSSJ084627.85+080051.8                    &  5.04  &  19.6  &  14 & 545 & 1342208484/1342208485  &    515 & 1342206216 \\
BWE910901+6942                             &  5.47  &  19.8  &  15 & 545 & 1342208518/1342208519  &    500 & 1342205085 \\
SDSSJ090245.77+085115.8                    &  5.22  &  20.6  &  14 & 545 & 1342208490/1342208491  &    515 & 1342206218 \\
SDSSJ091316.56+591921.5                    &  5.11  &  21.5  &  14 & 545 & 1342208514/1342208515  &    495 & 1342204961 \\
SDSSJ091543.64+492416.7                    &  5.20  &  19.3  &  14 & 546 & 1342209364/1342209365  &    515 & 1342206183 \\
SDSSJ092216.82+265359.1                    &  5.06  &  20.4  &  14 & 553 & 1342209457/1342209458  &    750 & 1342222126 \\
SDSSJ092721.82+200123.7                    &  5.77  &  19.9  &   8 & 553 & 1342209461/1342209462  &    522 & 1342206688 \\
SDSSJ095707.67+061059.5                    &  5.19  &  19.0  &  14 & 400 & 1342198559/1342198560  &    544 & 1342209293 \\
SDSSJ101336.33+424026.5                    &  5.06  &  19.4  &  14 & 545 & 1342208508/1342208509  &    395 & 1342198250 \\
SDSSJ103027.10+052455.0                    &  6.31  &  19.7  &   4 & 554 & 1342210454/1342210455  &    544 & 1342209290 \\
SDSSJ104433.04$-$012502.2                  &  5.78  &  19.2  &   4 & 415 & 1342199703/1342199704  &    411 & 1342199321 \\
SDSSJ104845.05+463718.3\tablenotemark{b}   &  6.23  &  19.2  &   6 & 554 & 1342210440/1342210441  &    402 & 1342198578 \\
SDSSJ111920.64+345248.2                    &  5.02  &  20.2  &  14 & 554 & 1342210464/1342210465  &    411 & 1342199334 \\
SDSSJ113246.50+120901.7                    &  5.17  &  19.4  &  14 & 418 & 1342199850/1342199851  &    411 & 1342199317 \\
SDSSJ113717.73+354956.9                    &  6.01  &  19.6  &   8 & 414 & 1342199595/1342199596  &    411 & 1342199335 \\
SDSSJ114657.79+403708.7                    &  5.01  &  19.7  &  14 & 414 & 1342199597/1342199598  &    411 & 1342199343 \\
SDSSJ114816.64+525150.3\tablenotemark{c}   &  6.43  &  19.0  &   6 & 403 & 1342187132/1342187133  &    395 & 1342198238 \\
RDJ1148+5253                               &  5.70  &  23.1  &  15 & 403 & 1342198852/1342198853  &    395 & 1342198239 \\
SDSSJ115424.74+134145.8                    &  5.08  &  20.9  &  14 & 418 & 1342199854/1342199855  &    411 & 1342199307 \\
SDSSJ120207.78+323538.8                    &  5.31  &  18.6  &  14 & 418 & 1342199857/1342199858  &    411 & 1342199337 \\
SDSSJ120441.73$-$002149.6                  &  5.03  &  19.1  &   2 & 607 & 1342212479/1342212480  &    423 & 1342200207 \\
SDSSpJ120823.82+001027.7                   &  5.27  &  20.5  &   3 & 757 & 1342222454/1342222455  &    393 & 1342198150 \\
SDSSJ122146.42+444528.0                    &  5.19  &  20.4  &  14 & 418 & 1342199859/1342199860  &    395 & 1342198242 \\
SDSSJ124247.91+521306.8                    &  5.05  &  20.6  &  14 & 554 & 1342210434/1342210435  &    395 & 1342198244 \\
SDSSJ125051.93+313021.9                    &  6.13  &  19.6  &   8 & 554 & 1342210466/1342210467  &    411 & 1342199339 \\
SDSSJ130608.26+035626.3                    &  6.02  &  19.6  &   4 & 615 & 1342213101/1342213102  &    438 & 1342201233 \\
SDSSJ133412.56+122020.7                    &  5.14  &  19.5  &  14 & 615 & 1342213095/1342213096  &    438 & 1342201227 \\
SDSSJ133550.81+353315.8                    &  5.90  &  19.9  &   8 & 554 & 1342210480/1342210481  &    411 & 1342199354 \\
SDSSJ133728.81+415539.9                    &  5.03  &  19.7  &  14 & 547 & 1342208823/1342208824  &    411 & 1342199357 \\
SDSSJ134015.04+392630.8                    &  5.07  &  19.6  &  14 & 554 & 1342210482/1342210483  &    411 & 1342199356 \\
SDSSJ134040.24+281328.2                    &  5.34  &  19.9  &  14 & 614 & 1342212806/1342212807  &    438 & 1342201226 \\
SDSSJ134141.46+461110.3                    &  5.01  &  21.3  &  14 & 547 & 1342208826/1342208827  &    411 & 1342199360 \\
SDSSJ141111.29+121737.4                    &  5.93  &  20.0  &   7 & 628 & 1342213592/1342213593  &    438 & 1342201228 \\
SDSSJ142325.92+130300.7                    &  5.08  &  19.6  &  14 & 629 & 1342213664/1342213665  &    586 & 1342211366 \\
FIRSTJ142738.5+331241                      &  6.12  &  20.3  &  12 & 629 & 1342213658/1342213659  &    438 & 1342201225 \\
SDSSJ143611.74+500706.9                    &  5.83  &  20.2  &   8 & 547 & 1342208828/1342208829  &    528 & 1342207034 \\
SDSSJ144350.67+362315.2                    &  5.29  &  20.3  &  14 & 629 & 1342213656/1342213657  &    438 & 1342201220 \\
SDSSJ151035.29+514841.0                    &  5.11  &  20.1  &  14 & 511 & 1342206005/1342206006  &    467 & 1342203598 \\
SDSSJ152404.10+081639.3                    &  5.08  &  20.6  &  15 & 483 & 1342204156/1342204157  &    434 & 1342201136 \\
SDSSJ160254.18+422822.9                    &  6.07  &  19.9  &   7 & 511 & 1342205994/1342205995  &    423 & 1342200199 \\
SDSSJ161425.13+464028.9                    &  5.31  &  20.3  &  14 & 539 & 1342208968/1342208969  &    423 & 1342200197 \\
SDSSJ162331.81+311200.5\tablenotemark{d}   &  6.25  &  20.1  &   7 & 501 & 1342205169/1342205170  &    495 & 1342204945 \\
SDSSJ162626.50+275132.4                    &  5.30  &  18.7  &  14 & 501 & 1342205173/1342205174  &    495 & 1342204946 \\
SDSSJ162629.19+285857.6\tablenotemark{e}   &  5.02  &  19.9  &  14 & 501 & 1342205171/1342205172  &    467 & 1342203594 \\
SDSSJ163033.90+401209.6                    &  6.07  &  20.6  &   6 & 511 & 1342205990/1342205991  &    495 & 1342204944 \\
SDSSJ165902.12+270935.1                    &  5.32  &  18.8  &  14 & 511 & 1342205986/1342205987  &    467 & 1342203591 \\
SDSSJ205406.49$-$000514.8                  &  6.04  &  20.6  &  10 & 545 & 1342208454/1342208455  &    544 & 1342209311 \\
SDSSJ211928.32+102906.6\tablenotemark{f}   &  5.18  &  20.6  &  15 & 545 & 1342208450/1342208451  &    544 & 1342209314 \\
SDSSJ222845.14$-$075755.2\tablenotemark{g} &  5.14  &  20.2  &  14 & 555 & 1342209648/1342209649  &    544 & 1342209308 \\
WFSJ2245+0024                              &  5.17  &  21.8  &   5 & 400 & 1342198517/1342198518  &    402 & 1342198588 \\
SDSSJ231546.57$-$002358.1                  &  6.12  &  21.3  &  10 & 400 & 1342198513/1342198514  &    411 & 1342199380 \\
\tableline
\end{tabular}
%\tablenotetext{a}{Re-observed under obsids 1342258790/1/2/3, 1342259270/1. $^{\rm b}$ Re-observed under obsids 1342255375/76/77/78/79/80}
%\tablenotetext{b}{Re-observed under obsids 1342255375/76/77/78/79/80}
%\tablenotetext{c}{Also observed under obsid 1342198854/5 at 70\,$\mu$m and 160\,$\mu$m.}
%\tablenotetext{d}{Re-observed under obsids 1342261336/37/38/39/40/41}
%\tablenotetext{e}{Re-observed under obsids 1342261330/1/2/3/4/5}
%\tablenotetext{f}{Re-observed under obsids 1342257397/398/939/400/401/402}
%\tablenotetext{g}{Re-observed under obsids 1342257619/20, 1342257767/68/69/70}
\tablecomments{Col.: (1) Full source name; (2) Redshift; (3) Apparent magnitude
at a rest frame wavelength of 1450\,\AA~(see text for details); 
(4) Reference for apparent magnitude at 1450\,\AA~(SDSS: value was 
derived from the SDSS spectrum. z-band: value 
derived from the SDSS QSO template spectrum scaled to the 
observed z-band flux). See text for details.; (5)-(6) Operational day (OD) and unique IDs 
of the observations (OBSID) with PACS. (7)-(8) Same for SPIRE.\newline Additional PACS observations:
$^{\rm a}$\,Re-observed under obsids 1342258790/1/2/3, 1342259270/1; 
$^{\rm b}$\,Re-observed under obsids 1342255375/76/77/78/79/80; 
$^{\rm c}$\,Also observed under obsid 1342198854/5 at 70\,$\mu$m and 160\,$\mu$m; 
$^{\rm d}$\,Re-observed under obsids 1342261336/37/38/39/40/41; 
$^{\rm e}$\,Re-observed under obsids 1342261330/1/2/3/4/5; 
$^{\rm f}$\,Re-observed under obsids 1342257397/398/939/400/401/402; 
$^{\rm g}$\,Re-observed under obsids 1342257619/20, 1342257767/68/69/70. }
\tablerefs{
(1) \citealt{fan99};    
(2) \citealt{fan00a};   
(3) \citealt{zhe00};   
(4) \citealt{fan01};   
(5) \citealt{sha01};   
(6) \citealt{fan03};   
(7) \citealt{fan04};
(8) \citealt{fan06};  
(9) \citealt{got06};   
(10) \citealt{jia08};   
(11) \citealt{wan08a};  
(12) \citealt{wan08b};  
(13) \citealt{jia10};   
(14) SDSS;              
(15) z-band.            
}
\end{center}
\end{table*}
%\end{center}

%% file: latex_table_nir_magnitudes.tex
\begin{center}
\begin{deluxetable*}{l r@{$\pm$}l c r@{$\pm$}l c r@{$\pm$}l c r@{$\pm$}l c r@{$\pm$}l c}
\tabletypesize{\scriptsize}
\tablecaption{The Dusty Young Universe: NIR photometry.\label{nir_fluxes}}
\tablehead{
\colhead{Source} &
\multicolumn{2}{c}{z} &
\colhead{ref} & 
\multicolumn{2}{c}{y} &
\colhead{ref} & 
\multicolumn{2}{c}{J} &
\colhead{ref} & 
\multicolumn{2}{c}{H} &
\colhead{ref} & 
\multicolumn{2}{c}{K} &
\colhead{ref} \\ 
\colhead{} &
\multicolumn{2}{c}{(mag)} &
\colhead{} &
\multicolumn{2}{c}{(mag)} &
\colhead{} &
\multicolumn{2}{c}{(mag)} &
\colhead{} &
\multicolumn{2}{c}{(mag)} &
\colhead{} &
\multicolumn{2}{c}{(mag)} &
\colhead{} \\
\colhead{(1)} &
\multicolumn{2}{c}{(2)} &
\colhead{(3)} &
\multicolumn{2}{c}{(4)} &
\colhead{(5)} &
\multicolumn{2}{c}{(6)} &
\colhead{(7)} &
\multicolumn{2}{c}{(8)} &
\colhead{(9)} &
\multicolumn{2}{c}{(10)} &
\colhead{(11)} \\
}
\startdata
J0002+2550&18.99&0.05&5&19.53&0.07&15&\multicolumn{2}{c}{\nodata}&\nodata&
\multicolumn{2}{c}{\nodata}&\nodata&\multicolumn{2}{c}{\nodata}&\nodata\\
J0005$-$0006&20.47&0.02&14&20.69&0.14&15&20.81&0.10&5&20.06&0.10&9&
\multicolumn{2}{c}{\nodata}&\nodata\\
J0017$-$1000&19.61&0.07&16&19.24&0.05&15&19.00&0.17&17&
\multicolumn{2}{c}{\nodata}&\nodata&\multicolumn{2}{c}{\nodata}&\nodata\\
J0054$-$0109&19.54&0.01&14&19.63&0.08&18&19.38&0.08&18&19.25&0.11&18&19.55&0.14&
18\\
J0133+0106&20.60&0.27&16&20.27&0.11&18&20.28&0.22&17&20.04&0.16&18&19.77&0.16&18
\\
J0203+0012&20.87&0.10&11&20.48&0.12&18&19.99&0.08&11&19.13&0.07&18&19.22&0.08&18
\\
J0231$-$0728&19.21&0.07&16&19.13&0.04&15&19.82&0.29&17&
\multicolumn{2}{c}{\nodata}&\nodata&\multicolumn{2}{c}{\nodata}&\nodata\\
J0303$-$0019&20.85&0.07&11&20.60&0.14&12&21.38&0.08&12&21.16&0.08&12&20.85&0.09&
12\\
J0338+0021&19.60&0.01&14&19.77&0.06&18&19.79&0.08&18&19.57&0.07&18&19.17&0.09&18
\\
J0353+0104&20.54&0.08&11&20.75&0.16&18&20.39&0.16&18&19.93&0.06&11&20.06&0.22&18
\\
J0731+4459&19.20&0.05&16&18.93&0.04&15&\multicolumn{2}{c}{\nodata}&\nodata&
\multicolumn{2}{c}{\nodata}&\nodata&\multicolumn{2}{c}{\nodata}&\nodata\\
J0756+4104&20.12&0.12&16&19.78&0.07&15&\multicolumn{2}{c}{\nodata}&\nodata&
\multicolumn{2}{c}{\nodata}&\nodata&\multicolumn{2}{c}{\nodata}&\nodata\\
J0818+1722&19.60&0.08&8&19.22&0.05&15&19.48&0.05&8&\multicolumn{2}{c}{\nodata}&
\nodata&\multicolumn{2}{c}{\nodata}&\nodata\\
J0833+2726&20.16&0.11&16&20.74&0.22&18&\multicolumn{2}{c}{\nodata}&\nodata&
20.07&0.22&18&19.61&0.15&18\\
J0836+0054&18.74&0.05&3&18.90&0.03&18&18.64&0.03&18&18.40&0.03&18&18.08&0.03&18
\\
J0840+5624&19.76&0.10&8&19.61&0.08&15&19.94&0.10&8&19.55&0.10&9&
\multicolumn{2}{c}{\nodata}&\nodata\\
J0841+2905&19.90&0.08&16&20.38&0.09&18&20.02&0.09&18&20.00&0.18&18&19.74&0.15&18
\\
J0842+1218&19.64&0.10&13&\multicolumn{2}{c}{\nodata}&\nodata&19.94&0.10&13&
\multicolumn{2}{c}{\nodata}&\nodata&\multicolumn{2}{c}{\nodata}&\nodata\\
J0846+0800&19.50&0.07&16&19.72&0.05&18&19.57&0.07&18&19.35&0.06&18&19.39&0.09&18
\\
J0901+6942&19.80&0.03&15&19.83&0.19&15&\multicolumn{2}{c}{\nodata}&\nodata&
\multicolumn{2}{c}{\nodata}&\nodata&\multicolumn{2}{c}{\nodata}&\nodata\\
J0902+0851&20.07&0.12&16&20.19&0.16&18&19.95&0.07&18&19.62&0.06&18&19.48&0.07&18
\\
J0913+5919&20.74&0.24&16&20.32&0.10&15&\multicolumn{2}{c}{\nodata}&\nodata&
\multicolumn{2}{c}{\nodata}&\nodata&\multicolumn{2}{c}{\nodata}&\nodata\\
J0915+4924&19.44&0.06&16&19.04&0.05&15&\multicolumn{2}{c}{\nodata}&\nodata&
\multicolumn{2}{c}{\nodata}&\nodata&\multicolumn{2}{c}{\nodata}&\nodata\\
J0922+2653&19.90&0.12&16&19.83&0.07&15&\multicolumn{2}{c}{\nodata}&\nodata&
\multicolumn{2}{c}{\nodata}&\nodata&\multicolumn{2}{c}{\nodata}&\nodata\\
J0927+2001&19.88&0.08&8&19.88&0.11&15&19.95&0.10&8&\multicolumn{2}{c}{\nodata}&
\nodata&\multicolumn{2}{c}{\nodata}&\nodata\\
J0957+0610&18.91&0.05&16&19.20&0.03&18&19.23&0.07&18&18.72&0.04&18&18.65&0.06&18
\\
J1013+4240&19.68&0.08&16&19.62&0.10&15&\multicolumn{2}{c}{\nodata}&\nodata&
\multicolumn{2}{c}{\nodata}&\nodata&\multicolumn{2}{c}{\nodata}&\nodata\\
J1030+0524&20.05&0.10&3&19.91&0.06&18&19.81&0.10&3&19.95&0.05&6&19.57&0.05&6\\
J1044$-$0125&19.26&0.07&16&19.51&0.05&18&19.25&0.05&18&19.30&0.12&18&18.92&0.04&
1\\
J1048+4637&19.82&0.08&16&19.49&0.12&15&19.34&0.05&4&19.21&0.05&6&19.02&0.05&6\\
J1119+3452&19.75&0.07&16&19.57&0.12&15&\multicolumn{2}{c}{\nodata}&\nodata&
\multicolumn{2}{c}{\nodata}&\nodata&\multicolumn{2}{c}{\nodata}&\nodata\\
J1132+1209&19.27&0.06&16&19.31&0.05&18&19.14&0.04&18&18.90&0.04&18&18.94&0.06&18
\\
J1137+3549&19.54&0.07&8&19.44&0.05&15&19.35&0.05&8&\multicolumn{2}{c}{\nodata}&
\nodata&\multicolumn{2}{c}{\nodata}&\nodata\\
J1146+4037&19.27&0.05&16&19.07&0.03&15&\multicolumn{2}{c}{\nodata}&\nodata&
\multicolumn{2}{c}{\nodata}&\nodata&\multicolumn{2}{c}{\nodata}&\nodata\\
J1148+5251&20.12&0.09&4&19.42&0.10&15&19.19&0.05&4&19.00&0.05&6&18.88&0.05&6\\
J1148+5253&23.00&0.30&7&\multicolumn{2}{c}{\nodata}&\nodata&22.39&0.06&7&
\multicolumn{2}{c}{\nodata}&\nodata&\multicolumn{2}{c}{\nodata}&\nodata\\
J1154+1341&20.14&0.12&16&20.07&0.09&18&19.86&0.10&18&19.66&0.08&18&19.53&0.10&18
\\
J1202+3235&18.44&0.05&16&18.65&0.02&15&\multicolumn{2}{c}{\nodata}&\nodata&
\multicolumn{2}{c}{\nodata}&\nodata&\multicolumn{2}{c}{\nodata}&\nodata\\
J1204$-$0021&18.99&0.04&16&19.21&0.06&18&18.97&0.07&18&18.88&0.08&18&18.95&0.09&
18\\
J1208+0010&20.13&0.11&16&20.42&0.15&18&20.37&0.10&2&\multicolumn{2}{c}{\nodata}&
\nodata&20.00&0.10&2\\
J1221+4445&19.97&0.07&16&19.61&0.05&15&\multicolumn{2}{c}{\nodata}&\nodata&
\multicolumn{2}{c}{\nodata}&\nodata&\multicolumn{2}{c}{\nodata}&\nodata\\
J1242+5213&20.01&0.14&16&19.74&0.12&15&\multicolumn{2}{c}{\nodata}&\nodata&
\multicolumn{2}{c}{\nodata}&\nodata&\multicolumn{2}{c}{\nodata}&\nodata\\
J1250+3130&19.53&0.08&8&20.18&0.10&18&19.86&0.11&18&19.74&0.19&18&19.33&0.11&18
\\
J1306+0356&19.47&0.05&3&19.88&0.09&18&19.71&0.10&3&20.07&0.21&18&19.24&0.10&18\\
J1334+1220&19.64&0.06&16&19.46&0.06&18&19.24&0.05&18&19.06&0.06&18&19.01&0.06&18
\\
J1335+3533&20.10&0.11&8&20.02&0.11&18&19.91&0.05&8&\multicolumn{2}{c}{\nodata}&
\nodata&19.51&0.14&18\\
J1337+4155&19.49&0.06&16&19.41&0.04&15&\multicolumn{2}{c}{\nodata}&\nodata&
\multicolumn{2}{c}{\nodata}&\nodata&\multicolumn{2}{c}{\nodata}&\nodata\\
J1340+3926&19.27&0.04&16&19.32&0.04&15&\multicolumn{2}{c}{\nodata}&\nodata&
\multicolumn{2}{c}{\nodata}&\nodata&\multicolumn{2}{c}{\nodata}&\nodata\\
J1340+2813&19.50&0.08&16&19.44&0.05&18&19.27&0.05&18&19.07&0.05&18&18.90&0.06&18
\\
J1341+4611&20.38&0.15&16&20.25&0.14&15&\multicolumn{2}{c}{\nodata}&\nodata&
\multicolumn{2}{c}{\nodata}&\nodata&\multicolumn{2}{c}{\nodata}&\nodata\\
J1411+1217&19.63&0.07&5&20.10&0.07&18&19.89&0.05&5&19.65&0.09&18&19.35&0.08&18\\
J1423+1303&19.43&0.08&16&19.34&0.05&18&19.32&0.06&18&19.00&0.04&18&18.91&0.05&18
\\
J1427+3312&21.15&0.15&15&\multicolumn{2}{c}{\nodata}&\nodata&20.62&0.05&10&
\multicolumn{2}{c}{\nodata}&\nodata&19.78&0.16&10\\
J1436+5007&20.00&0.12&8&20.24&0.08&15&19.98&0.10&8&\multicolumn{2}{c}{\nodata}&
\nodata&\multicolumn{2}{c}{\nodata}&\nodata\\
J1443+3623&19.49&0.06&16&19.08&0.03&15&19.15&0.25&17&\multicolumn{2}{c}{\nodata}
&\nodata&\multicolumn{2}{c}{\nodata}&\nodata\\
J1510+5148&20.04&0.08&16&19.41&0.05&15&19.41&0.24&17&\multicolumn{2}{c}{\nodata}
&\nodata&\multicolumn{2}{c}{\nodata}&\nodata\\
J1524+0816&20.52&0.11&16&20.79&0.19&18&\multicolumn{2}{c}{\nodata}&\nodata&
20.33&0.18&18&\multicolumn{2}{c}{\nodata}&\nodata\\
J1602+4228&19.89&0.10&5&\multicolumn{2}{c}{\nodata}&\nodata&19.40&0.05&5&
\multicolumn{2}{c}{\nodata}&\nodata&\multicolumn{2}{c}{\nodata}&\nodata\\
J1614+4640&19.70&0.07&16&19.74&0.06&15&19.57&0.25&17&\multicolumn{2}{c}{\nodata}
&\nodata&\multicolumn{2}{c}{\nodata}&\nodata\\
J1623+3112&20.09&0.10&5&20.35&0.18&18&20.09&0.10&5&19.83&0.11&18&19.76&0.13&18\\
J1626+2751&18.63&0.04&16&18.48&0.02&18&18.25&0.02&18&17.94&0.02&18&17.83&0.03&18
\\
J1626+2858&19.61&0.08&16&19.67&0.07&18&19.56&0.07&18&19.27&0.07&18&19.51&0.11&18
\\
J1630+4012&20.42&0.12&4&20.58&0.12&15&20.32&0.10&4&20.56&0.05&6&20.30&0.05&6\\
J1659+2709&18.82&0.04&16&18.77&0.03&15&18.60&0.14&17&\multicolumn{2}{c}{\nodata}
&\nodata&\multicolumn{2}{c}{\nodata}&\nodata\\
J2054$-$0005&20.72&0.09&11&20.66&0.17&15&20.12&0.06&11&
\multicolumn{2}{c}{\nodata}&\nodata&20.26&0.24&18\\
J2119+1029&20.55&0.15&16&20.01&0.12&15&20.13&0.16&17&\multicolumn{2}{c}{\nodata}
&\nodata&\multicolumn{2}{c}{\nodata}&\nodata\\
J2228$-$0757&19.66&0.12&16&19.77&0.06&15&19.49&0.25&17&
\multicolumn{2}{c}{\nodata}&\nodata&\multicolumn{2}{c}{\nodata}&\nodata\\
J2245+0024&21.86&0.11&14&20.62&0.21&15&22.24&0.12&14&\multicolumn{2}{c}{\nodata}
&\nodata&\multicolumn{2}{c}{\nodata}&\nodata\\
J2315$-$0023&20.88&0.08&11&\multicolumn{2}{c}{\nodata}&\nodata&20.88&0.08&11&
\multicolumn{2}{c}{\nodata}&\nodata&\multicolumn{2}{c}{\nodata}&\nodata
\enddata
\tablecomments{Col.: (1) Source name; (2)-(11) NIR photometry with references. All 
magnitudes are given in the AB system.}
\tablerefs{
(1) \citealt{fan00b}; 
(2) \citealt{zhe00}; 
(3) \citealt{fan01}; 
(4) \citealt{fan03}; 
(5) \citealt{fan04}; 
(6) \citealt{iwa04}; 
(7) \citealt{mah05}; 
(8) \citealt{fan06}; 
(9) \citealt{jia06}; 
(10) \citealt{mcg06}; 
(11) \citealt{jia08}; 
(12) \citealt{kur09}; 
(13) \citealt{jia10}; 
(14) \citealt{mcg13}; 
(15) Pan-STARRS; 
(16) SDSS; 
(17) This work; 
(18) UKIDSS. 
}
\end{deluxetable*}
\end{center}

%% file: latex_table_fluxes.tex
\begin{center}
\begin{deluxetable*}{l c r@{$\pm$}l r@{$\pm$}l r@{$\pm$}l r@{$\pm$}l r@{$\pm$}l r@{$\pm$}l r@{$\pm$}l r@{$\pm$}l r@{$\pm$}l r@{$\pm$}l}
\tabletypesize{\scriptsize}
\tablecaption{Source photometry.\label{tab_all}}
\tablehead{
\colhead{Source} &
\colhead{redshift} &
\multicolumn{2}{c}{$F_{3.6\mu{\rm m}}$} &
\multicolumn{2}{c}{$F_{4.5\mu{\rm m}}$} &
\multicolumn{2}{c}{$F_{5.8\mu{\rm m}}$} &
\multicolumn{2}{c}{$F_{8.0\mu{\rm m}}$} &
\multicolumn{2}{c}{$F_{24\mu{\rm m}}$} &
\multicolumn{2}{c}{$F_{100\mu{\rm m}}$} &
\multicolumn{2}{c}{$F_{160\mu{\rm m}}$} &
\multicolumn{2}{c}{$F_{250\mu{\rm m}}$} &
\multicolumn{2}{c}{$F_{350\mu{\rm m}}$} &
\multicolumn{2}{c}{$F_{500\mu{\rm m}}$} \\
\colhead{} &
\colhead{} &
\multicolumn{2}{c}{$\mu$Jy} &
\multicolumn{2}{c}{$\mu$Jy} &
\multicolumn{2}{c}{$\mu$Jy} &
\multicolumn{2}{c}{$\mu$Jy} &
\multicolumn{2}{c}{$\mu$Jy} &
\multicolumn{2}{c}{mJy} &
\multicolumn{2}{c}{mJy} &
\multicolumn{2}{c}{mJy} &
\multicolumn{2}{c}{mJy} &
\multicolumn{2}{c}{mJy} \\
\colhead{(1)} &
\colhead{(2)} &
\multicolumn{2}{c}{(3)} &
\multicolumn{2}{c}{(4)} &
\multicolumn{2}{c}{(5)} &
\multicolumn{2}{c}{(6)} &
\multicolumn{2}{c}{(7)} &
\multicolumn{2}{c}{(8)} &
\multicolumn{2}{c}{(9)} &
\multicolumn{2}{c}{(10)} &
\multicolumn{2}{c}{(11)} &
\multicolumn{2}{c}{(12)} \\
}
\startdata
J0002+2550&5.80&119&2&152&2&123&6&150&7&747&31&3.5&1.0&8.6&1.8&
\multicolumn{2}{c}{$<13.2$}&\multicolumn{2}{c}{$<13.8$}&
\multicolumn{2}{c}{$<15.6$}\\
J0005$-$0006&5.85&33&1&43&1&31&5&22&3&52&17&\multicolumn{2}{c}{$<2.4$}&
\multicolumn{2}{c}{$<4.8$}&\multicolumn{2}{c}{$<14.4$}&
\multicolumn{2}{c}{$<15.6$}&\multicolumn{2}{c}{$<16.5$}\\
J0017$-$1000&5.01&171&2&145&2&171&6&251&6&1580&33&3.8&0.5\tablenotemark{a}&
7.3&0.9\tablenotemark{a}&\multicolumn{2}{c}{$<13.5$}&\multicolumn{2}{c}{$<13.5$}
&\multicolumn{2}{c}{$<15.5$}\\
J0054$-$0109&5.09&70&1&59&1&69&5&85&5&376&43&4.3&1.4&\multicolumn{2}{c}{$<6.6$}&
\multicolumn{2}{c}{$<14.4$}&\multicolumn{2}{c}{$<13.8$}&
\multicolumn{2}{c}{$<16.2$}\\
J0133+0106&5.30&60&2&86&1&65&5&79&6&255&49&\multicolumn{2}{c}{$<3.9$}&
\multicolumn{2}{c}{$<6.6$}&\multicolumn{2}{c}{$<14.4$}&
\multicolumn{2}{c}{$<14.7$}&\multicolumn{2}{c}{$<17.4$}\\
J0203+0012&5.72&80&2&89&2&104&5&105&7&680&49&\multicolumn{2}{c}{$<3.0$}&
\multicolumn{2}{c}{$<6.0$}&\multicolumn{2}{c}{$<15.6$}&
\multicolumn{2}{c}{$<13.5$}&\multicolumn{2}{c}{$<18.0$}\\
J0231$-$0728&5.41&128&2&178&1&138&6&144&7&433&42&\multicolumn{2}{c}{$<4.2$}&
\multicolumn{2}{c}{$<5.7$}&\multicolumn{2}{c}{$<13.5$}&
\multicolumn{2}{c}{$<13.8$}&\multicolumn{2}{c}{$<16.2$}\\
J0303$-$0019&6.08&29&1&38&1&22&5&34&6&72&24&\multicolumn{2}{c}{$<3.0$}&
\multicolumn{2}{c}{$<5.4$}&\multicolumn{2}{c}{$<14.1$}&
\multicolumn{2}{c}{$<12.6$}&\multicolumn{2}{c}{$<15.3$}\\
J0338+0021&5.00&80&2&70&2&81&7&156&9&1186&51&11.4&1.1&22.3&2.5&19.6&5.9&18.5&6.2
&12.6&6.5\\
J0353+0104&6.07&61&2&71&2&56&9&76&14&368&92&\multicolumn{2}{c}{$<3.6$}&
\multicolumn{2}{c}{$<6.3$}&\multicolumn{2}{c}{$<18.9$}&
\multicolumn{2}{c}{$<21.9$}&\multicolumn{2}{c}{$<24.6$}\\
J0731+4459&5.01&165&3&133&3&138&7&226&8&1585&45&\multicolumn{2}{c}{$<3.6$}&
\multicolumn{2}{c}{$<5.7$}&\multicolumn{2}{c}{$<15.0$}&
\multicolumn{2}{c}{$<14.4$}&\multicolumn{2}{c}{$<18.3$}\\
J0756+4104&5.09&62&2&62&2&71&5&120&5&698&45&6.4&1.1&10.2&2.2&11.4&5.3&19.0&4.8&
19.9&5.0\\
J0818+1722&6.00&166&1&202&2&166&8&212&9&1004&36&\multicolumn{2}{c}{$<4.2$}&
\multicolumn{2}{c}{$<6.3$}&\multicolumn{2}{c}{$<14.7$}&
\multicolumn{2}{c}{$<13.8$}&\multicolumn{2}{c}{$<15.3$}\\
J0833+2726&5.02&67&2&53&2&55&7&76&6&429&48&\multicolumn{2}{c}{$<4.2$}&
\multicolumn{2}{c}{$<6.3$}&\multicolumn{2}{c}{$<13.2$}&
\multicolumn{2}{c}{$<13.8$}&\multicolumn{2}{c}{$<16.8$}\\
J0836+0054&5.81&258&2&418&1&282&5&303&6&929&52&6.3&1.3&
\multicolumn{2}{c}{$<6.3$}&\multicolumn{2}{c}{$<16.2$}&
\multicolumn{2}{c}{$<16.2$}&\multicolumn{2}{c}{$<16.2$}\\
J0840+5624&5.84&56&1&69&1&58&5&63&5&440&37&\multicolumn{2}{c}{$<3.3$}&
\multicolumn{2}{c}{$<6.9$}&\multicolumn{2}{c}{$<15.3$}&
\multicolumn{2}{c}{$<13.5$}&\multicolumn{2}{c}{$<15.3$}\\
J0841+2905&5.96&46&2&53&2&49&7&78&8&543&39&\multicolumn{2}{c}{$<3.6$}&
\multicolumn{2}{c}{$<6.0$}&\multicolumn{2}{c}{$<14.1$}&
\multicolumn{2}{c}{$<15.3$}&\multicolumn{2}{c}{$<16.8$}\\
J0842+1218&6.06&81&1&98&2&88&8&128&10&1292&75&5.9&1.3&16.1&2.3&
\multicolumn{2}{c}{$<19.8$}&\multicolumn{2}{c}{$<29.1$}&
\multicolumn{2}{c}{$<24.6$}\\
J0846+0800&5.04&74&2&61&2&66&6&108&9&510&56&\multicolumn{2}{c}{$<3.6$}&
\multicolumn{2}{c}{$<6.9$}&\multicolumn{2}{c}{$<16.2$}&
\multicolumn{2}{c}{$<16.2$}&\multicolumn{2}{c}{$<19.2$}\\
J0901+6942&5.47&61&1&79&1&71&4&89&5&532&35&\multicolumn{2}{c}{$<3.6$}&
\multicolumn{2}{c}{$<5.4$}&\multicolumn{2}{c}{$<15.6$}&
\multicolumn{2}{c}{$<13.8$}&\multicolumn{2}{c}{$<17.4$}\\
J0902+0851&5.22&64&1&84&1&66&4&85&5&399&22&\multicolumn{2}{c}{$<3.3$}&
\multicolumn{2}{c}{$<5.4$}&\multicolumn{2}{c}{$<14.1$}&
\multicolumn{2}{c}{$<15.6$}&\multicolumn{2}{c}{$<17.4$}\\
J0913+5919&5.11&35&1&39&1&37&1&53&2&280&32&\multicolumn{2}{c}{$<3.0$}&
\multicolumn{2}{c}{$<6.6$}&\multicolumn{2}{c}{$<14.4$}&
\multicolumn{2}{c}{$<13.8$}&\multicolumn{2}{c}{$<15.9$}\\
J0915+4924&5.20&87&2&108&2&89&6&115&7&583&53&\multicolumn{2}{c}{$<3.0$}&
\multicolumn{2}{c}{$<7.5$}&\multicolumn{2}{c}{$<14.1$}&
\multicolumn{2}{c}{$<13.8$}&\multicolumn{2}{c}{$<15.6$}\\
J0922+2653&5.06&57&2&51&2&49&6&82&7&388&27&\multicolumn{2}{c}{$<3.3$}&
\multicolumn{2}{c}{$<6.9$}&\multicolumn{2}{c}{$<14.7$}&
\multicolumn{2}{c}{$<15.3$}&\multicolumn{2}{c}{$<18.0$}\\
J0927+2001&5.77&47&2&50&2&43&7&74&7&639&47&\multicolumn{2}{c}{$<3.6$}&7.3&2.3&
13.1&5.3&15.3&5.0&19.5&5.8\\
J0957+0610&5.19&115&2&142&1&136&8&247&9&1148&51&5.0&1.3&11.3&2.1&14.0&5.0&
\multicolumn{2}{c}{$<15.3$}&\multicolumn{2}{c}{$<16.8$}\\
J1013+4240&5.06&61&2&53&2&51&6&70&6&302&36&\multicolumn{2}{c}{$<3.3$}&
\multicolumn{2}{c}{$<5.4$}&\multicolumn{2}{c}{$<14.4$}&
\multicolumn{2}{c}{$<12.9$}&\multicolumn{2}{c}{$<13.8$}\\
J1030+0524&6.31&74&3&90&2&52&7&84&9&425&60&\multicolumn{2}{c}{$<3.6$}&
\multicolumn{2}{c}{$<6.6$}&\multicolumn{2}{c}{$<14.7$}&
\multicolumn{2}{c}{$<14.7$}&\multicolumn{2}{c}{$<17.7$}\\
J1044$-$0125&5.78&106&2&125&2&109&7&190&8&1436&45&6.3&1.2&7.7&1.8&
\multicolumn{2}{c}{$<15.3$}&\multicolumn{2}{c}{$<12.6$}&
\multicolumn{2}{c}{$<16.5$}\\
J1048+4637&6.23&110&1&122&2&95&6&127&7&818&35&2.8&0.5\tablenotemark{a}&
5.7&1.0\tablenotemark{a}&\multicolumn{2}{c}{$<14.4$}&\multicolumn{2}{c}{$<14.1$}
&\multicolumn{2}{c}{$<18.6$}\\
J1119+3452&5.02&76&1&63&1&69&5&125&5&578&40&\multicolumn{2}{c}{$<3.3$}&
\multicolumn{2}{c}{$<6.9$}&\multicolumn{2}{c}{$<14.7$}&
\multicolumn{2}{c}{$<13.2$}&\multicolumn{2}{c}{$<17.4$}\\
J1132+1209&5.17&145&2&175&2&171&7&281&8&1176&49&7.0&1.0&
\multicolumn{2}{c}{$<7.2$}&\multicolumn{2}{c}{$<12.9$}&
\multicolumn{2}{c}{$<12.9$}&\multicolumn{2}{c}{$<15.6$}\\
J1137+3549&6.01&84&2&99&2&90&9&89&10&579&34&\multicolumn{2}{c}{$<3.9$}&
\multicolumn{2}{c}{$<6.3$}&\multicolumn{2}{c}{$<13.8$}&
\multicolumn{2}{c}{$<12.6$}&\multicolumn{2}{c}{$<17.1$}\\
J1146+4037&5.01&184&2&157&2&172&6&217&6&779&33&\multicolumn{2}{c}{$<3.0$}&
\multicolumn{2}{c}{$<6.3$}&\multicolumn{2}{c}{$<14.7$}&
\multicolumn{2}{c}{$<16.8$}&\multicolumn{2}{c}{$<17.7$}\\
J1148+5251&6.43&136&2&143&2&145&7&208&8&1349&49&4.1&0.9&7.4&1.9&21.0&5.3&
21.8&4.9&12.4&5.7\\
J1148+5253&5.70&11&1&13&1&\multicolumn{2}{c}{$<15$}&\multicolumn{2}{c}{$<15$}&
\multicolumn{2}{c}{$<105$}&\multicolumn{2}{c}{$<3.3$}&\multicolumn{2}{c}{$<5.4$}
&\multicolumn{2}{c}{$<15.0$}&\multicolumn{2}{c}{$<14.7$}&
\multicolumn{2}{c}{$<15.9$}\\
J1154+1341&5.08&77&1&64&1&68&4&107&5&470&47&\multicolumn{2}{c}{$<3.3$}&
\multicolumn{2}{c}{$<7.5$}&\multicolumn{2}{c}{$<12.6$}&
\multicolumn{2}{c}{$<13.2$}&\multicolumn{2}{c}{$<19.8$}\\
J1202+3235&5.31&125&2&147&2&150&6&233&7&1609&49&8.3&1.1&16.3&2.2&18.4&5.2&
24.6&5.2&13.7&5.6\\
J1204$-$0021&5.03&110&1&109&1&122&5&209&7&1312&28&11.6&1.2&14.7&2.3&30.8&4.6&
40.0&4.6&29.1&5.8\\
J1208+0010&5.27&23&1&30&1&23&4&23&4&\multicolumn{2}{c}{$<78$}&
\multicolumn{2}{c}{$<3.0$}&\multicolumn{2}{c}{$<6.0$}&
\multicolumn{2}{c}{$<12.9$}&\multicolumn{2}{c}{$<12.0$}&
\multicolumn{2}{c}{$<15.3$}\\
J1221+4445&5.19&97&1&127&1&106&5&163&5&689&37&\multicolumn{2}{c}{$<3.3$}&
\multicolumn{2}{c}{$<6.0$}&\multicolumn{2}{c}{$<13.2$}&
\multicolumn{2}{c}{$<13.5$}&\multicolumn{2}{c}{$<18.3$}\\
J1242+5213&5.05&114&1&92&1&92&5&84&6&291&31&\multicolumn{2}{c}{$<3.0$}&
\multicolumn{2}{c}{$<4.8$}&\multicolumn{2}{c}{$<14.1$}&
\multicolumn{2}{c}{$<14.1$}&\multicolumn{2}{c}{$<16.2$}\\
J1250+3130&6.13&84&1&108&1&92&7&140&7&1366&25&\multicolumn{2}{c}{$<3.6$}&
\multicolumn{2}{c}{$<5.4$}&\multicolumn{2}{c}{$<14.1$}&
\multicolumn{2}{c}{$<14.4$}&\multicolumn{2}{c}{$<20.7$}\\
J1306+0356&6.02&73&3&81&3&54&7&57&7&365&55&\multicolumn{2}{c}{$<3.3$}&
\multicolumn{2}{c}{$<5.4$}&\multicolumn{2}{c}{$<13.5$}&
\multicolumn{2}{c}{$<13.2$}&\multicolumn{2}{c}{$<18.3$}\\
J1334+1220&5.14&87&1&94&1&93&4&165&5&1089&56&5.4&1.0&6.5&1.7&
\multicolumn{2}{c}{$<14.4$}&\multicolumn{2}{c}{$<12.6$}&
\multicolumn{2}{c}{$<15.6$}\\
J1335+3533&5.90&66&1&70&1&55&4&58&5&456&19&\multicolumn{2}{c}{$<3.3$}&
\multicolumn{2}{c}{$<6.6$}&\multicolumn{2}{c}{$<13.5$}&
\multicolumn{2}{c}{$<14.1$}&\multicolumn{2}{c}{$<18.6$}\\
J1337+4155&5.03&95&1&68&1&66&5&99&5&564&45&\multicolumn{2}{c}{$<3.6$}&5.4&1.7&
\multicolumn{2}{c}{$<13.5$}&\multicolumn{2}{c}{$<14.1$}&
\multicolumn{2}{c}{$<18.6$}\\
J1340+3926&5.07&112&1&107&1&109&4&176&5&1267&36&6.1&1.1&7.5&1.9&
\multicolumn{2}{c}{$<13.8$}&\multicolumn{2}{c}{$<13.2$}&
\multicolumn{2}{c}{$<15.9$}\\
J1340+2813&5.34&129&1&169&1&171&6&250&6&1485&43&9.8&1.2&16.2&2.2&21.8&5.0&
22.4&4.9&\multicolumn{2}{c}{$<16.5$}\\
J1341+4611&5.01&68&1&53&1&55&4&81&4&492&44&\multicolumn{2}{c}{$<4.2$}&
\multicolumn{2}{c}{$<5.4$}&\multicolumn{2}{c}{$<14.4$}&
\multicolumn{2}{c}{$<12.9$}&\multicolumn{2}{c}{$<17.1$}\\
J1411+1217&5.93&87&2&137&2&88&6&97&7&168&52&\multicolumn{2}{c}{$<3.6$}&
\multicolumn{2}{c}{$<6.3$}&\multicolumn{2}{c}{$<14.1$}&
\multicolumn{2}{c}{$<15.0$}&\multicolumn{2}{c}{$<17.1$}\\
J1423+1303&5.08&118&1&104&1&104&5&167&6&947&23&\multicolumn{2}{c}{$<3.0$}&
\multicolumn{2}{c}{$<7.2$}&\multicolumn{2}{c}{$<15.0$}&
\multicolumn{2}{c}{$<14.7$}&\multicolumn{2}{c}{$<16.2$}\\
J1427+3312&6.12&58&1&71&2&75&6&62&7&411&58&\multicolumn{2}{c}{$<3.6$}&
\multicolumn{2}{c}{$<5.4$}&\multicolumn{2}{c}{$<16.2$}&
\multicolumn{2}{c}{$<15.0$}&\multicolumn{2}{c}{$<16.8$}\\
J1436+5007&5.83&44&1&49&1&35&5&59&12&365&35&\multicolumn{2}{c}{$<3.3$}&
\multicolumn{2}{c}{$<7.2$}&\multicolumn{2}{c}{$<12.6$}&
\multicolumn{2}{c}{$<13.8$}&\multicolumn{2}{c}{$<15.3$}\\
J1443+3623&5.29&146&2&191&2&204&6&396&6&3029&36&9.4&1.0&12.6&1.9&15.3&4.6&
\multicolumn{2}{c}{$<12.9$}&\multicolumn{2}{c}{$<18.9$}\\
J1510+5148&5.11&124&2&113&1&109&4&137&5&770&34&\multicolumn{2}{c}{$<3.6$}&
\multicolumn{2}{c}{$<5.7$}&\multicolumn{2}{c}{$<14.7$}&
\multicolumn{2}{c}{$<13.5$}&\multicolumn{2}{c}{$<15.6$}\\
J1524+0816&5.08&39&2&45&2&27&6&61&11&343&43&\multicolumn{2}{c}{$<3.6$}&
\multicolumn{2}{c}{$<6.3$}&\multicolumn{2}{c}{$<14.1$}&
\multicolumn{2}{c}{$<14.1$}&\multicolumn{2}{c}{$<18.9$}\\
J1602+4228&6.07&135&2&157&2&126&5&159&6&840&35&7.7&1.1&13.8&2.4&10.9&4.5&
10.5&4.6&\multicolumn{2}{c}{$<17.7$}\\
J1614+4640&5.31&138&1&184&1&161&5&228&5&998&51&5.1&1.4&
\multicolumn{2}{c}{$<7.2$}&\multicolumn{2}{c}{$<15.3$}&
\multicolumn{2}{c}{$<15.6$}&\multicolumn{2}{c}{$<21.0$}\\
J1623+3112&6.25&74&2&97&2&71&5&89&6&623&34&2.4&0.6\tablenotemark{a}&
\multicolumn{2}{c}{$<3.6$\tablenotemark{a}}&\multicolumn{2}{c}{$<15.6$}&
\multicolumn{2}{c}{$<14.7$}&\multicolumn{2}{c}{$<16.8$}\\
J1626+2751&5.30&325&1&395&1&367&6&498&6&2672&56&8.5&1.3&13.1&2.0&19.9&4.6&
28.4&5.7&19.9&6.2\\
J1626+2858&5.02&81&2&70&2&72&6&110&6&717&44&2.3&0.6\tablenotemark{a}&
\multicolumn{2}{c}{$<3.0$\tablenotemark{a}}&\multicolumn{2}{c}{$<15.9$}&
\multicolumn{2}{c}{$<15.3$}&\multicolumn{2}{c}{$<15.9$}\\
J1630+4012&6.07&37&2&43&2&26&5&37&6&148&21&\multicolumn{2}{c}{$<4.2$}&
\multicolumn{2}{c}{$<7.2$}&\multicolumn{2}{c}{$<14.4$}&
\multicolumn{2}{c}{$<15.0$}&\multicolumn{2}{c}{$<17.4$}\\
J1659+2709&5.32&135&3&163&2&163&6&234&8&1858&48&6.8&1.2&7.2&2.1&
\multicolumn{2}{c}{$<17.9$}&\multicolumn{2}{c}{$<19.8$}&
\multicolumn{2}{c}{$<20.1$}\\
J2054$-$0005&6.04&\multicolumn{2}{c}{\nodata}&\multicolumn{2}{c}{\nodata}&
\multicolumn{2}{c}{\nodata}&\multicolumn{2}{c}{\nodata}&
\multicolumn{2}{c}{\nodata}&3.1&1.0&10.5&2.0&15.2&5.4&12.0&4.9&
\multicolumn{2}{c}{$<19.5$}\\
J2119+1029&5.18&53&1&65&3&63&5&88&5&586&24&2.0&0.5\tablenotemark{a}&
\multicolumn{2}{c}{$<3.3$\tablenotemark{a}}&\multicolumn{2}{c}{$<14.4$}&
\multicolumn{2}{c}{$<13.6$}&\multicolumn{2}{c}{$<17.1$}\\
J2228$-$0757&5.14&93&2&120&2&90&7&88&7&250&22&
\multicolumn{2}{c}{$<1.5$\tablenotemark{a}}&
\multicolumn{2}{c}{$<3.9$\tablenotemark{a}}&\multicolumn{2}{c}{$<13.8$}&
\multicolumn{2}{c}{$<14.4$}&\multicolumn{2}{c}{$<15.0$}\\
J2245+0024&5.17&23&2&29&1&27&5&36&5&96&24&\multicolumn{2}{c}{$<3.9$}&
\multicolumn{2}{c}{$<6.6$}&\multicolumn{2}{c}{$<13.8$}&
\multicolumn{2}{c}{$<15.0$}&\multicolumn{2}{c}{$<15.6$}\\
J2315$-$0023&6.12&33&1&40&1&32&5&37&4&158&22&4.9&1.2&\multicolumn{2}{c}{$<6.3$}&
\multicolumn{2}{c}{$<14.4$}&\multicolumn{2}{c}{$<14.4$}&
\multicolumn{2}{c}{$<18.3$}
\enddata
\tablenotetext{a}{Based on the deeper observations available for these objects.}
\tablecomments{Col.: (1) Source name; (2) Redshift;  (3)-(7) 
Photometry in the {\it Spitzer} bands in $\mu$mJy. (8)-(12) Photometry in 
the {\it Herschel} bands in mJy.}
\end{deluxetable*}
\end{center}

%% file: latex_table_optical_nir_results_new_plus_ew.tex
\begin{table*}[t]
\scriptsize
\begin{center}
\caption{UV/optical and NIR properties.\label{tab_uvopt_ew}}
\begin{tabular}{l r@{.}l r@{.}l r@{.}l c r@{.}l r@{.}l c c c r@{.}l}
\tableline \tableline
Source & \multicolumn{2}{c}{$\alpha$} & \multicolumn{2}{c}{$L_{\rm UV/opt}$} & \multicolumn{2}{c}{${\nu}L_{\nu,{\rm 5100\AA}}$} & $T_{\rm NIR}$ &  \multicolumn{2}{c}{$L_{\rm NIR}$} & \multicolumn{2}{c}{${\nu}L_{\nu,6.7\mu {\rm m}}$} & EW Ly$\alpha$ & reference & EW H$\alpha$ & \multicolumn{2}{c}{F$_{{\nu}, {\rm cont}}$}\\
   & \multicolumn{2}{c}{} & \multicolumn{2}{c}{($10^{46}$erg\,s$^{-1}$)} & \multicolumn{2}{c}{($10^{46}$erg\,s$^{-1}$)} & (K) &  \multicolumn{2}{c}{($10^{46}$erg\,s$^{-1}$)} & \multicolumn{2}{c}{($10^{46}$erg\,s$^{-1}$)} & \AA &  & \AA & \multicolumn{2}{c}{}\\
(1) & \multicolumn{2}{c}{(2)} & \multicolumn{2}{c}{(3)} & \multicolumn{2}{c}{(4)} & (5) & \multicolumn{2}{c}{(6)} & \multicolumn{2}{c}{(7)} & (8) & (9) & (10) & \multicolumn{2}{c}{(11)}\\
 \tableline
J0002+2550                 &                      -0&45  &                      10&29  &                       3&23  &                       1076  &                       1&25  &                       3&78  &      60.0  &                6  &       336  &                       12&1  \\
J0005$-$0006               &                      -0&07  &                       3&04  &                       0&70  &       500\tablenotemark{a}  &                       0&02  &                    $<$0&76  &      81.5  &               10  &       455  &                        3&2  \\
J0017$-$1000               &                      -0&48  &                      10&29  &                       3&28  &                       1110  &                       2&41  &      4&41\tablenotemark{b}  &      55.2  &               11  &   \nodata  &\multicolumn{2}{c}{\nodata}  \\
J0054$-$0109               &                      -0&17  &                       6&67  &                       1&68  &                       1127  &                       0&60  &                       1&93  &      12.3  &               11  &   \nodata  &\multicolumn{2}{c}{\nodata}  \\
J0133+0106                 &                      -0&48  &                       4&44  &                       1&42  &                       1036  &                       0&29  &                    $<$1&66  &   \nodata  &          \nodata  &       689  &                        6&2  \\
J0203+0012                 &                      -0&20  &                      10&40  &                       2&68  &                       1079  &                       1&23  &                    $<$3&24  &      35.9  &               10  &       -21  &                        9&0  \\
J0231$-$0728               &                      -0&29  &                      12&25  &                       3&40  &       500\tablenotemark{a}  &                       0&36  &                    $<$2&49  &      83.8  &               11  &       500  &                       13&3  \\
J0303$-$0019               &                      -0&08  &                       3&50  &                       0&81  &                       1600  &                       0&26  &                    $<$1&16  &     139.4  &               10  &       714  &                        2&5  \\
J0338+0021                 &                      -0&29  &                       5&90  &                       1&64  &                       1205  &                       2&38  &                       5&43  &      42.5  &               11  &   \nodata  &\multicolumn{2}{c}{\nodata}  \\
J0353+0104                 &                      -0&67  &                       4&11  &                       1&49  &                       1042  &                       0&57  &                    $<$2&95  &   \nodata  &               10  &       307  &                        5&9  \\
J0731+4459                 &                      -0&17  &                      12&39  &                       3&12  &                       1184  &                       3&02  &                    $<$4&33  &      57.4  &               11  &   \nodata  &\multicolumn{2}{c}{\nodata}  \\
J0756+4104                 &                      -0&23  &                       5&92  &                       1&57  &                       1251  &                       1&55  &                       3&31  &      30.5  &               11  &   \nodata  &\multicolumn{2}{c}{\nodata}  \\
J0818+1722                 &                      -0&47  &                      14&70  &                       4&66  &                       1156  &                       2&05  &                    $<$5&23  &      10.0  &                8  &       309  &                       16&6  \\
J0833+2726                 &                      -0&63  &                       2&76  &                       0&98  &                       1039  &                       0&52  &                    $<$2&00  &   \nodata  &          \nodata  &   \nodata  &\multicolumn{2}{c}{\nodata}  \\
J0836+0054                 &                      -0&56  &                      19&84  &                       6&67  &       500\tablenotemark{a}  &                       0&59  &                       5&58  &      70.0  &                2  &       724  &                       26&9  \\
J0840+5624                 &                      -0&06  &                       8&19  &                       1&88  &                       1040  &                       0&78  &                    $<$3&91  &   \nodata  &          \nodata  &       281  &                        5&7  \\
J0841+2905                 &                      -0&35  &                       4&71  &                       1&37  &                       1344  &                       1&79  &                    $<$3&45  &      58.0  &                9  &       166  &                        4&7  \\
J0842+1218                 &                      -0&35  &                       9&93  &                       2&89  &                       1042  &                       2&53  &                       7&18  &   \nodata  &          \nodata  &       238  &                        8&4  \\
J0846+0800                 &                      -0&15  &                       6&02  &                       1&50  &                       1291  &                       1&19  &                    $<$2&14  &      28.4  &               11  &   \nodata  &\multicolumn{2}{c}{\nodata}  \\
J0901+6942                 &                      -0&29  &                       6&54  &                       1&82  &                       1115  &                       0&95  &                    $<$2&72  &      55.0  &                5  &       301  &                        6&5  \\
J0902+0851                 &                      -0&44  &                       4&57  &                       1&42  &                       1093  &                       0&59  &                    $<$1&95  &     109.6  &               11  &       598  &                        6&5  \\
J0913+5919                 &                      -0&16  &                       3&55  &                       0&89  &                       1210  &                       0&56  &                    $<$1&42  &     110.9  &               11  &   \nodata  &\multicolumn{2}{c}{\nodata}  \\
J0915+4924                 &                       0&01  &                      11&36  &                       2&46  &                       1203  &                       1&19  &                    $<$2&35  &      78.1  &               11  &       462  &                        8&8  \\
J0922+2653                 &                      -0&03  &                       5&31  &                       1&19  &                       1316  &                       0&97  &                    $<$1&76  &      57.8  &               11  &   \nodata  &\multicolumn{2}{c}{\nodata}  \\
J0927+2001                 &                       0&00  &                       6&13  &                       1&33  &                       1321  &                       2&04  &                    $<$3&47  &   \nodata  &          \nodata  &       141  &                        4&5  \\
J0957+0610                 &                      -0&28  &                      10&60  &                       2&91  &                       1360  &                       3&17  &                       4&32  &      51.4  &               11  &   \nodata  &\multicolumn{2}{c}{\nodata}  \\
J1013+4240                 &                       0&03  &                       6&36  &                       1&35  &                       1256  &                       0&66  &                    $<$1&50  &      41.5  &               11  &   \nodata  &\multicolumn{2}{c}{\nodata}  \\
J1030+0524                 &                      -0&12  &                       7&17  &                       1&73  &                       1550  &                       2&04  &                    $<$3&54  &      70.0  &                2  &       670  &                        6&3  \\
J1044$-$0125               &                      -0&34  &                       9&88  &                       2&85  &                       1329  &                       4&49  &                       7&00  &      26.0  &                1  &       213  &                       10&7  \\
J1048+4637                 &                      -0&28  &                      11&11  &                       3&05  &                       1311  &                       2&67  &      4&21\tablenotemark{b}  &      40.0  &                4  &       275  &                       10&3  \\
J1119+3452                 &                      -0&07  &                       6&73  &                       1&57  &                       1339  &                       1&50  &                    $<$2&22  &      33.8  &               11  &   \nodata  &\multicolumn{2}{c}{\nodata}  \\
J1132+1209                 &                      -0&49  &                      10&32  &                       3&33  &                       1315  &                       2&79  &                       4&94  &      40.8  &               11  &   \nodata  &\multicolumn{2}{c}{\nodata}  \\
J1137+3549                 &                      -0&20  &                      10&58  &                       2&72  &                        994  &                       0&96  &                    $<$3&79  &   \nodata  &          \nodata  &       202  &                        8&7  \\
J1146+4037                 &                      -0&40  &                      11&69  &                       3&52  &                       1098  &                       1&04  &                    $<$2&58  &      58.6  &               11  &   \nodata  &\multicolumn{2}{c}{\nodata}  \\
J1148+5251                 &                      -0&48  &                      14&04  &                       4&48  &                       1362  &                       4&80  &                       7&01  &      25.0  &                4  &        33  &                       14&0  \\
J1148+5253                 &                      -0&81  &                       0&56  &                       0&22  &      1100\tablenotemark{a}  &                    $<$0&18  &                    $<$1&17  &   \nodata  &          \nodata  &        29  &                        1&3  \\
J1154+1341                 &                      -0&37  &                       4&73  &                       1&39  &                       1273  &                       1&03  &                    $<$2&01  &      51.9  &               11  &   \nodata  &\multicolumn{2}{c}{\nodata}  \\
J1202+3235                 &                      -0&07  &                      17&16  &                       3&98  &                       1238  &                       3&78  &                       6&84  &      16.1  &               11  &       141  &                       13&6  \\
J1204$-$0021               &                      -0&24  &                       9&83  &                       2&63  &                       1224  &                       2&72  &                       5&93  &      53.9  &               11  &   \nodata  &\multicolumn{2}{c}{\nodata}  \\
J1208+0010                 &                       0&04  &                       3&23  &                       0&68  &       869\tablenotemark{a}  &                    $<$0&06  &                    $<$0&72  &   \nodata  &          \nodata  &       592  &                        2&3  \\
J1221+4445                 &                      -0&38  &                       7&55  &                       2&25  &                       1292  &                       1&59  &                    $<$2&69  &     105.8  &               11  &   \nodata  &\multicolumn{2}{c}{\nodata}  \\
J1242+5213                 &                      -0&32  &                       6&53  &                       1&85  &       500\tablenotemark{a}  &                       0&15  &                    $<$1&41  &      49.3  &               11  &   \nodata  &\multicolumn{2}{c}{\nodata}  \\
J1250+3130                 &                      -0&69  &                       7&10  &                       2&61  &                       1043  &                       2&63  &                    $<$5&99  &   \nodata  &          \nodata  &       334  &                        8&8  \\
J1306+0356                 &                      -0&17  &                       6&97  &                       1&75  &       692\tablenotemark{a}  &                       0&64  &                    $<$2&75  &      60.0  &                2  &       399  &                        6&3  \\
J1334+1220                 &                      -0&21  &                       8&04  &                       2&10  &                       1250  &                       2&48  &                       4&20  &      49.7  &               11  &   \nodata  &\multicolumn{2}{c}{\nodata}  \\
J1335+3533                 &                      -0&24  &                       6&15  &                       1&64  &       537\tablenotemark{a}  &                       0&69  &                    $<$2&94  &      -5.0  &                8  &       212  &                        6&1  \\
J1337+4155                 &                      -0&00  &                       7&70  &                       1&68  &                       1220  &                       1&16  &                    $<$2&26  &      78.8  &               11  &   \nodata  &\multicolumn{2}{c}{\nodata}  \\
J1340+3926                 &                      -0&25  &                       9&02  &                       2&42  &                       1177  &                       2&41  &                       4&68  &      59.6  &               11  &   \nodata  &\multicolumn{2}{c}{\nodata}  \\
J1340+2813                 &                      -0&59  &                      10&18  &                       3&50  &                       1159  &                       2&72  &                       7&06  &      69.3  &               11  &       241  &                       14&7  \\
J1341+4611                 &                      -0&36  &                       3&89  &                       1&13  &                       1142  &                       0&82  &                    $<$2&17  &     110.7  &               11  &   \nodata  &\multicolumn{2}{c}{\nodata}  \\
J1411+1217                 &                      -0&48  &                       6&69  &                       2&14  &       500\tablenotemark{a}  &                    $<$0&08  &                    $<$1&81  &     100.0  &                6  &       783  &                        8&7  \\
J1423+1303                 &                      -0&23  &                       8&83  &                       2&33  &                       1235  &                       2&01  &                    $<$3&01  &      48.4  &               11  &   \nodata  &\multicolumn{2}{c}{\nodata}  \\
J1427+3312                 &                      -0&26  &                       6&73  &                       1&82  &                        694  &                       0&32  &                    $<$3&20  &   \nodata  &                7  &       123  &                        6&5  \\
J1436+5007                 &                      -0&06  &                       4&57  &                       1&05  &                       1407  &                       1&33  &                    $<$2&52  &   \nodata  &          \nodata  &       316  &                        4&0  \\
J1443+3623                 &                      -0&46  &                      13&16  &                       4&15  &                       1253  &                       7&22  &                      10&43  &      28.3  &               11  &       221  &                       17&1  \\
J1510+5148                 &                      -0&32  &                       8&61  &                       2&44  &                       1056  &                       1&05  &                    $<$2&87  &      72.4  &               11  &   \nodata  &\multicolumn{2}{c}{\nodata}  \\
J1524+0816                 &                      -0&14  &                       2&27  &                       0&56  &                       1351  &                       0&95  &                    $<$1&70  &   \nodata  &          \nodata  &   \nodata  &\multicolumn{2}{c}{\nodata}  \\
J1602+4228                 &                      -0&61  &                       9&60  &                       3&35  &                       1067  &                       1&44  &                       6&53  &   \nodata  &          \nodata  &       292  &                       13&1  \\
J1614+4640                 &                      -0&72  &                       8&15  &                       3&04  &                       1147  &                       1&58  &                       4&22  &      61.7  &               11  &       424  &                       14&8  \\
J1623+3112                 &                      -0&57  &                       5&97  &                       2&03  &                       1063  &                       1&22  &      3&43\tablenotemark{b}  &     150.0  &                6  &       499  &                        7&3  \\
J1626+2751                 &                      -0&52  &                      23&54  &                       7&74  &                       1117  &                       4&34  &                       9&33  &      45.6  &               11  &       268  &                       34&4  \\
J1626+2858                 &                      -0&20  &                       6&33  &                       1&63  &                       1171  &                       1&32  &      2&23\tablenotemark{b}  &      18.9  &               11  &   \nodata  &\multicolumn{2}{c}{\nodata}  \\
J1630+4012                 &                      -0&09  &                       3&59  &                       0&84  &                       1505  &                       0&61  &                    $<$1&97  &      70.0  &                4  &       539  &                        3&1  \\
J1659+2709                 &                      -0&20  &                      16&00  &                       4&13  &                       1157  &                       3&67  &                       6&92  &      30.3  &               11  &       181  &                       14&7  \\
J2054$-$0005               &\multicolumn{2}{c}{\nodata}  &\multicolumn{2}{c}{\nodata}  &\multicolumn{2}{c}{\nodata}  &                    \nodata  &\multicolumn{2}{c}{\nodata}  &\multicolumn{2}{c}{\nodata}  &      17.0  &               10  &   \nodata  &\multicolumn{2}{c}{\nodata}  \\
J2119+1029                 &                      -0&31  &                       5&07  &                       1&43  &                       1130  &                       1&01  &      2&00\tablenotemark{b}  &   \nodata  &          \nodata  &   \nodata  &\multicolumn{2}{c}{\nodata}  \\
J2228$-$0757               &                      -0&35  &                       6&36  &                       1&85  &       500\tablenotemark{a}  &                       0&13  &   $<$1&04\tablenotemark{b}  &     115.7  &               11  &   \nodata  &\multicolumn{2}{c}{\nodata}  \\
J2245+0024                 &                      -0&14  &                       2&72  &                       0&67  &                       1371  &                       0&24  &                    $<$0&84  &     115.0  &                3  &   \nodata  &\multicolumn{2}{c}{\nodata}  \\
J2315$-$0023               &                      -0&36  &                       3&29  &                       0&96  &                       1145  &                       0&32  &                       2&26  &     126.8  &               10  &       332  &                        3&3  \\
\tableline
\end{tabular}
\tablecomments{(1) source name; 
(2) UV/optical power-law slope ($F_{\nu}$\,$\propto$\,${\nu}^{\alpha}$). Typical uncertainty is $\pm0.05$; 
(3) UV/optical luminosity determined by integrating the power-law component between 0.1\,$\mu$m and 1\,$\mu$m {\bf (not corrected for extinction)}. Typical uncertainty is $\pm10\%$; 
(4) monochromatic luminosity at 5100\,\AA (rest frame) determined from the power-law component {\bf (not corrected for extinction)}; 
(5) temperature of the fitted blackbody in the NIR. Parameter limits during fitting were 500\,K and 1600\,K. Typical uncertainty is $\pm50$\,K; 
(6) NIR luminosity determined by integrating the blackbody component between 1\,$\mu$m and 3\,$\mu$m. Typical uncertainty is $\pm10\%$; 
(7) monochromatic luminosity at 6.7$\mu$m (rest frame) determined from the power-law interpolation (in  $\nu F_{\nu}$) between the observed bands at 24\,$\mu$m and 100\,$\mu$m; 
(8) Ly$\alpha$ equivalent width; 
(9) reference for Ly$\alpha$ equivalent width; 
(10) H$\alpha$ equivalent width (in \AA) determined from the photometry as described in the text; 
(11) interpolated continuum flux at the position of the redshifted H$\alpha$ line given in 10$^{-28}$\,erg\,s$^{-1}$\,cm$^{-2}$\,Hz$^{-1}$. \newline $^{\rm a}$\,For these objects, $T_{\rm NIR}$ is not well defined (see text for details); $^{\rm b}$\,Based on the deeper {\it Herschel} observations available for these objects.}
\tablerefs{
(1) \citealt{fan00b}; 
(2) \citealt{fan01}; 
(3) \citealt{sha01}; 
(4) \citealt{fan03}; 
(5) \citealt{rom04}; 
(6) \citealt{fan04}; 
(7) \citealt{mcg06}; 
(8) \citealt{fan06}; 
(9) \citealt{got06}; 
(10) \citealt{jia08}; 
(11) \citealt{dia09}.
}
\end{center}
\end{table*}

%% file: dusty_young_universe_catalog_revised.bbl
\begin{thebibliography}{}

\bibitem[Balog et al.(2013)]{bal13} Balog, Z., M{\"u}ller, T., Nielbock, M., et al.\ 2013, Experimental Astronomy, 38 

\bibitem[Barvainis(1987)]{bar87} Barvainis, R.\ 1987, \apj, 320, 537 

\bibitem[Beelen et al.(2006)]{bee06} Beelen, A., Cox, P., Benford, D.~J., et al.\ 2006, \apj, 642, 694

\bibitem[Bendo et al.(2013)]{ben13} Bendo, G.~J., Griffin, M.~J., Bock, J.~J., et al.\ 2013, \mnras, 433, 3062 

\bibitem[Berta et al.(2011)]{ber11} Berta, S., Magnelli, B., Nordon, R., et al.\ 2011, \aap, 532, A49 

\bibitem[Bertoldi et al.(2003)]{ber03} Bertoldi, F., Carilli, C.~L., Cox, P., et al.\ 2003, \aap, 406, L55 

\bibitem[Blain et al.(2013)]{bla13} Blain, A.~W., Assef, R., Stern, D., et al.\ 2013, \apj, 778, 113 

\bibitem[Carilli et al.(2001)]{car01} Carilli, C.~L., Bertoldi, F., Rupen, M.~P., et al.\ 2001, \apj, 555, 625 

\bibitem[Dai et al.(2012)]{dai12} Dai, Y.~S., Bergeron, J., Elvis, M., et al.\ 2012, \apj, 753, 33 

\bibitem[De Rosa et al.(2011)]{der11} De Rosa, G., Decarli, R., Walter, F., et al.\ 2011, \apj, 739, 56 

\bibitem[Diamond-Stanic et al.(2009)]{dia09} Diamond-Stanic, A.~M., Fan, X., Brandt, W.~N., et al.\ 2009, \apj, 699, 782 

\bibitem[Elbaz et al.(2011)]{elb11} Elbaz, D., Dickinson, M., Hwang, H.~S., et al.\ 2011, \aap, 533, A119 

\bibitem[Fan et al.(1999)]{fan99} Fan, X., Strauss, M.~A., Schneider, D.~P., et al.\ 1999, \aj, 118, 1 

\bibitem[Fan et al.(2000a)]{fan00a} Fan, X., Strauss, M.~A., Schneider, D.~P., et al.\ 2000, \aj, 119, 1 

\bibitem[Fan et al.(2000b)]{fan00b} Fan, X., White, R.~L., Davis, M., et al.\ 2000, \aj, 120, 1167 

\bibitem[Fan et al.(2001)]{fan01} Fan, X., Narayanan, V.~K., Lupton, R.~H., et al.\ 2001, \aj, 122, 2833 

\bibitem[Fan et al.(2003)]{fan03} Fan, X., Strauss, M.~A., Schneider, D.~P., et al.\ 2003, \aj, 125, 1649 
\bibitem[Fan et al.(2004)]{fan04} Fan, X., Hennawi, J.~F., Richards, G.~T., et al.\ 2004, \aj, 128, 515

\bibitem[Fan et al.(2006)]{fan06} Fan, X., Strauss, M.~A., Richards, G.~T., et al.\ 2006, \aj, 131, 1203 

\bibitem[Fazio et al.(2004)]{faz04} Fazio, G.~G., Hora, J.~L., Allen, L.~E., et al.\ 2004, \apjs, 154, 10 

\bibitem[Freudling et al.(2003)]{fre03} Freudling, W., Corbin, M.~R., \& Korista, K.~T.\ 2003, \apjl, 587, L67 
\bibitem[Fruchter \& Hook(2002)]{fru02} Fruchter, A.~S., \& Hook, R.~N.\ 2002, \pasp, 114, 144 

\bibitem[Goto(2006)]{got06} Goto, T.\ 2006, \mnras, 371, 769 

\bibitem[Griffin et al.(2010)]{gri10} Griffin, M.~J., Abergel, A., Abreu, A., et al.\ 2010, \aap, 518, L3 

\bibitem[Hao et al.(2011)]{hao11} Hao, H., Elvis, M., Civano, F., \& Lawrence, A.\ 2011, \apj, 733, 108 

\bibitem[Hasinger(2008)]{has08} Hasinger, G.\ 2008, \aap, 490, 905 

\bibitem[Hewett et al.(2006)]{hew06} Hewett, P.~C., Warren, S.~J., Leggett, S.~K., \& Hodgkin, S.~T.\ 2006, \mnras, 367, 454 

\bibitem[Hines et al.(2006)]{hin06} Hines, D.~C., Backman, D.~E., Bouwman, J., et al.\ 2006, \apj, 638, 1070 

\bibitem[H{\"o}nig \& Kishimoto(2010)]{hon10} H{\"o}nig, S.~F., \& Kishimoto, M.\ 2010, \aap, 523, A27

\bibitem[Houck et al.(2004)]{hou04} Houck, J.~R., Roellig, T.~L., van Cleve, J., Forrest, W.~J., Herter, T., Lawrence, C.~R., Matthews, K., et al.\ 2004, \apjs, 154, 18

\bibitem[Isobe et al.(1986)]{iso86} Isobe, T., Feigelson, E.~D., \& Nelson, P.~I.\ 1986, \apj, 306, 490 

\bibitem[Iwamuro et al.(2004)]{iwa04} Iwamuro, F., Kimura, M., Eto, S., et al.\ 2004, \apj, 614, 69 

\bibitem[Jiang et al.(2006)]{jia06} Jiang, L., Fan, X., Hines, D.~C., et al.\ 2006, \aj, 132, 2127 
\bibitem[Jiang et al.(2007)]{jia07} Jiang, L., Fan, X., Vestergaard, M., et al.\ 2007, \aj, 134, 1150 

\bibitem[Jiang et al.(2008)]{jia08} Jiang, L., Fan, X., Annis, J., et al.\ 2008, \aj, 135, 1057 

\bibitem[Jiang et al.(2010)]{jia10} Jiang, L., Fan, X., Brandt, W.~N., et al.\ 2010, \nat, 464, 380 

\bibitem[Juarez et al.(2009)]{jua09} Juarez, Y., Maiolino, R., Mujica, R., et al.\ 2009, \aap, 494, L25 

\bibitem[Jun \& Im(2013)]{jun13} Jun, H.~D., \& Im, M.\ 2013, arXiv:1310.3034 


\bibitem[Kaiser et al.(2010)]{kai10} Kaiser, N., Burgett, W., Chambers, K., et al.\ 2010, \procspie, 7733,  
\bibitem[Kennicutt(1998)]{ken98} Kennicutt, R.~C., Jr.\ 1998, \araa, 36, 189

\bibitem[Kurk et al.(2007)]{kur07} Kurk, J.~D., Walter, F., Fan, X., et al.\ 2007, \apj, 669, 32 
\bibitem[Kurk et al.(2009)]{kur09} Kurk, J.~D., Walter, F., Fan, X., et al.\ 2009, \apj, 702, 833 

\bibitem[Lawrence(1991)]{law91} Lawrence, A.\ 1991, \mnras, 252, 586 

\bibitem[Lawrence et al.(2007)]{law07} Lawrence, A., Warren, S.~J., Almaini, O., et al.\ 2007, \mnras, 379, 1599

\bibitem[Leipski et al.(2013)]{lei13} Leipski, C., Meisenheimer, K., Walter, F., et al.\ 2013, \apj, 772, 103 


\bibitem[Lusso et al.(2013)]{lus13} Lusso, E., Hennawi, J.~F., Comastri, A., et al.\ 2013, \apj, 777, 86 

\bibitem[Lutz et al.(2008)]{lut08} Lutz, D., Sturm, E., Tacconi, L.~J., et al.\ 2008, \apj, 684, 853 
\bibitem[Lutz et al.(2011)]{lut11} Lutz, D., Poglitsch, A., Altieri, B., et al.\ 2011, \aap, 532, A90 

\bibitem[Mahabal et al.(2005)]{mah05} Mahabal, A., Stern, D., Bogosavljevi{\'c}, M., Djorgovski, S.~G., \& Thompson, D.\ 2005, \apjl, 634, L9 

\bibitem[Maiolino et al.(2003)]{mai03} Maiolino, R., Juarez, Y., Mujica, R., Nagar, N.~M., \& Oliva, E.\ 2003, \apjl, 596, L155 

\bibitem[Maiolino et al.(2007)]{mai07} Maiolino, R., Shemmer, O., Imanishi, M., et al.\ 2007, \aap, 468, 979 

\bibitem[McGreer et al.(2006)]{mcg06} McGreer, I.~D., Becker, R.~H., Helfand, D.~J., \& White, R.~L.\ 2006, \apj, 652, 157 

\bibitem[McGreer et al.(2013)]{mcg13} McGreer, I.~D., Jiang, L., Fan, X., et al.\ 2013, \apj, 768, 105 

\bibitem[Mor et al.(2009)]{mor09} Mor, R., Netzer, H., \& Elitzur, M.\ 2009, \apj, 705, 298 

\bibitem[Mor \& Trakhtenbrot(2011)]{mor11} Mor, R., \& Trakhtenbrot, B.\ 2011, \apjl, 737, L36 

\bibitem[Mor \& Netzer(2012)]{mor12} Mor, R., \& Netzer, H.\ 2012, \mnras, 420, 526 

\bibitem[Nenkova et al.(2008)]{nen08} Nenkova, M., Sirocky, M.~M., Nikutta, R., Ivezi{\'c}, {\v Z}., \& Elitzur, M.\ 2008, \apj, 685, 160 

\bibitem[Netzer et al.(2007)]{net07} Netzer, H., Lutz, D., Schweitzer, M., et al.\ 2007, \apj, 666, 806 

\bibitem[Netzer et al.(2013)]{net13} Netzer, H., Mor, R., Trakhtenbrot, B., Shemmer, O., \& Lira, P.\ 2013, arXiv:1308.0012 

\bibitem[Nguyen et al.(2010)]{ngu10} Nguyen, H.~T., Schulz, B., Levenson, L., et al.\ 2010, \aap, 518, L5 

\bibitem[Omont et al.(1996)]{omo96} Omont, A., McMahon, R.~G., Cox, P., et al.\ 1996, \aap, 315, 1 


\bibitem[Ott(2010)]{ott10} Ott, S.\ 2010, Astronomical Data Analysis Software and Systems XIX, 434, 139 

\bibitem[Pascale et al.(2011)]{pas11} Pascale, E., Auld, R., Dariush, A., et al.\ 2011, \mnras, 415, 911 

\bibitem[Petric et al.(2003)]{pet03} Petric, A.~O., Carilli, C.~L., Bertoldi, F., et al.\ 2003, \aj, 126, 15

\bibitem[Pilbratt et al.(2010)]{pil10} Pilbratt, G.~L., Riedinger, J.~R., Passvogel, T., et al.\ 2010, \aap, 518, L1

\bibitem[Priddey et al.(2003)]{pri03} Priddey, R.~S., Isaak, K.~G., McMahon, R.~G., Robson, E.~I., \& Pearson, C.~P.\ 2003, \mnras, 344, L74 

\bibitem[Priddey et al.(2008)]{pri08} Priddey, R.~S., Ivison, R.~J., \& Isaak, K.~G.\ 2008, \mnras, 383, 289

\bibitem[Poglitsch et al.(2010)]{pog10} Poglitsch, A., Waelkens, C., Geis, N., et al.\ 2010, \aap, 518, L2 

\bibitem[Popesso et al.(2012)]{pop12} Popesso, P., Magnelli, B., Buttiglione, S., et al.\ 2012, arXiv:1211.4257 

\bibitem[Richards et al.(2006)]{ric06} Richards, G.~T., Lacy, M., Storrie-Lombardi, L.~J., et al.\ 2006, \apjs, 166, 470

\bibitem[Rieke et al.(2004)]{rie04} Rieke, G.~H., Young, E.~T., Engelbracht, C.~W., et al.\ 2004, \apjs, 154, 25 

\bibitem[Robson et al.(2004)]{rob04} Robson, I., Priddey, R.~S., Isaak, K.~G., \& McMahon, R.~G.\ 2004, \mnras, 351, L29

\bibitem[Romani et al.(2004)]{rom04} Romani, R.~W., Sowards-Emmerd, D., Greenhill, L., \& Michelson, P.\ 2004, \apjl, 610, L9 

\bibitem[Roseboom et al.(2010)]{ros10} Roseboom, I.~G., Oliver, S.~J., Kunz, M., et al.\ 2010, \mnras, 409, 48 

\bibitem[Sanders et al.(1989)]{san89} Sanders, D.~B., Phinney, E.~S., Neugebauer, G., Soifer, B.~T., \& Matthews, K.\ 1989, \apj, 347, 29 

\bibitem[Savage \& Oliver(2007)]{sav07} Savage, R.~S., \& Oliver, S.\ 2007, \apj, 661, 1339 

\bibitem[Schartmann et al.(2008)]{sch08} Schartmann, M., Meisenheimer, K., Camenzind, M., et al.\ 2008, \aap, 482, 67 

\bibitem[Schlegel et al.(1998)]{sch98} Schlegel, D.~J., Finkbeiner, D.~P., \& Davis, M.\ 1998, \apj, 500, 525 

\bibitem[Sharp et al.(2001)]{sha01} Sharp, R.~G., McMahon, R.~G., Irwin, M.~J., \& Hodgkin, S.~T.\ 2001, \mnras, 326, L45 

\bibitem[Shen et al.(2011)]{she11} Shen, Y., Richards, G.~T., Strauss, M.~A., et al.\ 2011, \apjs, 194, 45 

\bibitem[Stalevski et al.(2012)]{sta12} Stalevski, M., Fritz, J., Baes, M., Nakos, T., \& Popovi{\'c}, L.~{\v C}.\ 2012, \mnras, 420, 2756 

\bibitem[Treister \& Urry(2006)]{tre06} Treister, E., \& Urry, C.~M.\ 2006, \apjl, 652, L79 


\bibitem[Treister et al.(2008)]{tre08} Treister, E., Krolik, J.~H., \& Dullemond, C.\ 2008, \apj, 679, 140 

\bibitem[Ueda et al.(2003)]{ued03} Ueda, Y., Akiyama, M., Ohta, K., \& Miyaji, T.\ 2003, \apj, 598, 886 


\bibitem[Valiante et al.(2011)]{val11} Valiante, R., Schneider, R., Salvadori, S., \& Bianchi, S.\ 2011, \mnras, 416, 1916 

\bibitem[Vanden Berk et al.(2001)]{van01} Vanden Berk, D.~E., Richards, G.~T., Bauer, A., et al.\ 2001, \aj, 122, 549 

\bibitem[Venemans et al.(2012)]{ven12} Venemans, B.~P., McMahon, R.~G., Walter, F., et al.\ 2012, \apjl, 751, L25 

\bibitem[Walter et al.(2009)]{wal09} Walter, F., Riechers, D., Cox, P., et al.\ 2009, \nat, 457, 699 

\bibitem[Wang et al.(2007)]{wan07} Wang, R., Carilli, C.~L., Beelen, A., et al.\ 2007, \aj, 134, 617 
\bibitem[Wang et al.(2008a)]{wan08a} Wang, R., Wagg, J., Carilli, C.~L., et al.\ 2008a, \aj, 135, 1201 % 350mu

\bibitem[Wang et al.(2008b)]{wan08b} Wang, R., Carilli, C.~L., Wagg, J., et al.\ 2008b, \apj, 687, 848 % 1mm

\bibitem[Wang et al.(2010)]{wan10} Wang, R., Carilli, C.~L., Neri, R., et al.\ 2010, \apj, 714, 699 
\bibitem[Wang et al.(2011)]{wan11} Wang, R., Wagg, J., Carilli, C.~L., et al.\ 2011, \aj, 142, 101 
\bibitem[Wang et al.(2013)]{wan13} Wang, R., Wagg, J., Carilli, C.~L., et al.\ 2013, \apj, 773, 44 

\bibitem[Werner et al.(2004)]{wer04} Werner, M.~W., Roellig, T.~L., Low, F.~J.,
et al.\ 2004, \apjs, 154, 1

\bibitem[Willott et al.(2003)]{wil03} Willott, C.~J., McLure, R.~J.,  \& Jarvis, M.~J.\ 2003, \apjl, 587, L15
\bibitem[Willott et al.(2013)]{wil13} Willott, C.~J., Omont, A., \& Bergeron, J.\ 2013, \apj, 770, 13 

 
\bibitem[Zheng et al.(2000)]{zhe00} Zheng, W., Tsvetanov, Z.~I., Schneider, D.~P., et al.\ 2000, \aj, 120, 1607 


\end{thebibliography}
